\documentclass[acmsmall]{acmart}

\usepackage{hyperref}    
\usepackage{url}            
\usepackage{booktabs}       
\usepackage{amsfonts}       
\usepackage{xfrac}       
\usepackage{microtype}      
\usepackage{bbm}
\usepackage{adjustbox}
\usepackage{mathtools}
\usepackage{amsmath,amsthm}
\usepackage{footnote}

\usepackage{graphicx}
\graphicspath{{./Figures/}}
\usepackage{multirow}
\usepackage{pifont}

\usepackage{subcaption}
\usepackage{array}

\newcommand{\Upl}{\operatorname{Upl}}
\newcommand{\bigCI}{\mathrel{\text{\scalebox{1.07}{$\perp\mkern-10mu\perp$}}}}

\newtheorem{assumption}{Assumption}

\AtBeginDocument{%
	\providecommand\BibTeX{{%
			\normalfont B\kern-0.5em{\scshape i\kern-0.25em b}\kern-0.8em\TeX}}}

\setcopyright{acmcopyright}
\copyrightyear{2021}
\acmYear{2021}
\acmDOI{10.1145/12345}

\acmJournal{CSUR}
\acmVolume{xx}
\acmNumber{x}
\acmArticle{123}
\acmMonth{8}

\begin{document}
\title{A unified survey of treatment effect heterogeneity modelling and uplift modelling}

\author{Weijia Zhang}

\orcid{0001-8103-5325}
\affiliation{%
	\department{School of Computer Science and Engineering}
	\institution{Southeast University}
	\city{Nanjing}
	\postcode{210096}
	\country{China}
}
\email{weijia.zhang.xh@gmail.com}
\affiliation{%
	\department{UniSA STEM}
	\institution{University of South Australia}
	\city{Adelaide}
	\postcode{5095}
	\country{Australia}
}

\author{Jiuyong Li}
\affiliation{%
	\department{UniSA STEM}
	\institution{University of South Australia}
	\city{Adelaide}
	\postcode{5095}
	\country{Australia}}
\email{jiuyong.li@unisa.edu.au}

\author{Lin Liu}
\affiliation{%
	\department{UniSA STEM}
	\institution{University of South Australia}
	\city{Adelaide}
	\postcode{5095}
	\country{Australia}}
\email{lin.liu@unisa.edu.au}

\thanks{This work is partially supported by the Australian Research Council Discovery Projects Grant DP170101306.}

\begin{abstract}
	A central question in many fields of scientific research is to determine how an outcome is affected by an action, i.e., to estimate the causal effect or treatment effect of an action.
	In recent years, in areas such as personalised healthcare, sociology, and online marketing, a need has emerged to estimate heterogeneous treatment effects with respect to individuals of different characteristics.
	To meet this need, two major approaches have been taken: treatment effect heterogeneity modelling and uplifting modelling. Researchers and practitioners in different communities have developed algorithms based on these approaches to estimate the effect of heterogeneous treatment.
	In this paper, we present a unified view of these two seemingly disconnected yet closely related approaches under the potential outcome framework. 
	We provide a structured survey of existing methods following either of the two approaches, emphasising their inherent connections and using unified notation to facilitate comparisons. 
	We also review the main applications of the surveyed methods in personalised marketing, personalised medicine, and sociology.
	Finally, we summarise and discuss the available software packages and source codes in terms of their coverage of different methods and applicability to different datasets, and we provide general guidelines for method selection.
\end{abstract}


\maketitle

\section{Introduction}
A fundamental question for scientific research and applications in many disciplines is to determine whether and to what extent changing the value of one variable (i.e., a treatment) affects the value of another variable (i.e., an outcome), which is one of the main tasks in causal inference \cite{Imbens2015,Morgan2015}.
To answer this question, we need to estimate the causal effect of the treatment on the outcome. 
For example, an oncologist might estimate the average causal effect that a cancer therapy has on prognosis outcomes, such as the expected survival time after treatment. 
An employment assistance project might study the average causal effect of a job training program on employment prospects, i.e., whether the program reduces unemployment. 
An online retailer might want to model the average causal effect that an advertisement will have on sales.

However, it is often insufficient merely to determine the average causal effect.
For example, a cancer patient may be more interested in individual-level causal effects, asking ``Would this treatment be effective for a patient like me with a specific gene mutation?'', since many cancer treatments are known to be effective for those with certain gene expression patterns \citep{Bellon2015}.
From the perspective of policymakers, it is more reasonable to offer a job training program to those who will benefit from it, since the program's effectiveness may depend on the education background, experience, and employment history of the participants. 
For an online retailer, it is also preferable to target only persuadable customers in order to reduce advertisement costs, and to avoid disturbing customers who do not wish to receive unsolicited advertisements \citep{Rzepakowski2012_UpliftMarketing}.


Motivated by the different application scenarios, researchers from two closely related yet surprisingly isolated research communities---the treatment effect heterogeneity modelling and the uplift modelling communities---have contributed significantly  to solving this causal inference problem regarding conditional causal effects, i.e., causal effects in different sub-populations.
Many methods have been developed by the treatment effect heterogeneity modelling community \citep{Su2009,Hill2011,Su2012,Imai2013,Athey2015,Louizos2017,Atan2018,Wager2018,Zhang2017,Hassanpour2018,Kuenzel2019} and the uplift modelling community \citep{Hansotia2002,DiemertEustacheBetleiArtem,Radcliffe2007,Rzepakowski2010_DTuplift,Guelman2012,Zaniewicz2013,Soltys2015,Guelman2015,Gutierrez2017,Yamane2018}.

\begin{figure}
	\begin{adjustbox}{minipage=\linewidth,scale=1}
	\includegraphics[width=\linewidth]{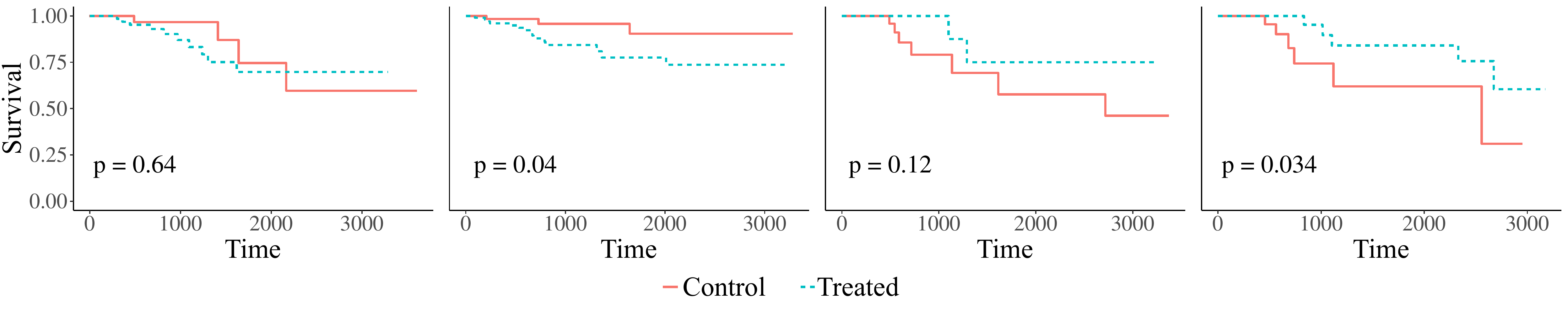}
	\end{adjustbox}
	\caption{
	An example of the heterogeneous treatment effects of radiotherapy on the survival of breast cancer patients \citep{Zhang2017}. Patients in four different subgroups are distinguished by their gene expressions and have different survival times among treated patients (dashed lines) and control patients (solid lines). This means that patients in different subgroups may respond differently to radiotherapy.
}
\label{bio-example}
\end{figure}

An exemplar application of treatment effect heterogeneity modelling is personalised medical treatment, where the goal is to determine whether a treatment is effective for a patient with certain characteristics, such as a certain gene expression profile.
A causal tree-based method was developed to model the heterogeneous effects of radiotherapy on the survival time of breast cancer patients~\citep{Zhang2017}. Four subgroups of patients characterised by their gene expressions responded differently to the radiotherapy treatment. The Kaplan--Meier survival curves~\citep{Kaplan1958} of the subgroups are shown in Figure \ref{bio-example}. From the curves, we can see that the effects of the treatment are positive for patients in the third and fourth subgroups, negative for those in the second subgroup, and marginal for patients in the first subgroup. Correctly identifying treatment effect heterogeneity is beneficial for both patients and healthcare systems. 

A typical application of uplift modelling is targeted advertising in online marketing, where the goal is to predict whether a promotion will be effective for customers with a certain purchase history and certain preferences. Uplift is a term in the marketing application and refers to the differences in the purchasing behaviour between customers who are offered the promotion (treated) and those who are not (control). 
As we will discuss in Section 2, with some assumptions, uplift is the conditional average causal effect of a treatment (promotion) on the outcome (purchase behaviour) for a given sub-population of customers with certain characteristics (e.g. purchase history and preferences).

\begin{figure}[!t]
	\centering
	\begin{adjustbox}{minipage=\linewidth,scale=0.7}
	\includegraphics[width=0.95\linewidth]{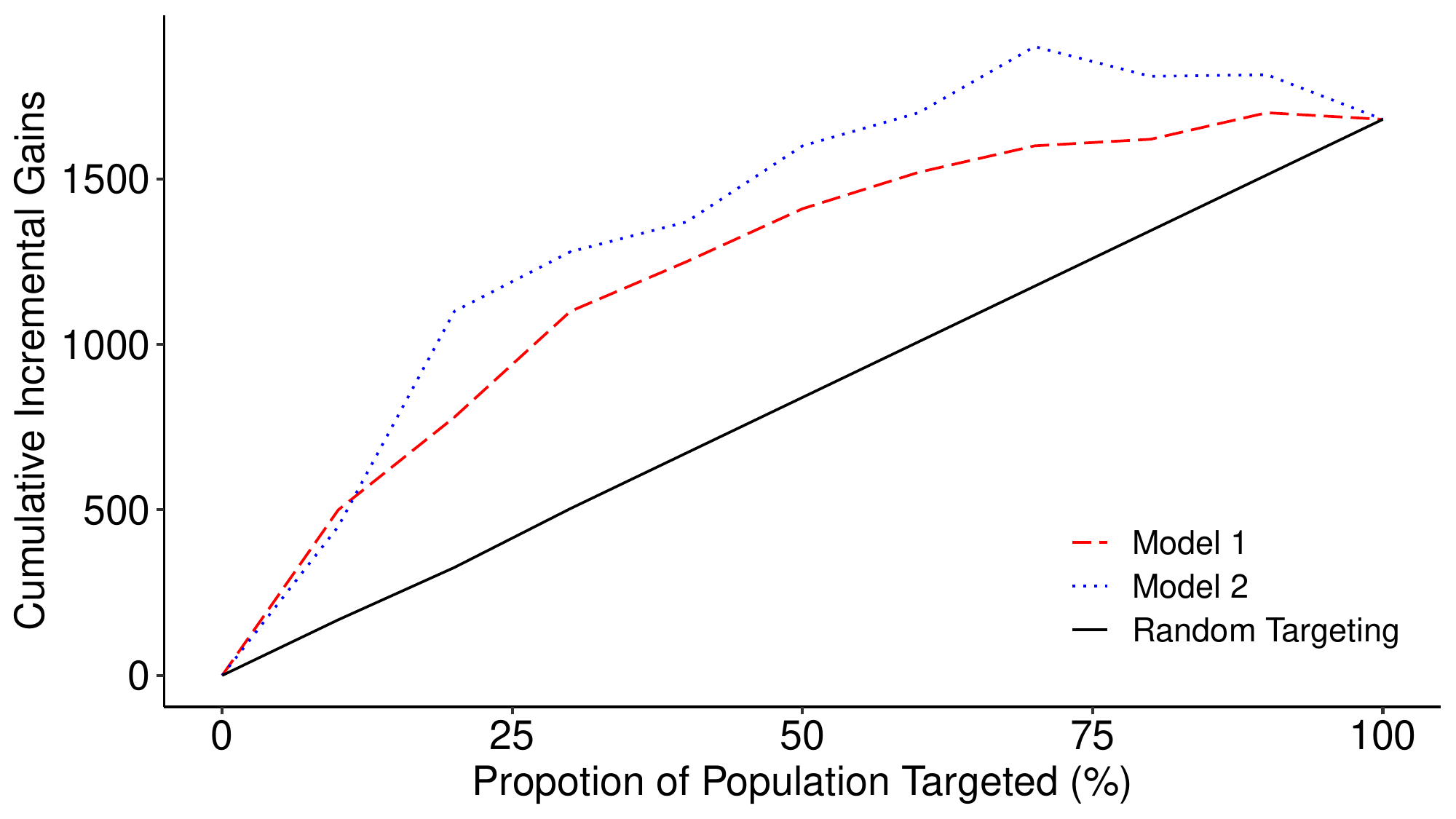}
	\end{adjustbox}
	\caption{Incremental gain curves showing the increase in sales from targeting different proportions of customers with two uplift modelling models versus random targeting using data from an online retailer~\citep{Gubela2019}. 
	}
	\label{uplift-example}
\end{figure}

Figure \ref{uplift-example} shows the incremental gain curves of smartphone sales by promotional emails from an e-commerce company with two uplift modelling methods~\citep{Gubela2019}. We can see that when targeting the same proportion (10\% to 100\%) of customers, both uplift models achieved higher sales increases over random targeting. In Figure 2, Model 2 (blue dotted line) performs better than Model 1 (red dashed line) except when targeting only the first 10\% of the customers.

Treatment effect heterogeneity modelling and uplift modelling share a common objective: to estimate the change in the outcome caused by the change of the treatment for some given subjects---e.g., changes in the survival time that result from radiotherapy treatment, or changes in purchasing behaviour that result from promotional emails.
However, a distinct difference among the algorithms from the two communities is that those from the uplift modelling community are implicitly designed for data from randomised experiments, and no assumptions for using the methods are explicitly discussed. By contrast, methods developed in the treatment effect heterogeneity modelling community are explicitly specified with assumptions and they can be used for experimental and observational data that satisfy these assumptions.

Over the last few years, several surveys have been published in the uplift modelling community \citep{Gutierrez2017,Devriendt2018,Gubela2019}. Gutierrez and G\'erardy \citep{Gutierrez2017} were the first to briefly discuss the link between uplift modelling and treatment effect heterogeneity modelling. However, to the best of our knowledge, no literature explicitly discusses the necessary assumptions for unifying the methods from the two communities.
Furthermore, none of the existing work has discussed the connections between the methods developed by the two communities using unified notation.
Researchers from the two communities have been working in parallel, and the progress in one community is not contributing to the progress of the other.
This article contributes to bridging the gap between the two communities and accelerating the research progress
by showing that the methods, benchmarks, evaluation metrics, and applications in both communities can be used to cross-fertilize each other.

Causal inference is a diverse research area, where the potential outcome framework \citep{Rubin1974} and the structural causal models \citep{Pearl2009} are two of the most influential paradigms. 
This work follows the potential outcome framework for estimating the causal effects of a treatment on an outcome, which is a central task in causal inference. 
Treatment effects can be estimated from either experimental or observational data \citep{Imbens2015,Morgan2015}.
We focus on estimating the conditional average treatment effect (CATE) within different sub-populations, whereas most treatment effect estimation methods estimate the average treatment effect in a general population.
This article presents a unified survey of the CATE estimation methods developed by the treatment effect heterogeneity modelling community and the uplift modelling community.
For more general topics on data-driven causal inference, including learning structural causal models and estimating average causal effects, please refer to the surveys of \citet{Knaus2020} and \citet{Guo2020}, and the textbook of \citet{Peters2019}.

The survey contributes to the literature in the following ways. 
Firstly, we explicitly present and discuss the fundamental assumptions needed for unifying the methods proposed by the two communities under the potential outcome framework \citep{Rubin1974}.
Secondly, we survey and discuss the methods proposed by both communities using unified notation, which allows us to clarify the connections and distinctions between the methods developed by the two communities.
Thirdly, we discuss the feasibility of the methods by evaluating software packages developed by both communities, and we outline the challenges for both treatment effect heterogeneity modelling and uplift modelling.


\section{Unifying Treatment Effect Heterogeneity Modelling and Uplift Modelling}

In this section, we discuss and unify the definitions, assumptions, and objectives for treatment effect heterogeneity modelling and uplift modelling under the potential outcome framework \citep{Rubin1974}.

\subsection{Objectives of Treatment Effect Heterogeneity Modelling and Uplift Modelling}
We use $T \in \{0,1\}$ to denote a binary treatment variable,
$T = 0$ for having no treatment (control), and $T= 1$ for having treatment (treated). 
Under the potential outcome framework, every subject $i$ has two \textit{potential outcomes}:  
$Y_i(0)$, the potential outcome if the subject had received no treatment; and $Y_i(1)$, the potential outcome if the subject had received the treatment. 
Potential outcomes can be either continuous or discrete. 
For example, in cancer treatment, $Y_i(0)$ corresponds to a continuous variable indicating the number of years the patient would have survived without treatment, and $Y_i(1)$ corresponds to the number of years the patient would have survived with the treatment.
In marketing, binary potential outcomes are used to indicate whether a customer would purchase a product if that customer were offered a promotion ($Y_i(1)$) or not offered the promotion ($Y_i(0)$). 

For subject $i$, the individual treatment effect (ITE) of a treatment $T$, denoted as $\tau_i$, is defined as the difference between the two potential outcomes:
\begin{align}
\tau_i \vcentcolon= Y_i(1) - Y_i(0).
\label{ITE}
\end{align}

In an ideal world, if we could determine the ITE $\tau_i$ as defined in Equation (\ref{ITE}), then the ultimate goal of both treatment effect heterogeneity modelling and uplift modelling, i.e., predicting individual treatment effect, would be fulfilled, and we would know exactly whether an individual should be prescribed a treatment.



Unfortunately, only one of the two potential outcomes can be observed for any subject. For example, if we observe the potential outcome of a cancer patient who receives a radiotherapy treatment, it will be impossible for us to observe the potential outcome of the same patient receiving no treatment.
The unobserved potential outcome is often referred as the \textit{counterfactual} of the observed outcome.
We use $Y_i$ to denote the observed outcome of subject $i$, which can be expressed using the interaction between the treatment and the two potential outcomes:
\begin{equation}
Y_i = T Y_i(1) + (1-T)  Y_i(0). 
\end{equation}
Hereafter, when the context is clear, we drop the subscript $i$ for simplicity of notation. 

Since we never observe both potential outcomes for any subject, ITE $\tau_i$ is not identifiable.
With some adequate assumptions, however, it is possible to estimate the conditional average treatment effect (CATE), which is the average treatment effect conditioning on a set of covariates that describe the subjects. 
Specifically, let $X = \pmb{x} \in\mathbb{R}^k$ be a $k$-dimensional covariate vector describing the pre-treatment characteristics of a group of subjects. The CATE, denoted $\tau(\pmb{x})$, is defined as:
\begin{align}
\tau (\pmb{x}) \vcentcolon= \mathbb{E}[\tau \mid X=\pmb{x}] = \mathbb{E}[Y(1)-Y(0) \mid X=\pmb{x}].
\label{CATE}
\end{align}
Equation (\ref{CATE}) is frequently used as the objective of treatment effect heterogeneity modelling methods \citep{Athey2016,Zhang2017,Kuenzel2019}. Although the CATE is not the same as the ITE, it has been shown that the CATE is the best estimator for the ITE in terms of the mean squared error \citep{Kuenzel2019}.

Uplift modelling techniques assume that data are obtained from experiments with randomised treatment assignment, and have the following objective:
\begin{align}
\Upl (\pmb{x}) = \mathbb{E}(Y \mid T=1,X=\pmb{x}) - \mathbb{E}(Y \mid T=0,X=\pmb{x}).
\label{uplift_In_data_2}
\end{align}

From Equation (\ref{CATE}), the objective of treatment effect heterogeneity modelling involves the conditional expectations of two potential outcomes, and thus cannot be directly estimated from data. 
By contrast, the objective of uplift modelling (Equation (\ref{uplift_In_data_2})) involves two conditional expectations of the observed outcomes and thus can be estimated from data, but assuming the use of randomised experiment data for unbiased estimation. 
In the next section, we discuss the link between the objectives and the assumptions of treatment effect heterogeneity modelling and uplift modelling. 

\subsection{Linking the Objectives}

We can re-arrange the objective of treatment effect heterogeneity modelling, i.e., Equation (\ref{CATE}), using rules of conditional probability and conditional expectations, as follows: 
\begin{align}
\tau(\pmb{x}) = & \mathbb{E}[Y(1) - Y(0) \mid X=\pmb{x}] \nonumber\\
=& \mathbb{E}[Y(1)  \mid T=1,X=\pmb{x}] P(T=1 \mid X=\pmb{x}) + \mathbb{E}[Y(1) \mid T=0,X=\pmb{x}]P(T=0 \mid X=\pmb{x}) \nonumber\\
& -  \mathbb{E}(Y(0) \mid T=1,X=\pmb{x})P(T=1\mid X=\pmb{x}) - \mathbb{E}[Y(0) \mid T=0,X=\pmb{x}]P(T=0\mid X=\pmb{x}) \nonumber\\
=& \mathbb{E}[Y(1)  \mid T=1,X=\pmb{x}] P(T=1\mid X=\pmb{x}) + \mathbb{E}[Y(1) \mid T=0,X=\pmb{x}]P(T=0\mid X=\pmb{x}) \nonumber\\
& -  \mathbb{E}(Y(0) \mid T=1,X=\pmb{x})P(T=1\mid X=\pmb{x}) - \mathbb{E}[Y(0) \mid T=0,X=\pmb{x}]P(T=0\mid X=\pmb{x}) \nonumber\\
& + \mathbb{E}[Y(1)  \mid T=1,X=\pmb{x}] P(T=0\mid X=\pmb{x}) - \mathbb{E}[Y(1)  \mid T=1,X=\pmb{x}] P(T=0\mid X=\pmb{x})\nonumber\\
& + \mathbb{E}[Y(0) \mid T=0,X=\pmb{x}]P(T=1\mid X=\pmb{x}) - \mathbb{E}[Y(0) \mid T=0,X=\pmb{x}]P(T=1\mid X=\pmb{x})\nonumber\\
=& \mathbb{E}[Y(1)  \mid T=1,X=\pmb{x}] P(T=1\mid X=\pmb{x}) + \mathbb{E}[Y(1)  \mid T=1,X=\pmb{x}] P(T=0\mid X=\pmb{x})\nonumber\\
& - \mathbb{E}[Y(0) \mid T=0,X=\pmb{x}]P(T=0\mid X=\pmb{x}) - \mathbb{E}[Y(0) \mid T=0,X=\pmb{x}]P(T=1\mid X=\pmb{x})\nonumber\\
& + \mathbb{E}[Y(0) \mid T=0,X=\pmb{x}]P(T=1\mid X=\pmb{x})  -  \mathbb{E}(Y(0) \mid T=1,X=\pmb{x})P(T=1\mid X=\pmb{x}) \nonumber\\
& + \mathbb{E}[Y(1) \mid T=0,X=\pmb{x}]P(T=0\mid X=\pmb{x}) - \mathbb{E}[Y(1)  \mid T=1,X=\pmb{x}] P(T=0\mid X=\pmb{x}),\nonumber
\end{align}
\noindent which gives us:

\begin{align}
\tau(\pmb{x})=& \underbrace{\mathbb{E}[Y(1) \mid T=1,X=\pmb{x}] - \mathbb{E}[Y(0) \mid T=0,X=\pmb{x}]}_\text{observed} 
\nonumber\\
&+  P(T=1\mid X=\pmb{x}) \{ \underbrace{ \mathbb{E}[Y(0) \mid T=0,X=\pmb{x}]}_\text{observed} - \underbrace{\mathbb{E}[Y(0) \mid T=1,X=\pmb{x}] }_\text{unobserved} \} 
\nonumber \\
& + P(T=0\mid X=\pmb{x}) \{\underbrace{ \mathbb{E}[Y(1) \mid T=0,X=\pmb{x}]}_\text{unobserved} - \underbrace{\mathbb{E}[Y(1) \mid T=1,X=\pmb{x}]}_\text{observed} \}.
\label{CATE-decompose}
\end{align}

The above process decomposes the objective into three components. 
For the two conditional expectations in the first line of Equation (\ref{CATE-decompose}), observing that the potential outcome of treatment (or control) equals the observed outcome when conditioning on $T=1$ (or $T=0$), we have:
\begin{align}
\mathbb{E}[Y(1) \mid T=1,X=\pmb{x}] - \mathbb{E}[Y(0) \mid T=0,X=\pmb{x}] = \mathbb{E}[Y \mid T=1,X=\pmb{x}] - \mathbb{E}[Y \mid T=0,X=\pmb{x}].
\label{CATE_observed}
\end{align}
It is worth noting that the expectations on the right-hand side of Equation (\ref{CATE_observed}) are the same as the uplift modelling objective in Equation (\ref{uplift_In_data_2}). 
Furthermore, both expectations only involve observed outcomes without counterfactuals. Their values can be estimated without bias using the overlap, the stable unit treatment value (SUTV), and the unconfoundedness assumption described below. 

\begin{assumption}
	(\textbf{Overlap)} Any subject has a non-zero probability of receiving the treatment and the control. In other words, for all $\pmb{x}$ in the support of $X$, we have:
	\begin{align}
	0< P(T=1 \mid X=\pmb{x})<1.
	\end{align}
	\label{overslap}
\end{assumption}
The overlap assumption states that the probability of any subject with covariates $\pmb{x}$ being treated is bounded away from $0$ and $1$. 
This ensures that all types of individuals have been observed in both treatment and control groups. This is necessary because if subjects with some covariate value $\pmb{x}$ always receive treatment (or control) in the data, the expectations cannot be estimated. 
\begin{assumption}
	(\textbf{SUTV)} The stable unit treatment value (SUTV) assumption states that the individuals do not interfere with each other. In other words, a treatment applied to one subject does not affect the outcome of other subjects. 
	\label{SUTV}
\end{assumption}
The SUTV assumption is usually satisfied in health and bioinformatics applications since it is reasonable to assume that giving radiotherapy to a patient will not affect the life expectancy of other patients.
However, the assumption should be considered carefully in marketing where treatment to one subject may affect
the outcomes of other subjects. For example, in online advertising, there is no guarantee that sending an email promotion (treatment) to an individual will not affect other individuals' knowledge of the promotion. For example, a treated individual could forward the email to friends and thus their decision to purchase the product might change.

At this point, we have seen that the differences between the objective of treatment effect heterogeneity modelling and the objective of uplift modelling lie in the last two components of Equation (\ref{CATE-decompose}). 
These components are not estimable from data, since each of them involves an unobservable potential outcome (counterfactual), $\mathbb{E}[Y(0) \mid T=1,X=\pmb{x}]$, the conditional average of potential outcome $Y(0)$ in the treatment group, and $\mathbb{E}[Y(1) \mid T=0,X=\pmb{x}]$, the conditional average of potential outcome $Y(1)$ in the control group.
To overcome the counterfactual problem, methods in the treatment effect heterogeneity modelling literature have introduced an important assumption, namely, that potential outcomes are independent of treatment assignment when conditioning on a set of covariates. This is known as the unconfoundedness assumption:

\begin{assumption}
	(\textbf{Unconfoundedness)} The distribution of treatment is independent of the distribution of potential outcomes when conditioning on a set of observed variables. Formally, we have:
	\begin{align}
	(Y(0), Y(1)) \bigCI T  \mid  X=\pmb{x}.
	\end{align}
	\label{unconfound}
\end{assumption}
\vspace{-15pt}
The unconfoundedness assumption is also often referred as the strong ignorability assumption in the literature.
By applying the unconfoundedness assumption to the last two components of Equation (\ref{CATE-decompose}), we get the following results:
\begin{align}
\mathbb{E}[Y(0) \mid T=0,\pmb{x}] - \mathbb{E}[Y(0) \mid T=1,\pmb{x}] =  \mathbb{E}[Y(0) \mid \pmb{x}] - \mathbb{E}[Y(0) \mid \pmb{x}] = 0,
\\
\mathbb{E}[Y(1) \mid T=0,\pmb{x}] - \mathbb{E}[Y(1) \mid T=1,\pmb{x}] = \mathbb{E}[Y(1) \mid \pmb{x}] - \mathbb{E}[Y(1) \mid \pmb{x}] = 0,
\label{bias_elimination}
\end{align}

\noindent and  
the objective of treatment effect heterogeneity can be written as:
\begin{align}
\tau (\pmb{x}) = \mathbb{E}[Y \mid T=1,X=\pmb{x}] - \mathbb{E}[Y \mid T=0,X=\pmb{x}],
\label{CATE_In_data}
\end{align}

\noindent which is the same as the objective of uplift modelling given in Equation  (\ref{uplift_In_data_2}). 

Therefore, when the assumptions of overlap, SUTV, and unconfoundedness are all satisfied, the objective of treatment effect heterogeneity modelling (Equation (\ref{CATE})) and the objective of uplift modelling (Equation (\ref{uplift_In_data_2})) are the same.
However, it is worth mentioning that the above three assumptions are not explicitly stated or discussed in the majority of uplift modelling literature. 
If the data do not satisfy these assumptions, the estimated uplift will be biased, since Equation (\ref{uplift_In_data_2}) does not correspond to the true causal effect of an action on the outcome.



In practice, it is challenging to determine whether the unconfoundedness assumption is satisfied, as it is untestable from data. In other words, it is hard to know if the covariate set used is correct when exploring treatment heterogeneity or building uplift models.
This is an important related topic, but one that is outside the scope of our survey. 
A simple criterion for covariate selection is to have all direct causes of outcome $Y$ as covariates and exclude all effect variables of $Y$ from the covariate set, as shown in~\citep{generalframework2020}. For an in-depth discussion of the topic, we refer the reader to \citet{VanderWeele-NewCriteria,de2011covariate,entner2013data,maathuis2015generalized}.


\section{Methods for CATE Estimation}
In this section, assuming that the data satisfy the overlap, SUTV and unconfoundedness assumptions (i.e., Assumptions \ref{overslap}, \ref{SUTV}, and \ref{unconfound}), we provide an extensive survey of the existing treatment effect heterogeneity modelling and uplift modelling algorithms using unified notation. 
We categorise the methods for treatment effect heterogeneity modelling and uplift modelling into two major categories. The first category consists of methods that extend existing supervised learning methods for CATE estimation, and the second category consists of tailored methods for treatment effect heterogeneity modelling or uplift modelling.

\subsection{Methods Extending Existing Supervised Learning Models}
\label{sec_meta}
\subsubsection{The Single-model Approach}
\label{sec_single}

Estimating the CATE in Equation (\ref{CATE_In_data}) can be achieved by estimating conditional expectations (or probabilities) from data. The problem reduces to a regression or classification problem. 
Specifically, given a dataset $D$ for $(Y,X,T)$ that satisfies Assumptions 1--3, the single-model approach uses the concatenation of treatment and covariates $[T,X]$ as the features, and $Y$ as the target to train a supervised model, where $Y = \hat{\mu}(T, X)$ from $D$, e.g., a regression model if $Y$ is a continuous variable, or a classification model if $Y$ is a categorical variable. Then, the trained model is used to predict the CATE for a subject described by covariates $\pmb{x}$ as follows:
\begin{equation}
\hat{\tau}(\pmb{x}) = \hat{\mu}([T=1,\pmb{x}]) -\hat{\mu}([T =0,\pmb{x}]),
\end{equation}
where $\pmb{x}$ is the short-hand form for $X=\pmb{x}$. 

Lo~\citep{Lo2002_TrueLiftModel} used two estimators, a logistic regression model and a neural network with one hidden layer, to estimate the uplift in a market experimental dataset. Lo \citep{Lo2002_TrueLiftModel} also proposed that standard supervised methods such as linear regression, regression tree, and spline regression could be used as the estimators.
Athey and Imbens~\citep{Athey2015} implemented a single-model method using a regression tree for treatment effect heterogeneity modelling. 

The single-model approach is simple, easy to implement, and has the flexibility of being able to use any off-the-shelf supervised learning algorithm. 
However, a major drawback to this approach is that a single model may not model both  potential outcomes well, and hence the estimation of CATE may be biased. Another problem with the single-model approach is that $T$ may not be selected by a model that only uses a subset of the features for prediction (such as a tree model), and thus the CATE will be estimated as zero for all subjects. 


\subsubsection{The Two-Model Approach}
\label{sec_two}
An improvement to the single-model approach is to model the two potential outcomes using two separate models.
The two-model approach trains two models on the control and treated subjects separately, and then uses the difference of the two predictions as the estimated conditional average causal effect or uplift.

Specifically, given a dataset $D$ for $(Y, X,T)$ that satisfies Assumptions 1--3, the two-model approach estimates the CATE as:
\begin{align*}
\hat{\tau}(\pmb{x}) & = \mathbb{E}[Y|T=1,X =\pmb{x}] - \mathbb{E}[Y \mid T=0,X=\pmb{x}] \\
&= \hat{\mu}_1(\pmb{x}) - \hat{\mu}_0(\pmb{x}),
\end{align*}
where $\hat{\mu}_1$ is trained using the sub-dataset of $D$ containing the samples of  treated subjects only, and $\hat{\mu}_0$ is trained using the sub-dataset of $D$ containing the samples of control subjects only. 

Any off-the-shelf estimator can be used to estimate $\hat{\mu}_1(\pmb{x})$ and $\hat{\mu}_0(\pmb{x})$. Popular choices include linear regression, as in ~\citep{Hansotia2002,TreatmentSElection2011}; regression trees \citep{Breiman1984}, as in the two-tree method in~\citep{Athey2015}; decision trees~\citep{Quinlan1993_C45}, as in~\citep{Soltys2015}; rule-based methods, as in \citep{Nassif2012,Nassif2013}; gradient boosting trees, as in \citep{Kane2014}; and Bayesian additive regression trees (BART) \citep{Chipman2010}, as in \citep{Hill2011}. 


The two-model approach is simple and flexible. The models for the control and treated subjects can be built straightforwardly using a wide range of off-the-shelf estimators. 
The freedom of choice in the estimators also provides flexibility for modelling various treatment and outcome relationships. 
However, as the two models are built separately, they do not utilise the information shared by the control and treated subjects. Furthermore, they cannot mitigate the impact of the disparity in covariate distributions between the treatment and control groups on CATE estimation \citep{Athey2015}.

\subsubsection{X-Learner}
\label{sec_X}
\citet{Kuenzel2019} proposed an improvement to the two-model approach called the X-Learner. 
A motivation for the X-Learner approach is that the sample size in the treatment group is usually very small, and thus the estimator $\hat{\mu}_1(\pmb{x})$ may not be modelled accurately. 
The X-Learner addresses this problem by information crossover between treatment and control groups.

Specifically, the X-Learner consists of three steps. First, two separate estimators, $\hat{\mu}_1(\pmb{x})$ and $\hat{\mu}_0(\pmb{x})$, are built using the subjects from the treatment and control groups, respectively, similar to the model building process in the two-model approach. 
Then, the treatment effect for a subject in the treatment group, denoted as $\hat{\tau}_{1i}$, is inputted using the observed outcome and the estimator $\hat{\mu}_0(\pmb{x})$ from the control group. The treatment effects for a subject in the control group, denoted as $\hat{\tau}_{0i}$, is inputted using the observed outcome and the estimator  $\hat{\mu}_1(\pmb{x})$ from the treatment group. 
That is, 
\begin{align*}
\hat{\tau}_{1i} &= Y_i - \hat{\mu}_0(\pmb{x}_i), \mbox{for subject $i$ belonging to the treated group, and}\\
\hat{\tau}_{0i} &=  \hat{\mu}_1(\pmb{x}_i) - Y_i, \mbox{for subject $i$ belonging to the control group}.
\end{align*}
Now we have two sets of inputted CATE estimations: $\hat{\tau}_1$, corresponding to the CATE estimations for subjects in the treatment group; and $\hat{\tau}_0$, corresponding to the CATE estimations for subjects in the control group.
Using these two sets of inputted CATE estimations, the X-Learner builds two estimators, $\hat{\tau}_1(\pmb{x})$ and $\hat{\tau}_0(\pmb{x})$, using covariates $\pmb{x}$, with $\hat{\tau}_{1}$ and $\hat{\tau}_{0}$, respectively.  
Finally, the CATE is estimated using a weighted average of the two estimators:
\begin{equation}
\hat{\tau}(\pmb{x}) = e(\pmb{x})\hat{\tau}_0(\pmb{x}) + (1-e(\pmb{x}))\hat{\tau}_1(\pmb{x}),
\label{formula_xlearner}
\end{equation}
where $e(\pmb{x}) \in [0,1]$ is a weight function defined as the estimated probability of a subject receiving the treatment, i.e., $\hat{e}(\pmb{x})=P(T=1|\pmb{x})$, as suggested in \citet{Kuenzel2019}.
This probability is referred to as the propensity score \cite{Rubin1997_PropensityScore} and in practice is often estimated using logistic regression, decision trees, neural networks \citep{Setoguchi2008}, or boosting algorithms \citep{McCaffrey2004}. For a comprehensive discussion of propensity score, see Austin \citep{Austin2011}.

An advantage of the X-Learner is that it cross-references the data in the treatment and control groups, and thus can perform better than the two-model approach when the number of subjects in the treatment group is significantly smaller than that in the control group. However, the X-Learner requires building four estimators, twice as many as the two-model approach. This increases the risk of overfitting and the difficulty tuning parameters.

\subsubsection{R-Learner}

Recently, \citet{Nie2020} proposed the R-Learner, which is a class of two-step algorithms designed for treatment effect heterogeneity modelling.
R-Learner first estimates the treatment outcome and control outcomes $\hat{\mu}(\pmb{x})$ along with the propensity score $\hat{e}(\pmb{x})$ using a base learner trained by $k$-fold cross-validation, e.g., penalised regression, neural networks, or boosting.
Then, it estimates the CATE by minimising the following loss function:
\begin{align}
	L[\hat{\tau}(\pmb{x})] = \frac{1}{n} \sum\limits_{i=1}^N \{[(Y_i - \hat{\mu}^{(-i)}(\pmb{x}_i)] - [T_i - \hat{e}^{(-i)}(\pmb{x}_i)] \hat{\tau}(\pmb{x}_i)\}^2,
\end{align}
where $\hat{\mu}^{(-i)}(\pmb{x}_i)$ and $\hat{e}^{(-i)}(\pmb{x}_i)$ denote the predictions made from the samples excluding the cross-validation fold that the $i$-th training sample belongs.

An advantage of the R-Learner is that it has guaranteed error bounds when using penalised kernel regression as the base learner. \citet{Nie2020} showed that the asymptotic error bound of the R-Learner is the same as the bound of an oracle learner that has access to how the outcomes and treatments are generated, but not to the ground-truth treatment effects.

\subsubsection{Transformed outcome Approach}
\label{sec_OutcomeTransform}
The transformed outcome approach transforms the observed outcome $Y$ to $Y^*$ such that the CATE equals the conditional expectation of the transformed outcome $Y^*$. 
After the transformation, an off-the-shelf estimator can be  applied to the dataset containing the original covariates and the transformed outcomes for estimating the CATE.

For example, the following transformed outcome for continuous outcomes is used in~\citep{Athey2015}:
\begin{equation}
Y^* =  \frac{Y}{e(\pmb{x})}\cdot T -  \frac{Y}  {(1-e(\pmb{x}))} \cdot (1-T),
\label{TOT}
\end{equation}
where  $Y^*$ is the transformed outcome, and $e(\pmb{x})$ is the probability of a subject with covariates $\pmb{x}$ receiving the treatment, i.e., $e(\pmb{x}) = P(T=1|\pmb{x})$.

Using the transformed outcome described above, it is straightforward to derive that the conditional expectation of the transformed outcome equals the CATE, i.e., $\mathbb{E}[Y^* |X=\pmb{x}] = \tau(\pmb{x})$.
Therefore, an off-the-shelf regression algorithm can be used to estimate the CATE by using $Y^*$ as the target and $X$ as the covariates. In the treatment effect heterogeneity modelling community, an instantiation of this approach using regression trees is discussed by \citet{Athey2015}. 

For a binary outcome, \citet{Jaskowski2012} proposed the following transformation:
\begin{equation}
Y^* = Y T + (1-Y)(1-T),
\end{equation}
where the transformed outcome $Y^*$ corresponds to one of the following cases: $Y^*=1$ when ($T=1$ and $Y = 1$) or ($T=0$ and $Y=0$); and $Y^*=0$ otherwise.
Under the assumption that $e(\pmb{x}) = 0.5$ for all $\pmb{x}$, i.e., that a subject has an equal chance to be in the treatment or the control group, \citet{Jaskowski2012} proved that the CATE can be estimated as:
\begin{equation}
\tau(\pmb{x}) = 2P(Y^*_i = 1|\pmb{x}) - 1.
\label{uplift-TOT}
\end{equation}
The transformed outcome in Equation (\ref{uplift-TOT}) can be viewed as a special case of the transformation in Equation (\ref{TOT}), where $Y$ is binary and $e(\pmb{x}) =0.5$ for all $\pmb{x}$. \citet{Jaskowski2012} built a logistic regression model using the transformed outcomes.  \citet{Weisberg2015} and \citet{incrementalityandrevenue13} also used the same transformation and logistic regression as the base learner. 

%


A main advantage of the transformed outcome approach is that, after transformation, the CATE can be modelled directly. Furthermore, it provides the flexibility for choosing any existing off-the-shelf supervised methods for CATE estimation. 
However, the transformed outcome approach relies heavily on the accurate estimation of $e(\pmb{x})$, the propensity score. In Equation (\ref{TOT}) the estimated $e(\pmb{x})$ appears in the denominator, and thus a small variation in the estimation of $e(\pmb{x})$ will lead to a large variation in the transformed outcomes.

\subsubsection{Deep Learning-Based Methods}
\label{sec_deep_learning_1}
Recently, several deep learning-based treatment effect heterogeneity modelling algorithms have been proposed~\citep{Shalit2016,Louizos2017,Yao2018_Twin,Hassanpour2018}. Here, we introduce three types of deep learning-based methods that extend the single-model approach, the two-model approach, and the X-Learner approach. Some other deep learning-based algorithms that do not fall into the above three categories, i.e., those specifically designed as neural network-based methods, are discussed later in Section \ref{sec_deep2}.

The main advantages of deep learning-based methods are that they can model complex non-linear relationships between the treatment, covariates, and the outcome, and can handle high-dimensional and large data. However, deep learning-based methods are difficult to interpret and they offer no convergence guarantees or error bounds.
Furthermore, these methods are sensitive to the selection of parameters and their parameter tuning procedures are difficult.

\paragraph{Deep-Treat~\citep{Atan2018}} 
Deep-Treat is a deep learning-based single-model approach that consists of two stages.
The first stage takes the covariates and the treatment as input, and utilises a debiasing auto-encoder \citep{Vincent2010} to learn a representation $\Phi(\pmb{x})$ of the original covariates $\pmb{x}$ such that the treatment and control groups are balanced.
Balancing of the learned representation $\Phi(\pmb{x})$ is measured by the cross-entropy loss between the marginal treatment distribution $P(T)$ and the conditional treatment distribution, given the learned representation $P(T|\Phi(\pmb{x}))$. 
In other words, the learned representations are considered balanced if the cross-entropy loss between the marginal distribution and the conditional distribution is minimised.
In the second stage, Deep-Treat uses the learned representation $\Phi(\pmb{x})$, the treatment $T$, and the outcome $Y$ as inputs, and trains a single neural network to predict the outcome $Y$ using the concatenated features $[T,\Phi(\pmb{x})]$.
The main advantage of Deep-Treat over the single-model is that Deep-Treat learns a balanced encoding for the control and treated subjects. However, it also inherits the disadvantages of the single-model approach.

\paragraph{Counterfactual Regression (CFR) \citep{Johansson:2016:LRC:3045390.3045708,Shalit2016}}
\label{sec_cfr}
CFR is a deep learning-based method that extends the two-model approach. 
CFR estimates $\tau(\pmb{x})$ by learning two functions parameterised by two neural networks (similar to the two-model approach). 
Before learning the two functions, CFR utilises representation learning to minimise the discrepancy between the two distributions, $P(\pmb{x} \mid T=0)$ and $P(\pmb{x}|T=1)$, measured by either the maximum mean discrepancy or the Wasserstein distance.

Several works have been proposed to improve upon CFR. In particular, CFR with importance sampling weights (CFR-ISW) \citep{Hassanpour2018} is based on the problem that the representation learned by CFR cannot completely eliminate bias. Thus, CFR-ISW adds a propensity network to alleviate this problem. The similarity-preserved individual treatment effect (SITE) estimation algorithm  \citep{Yao2018_Twin} improves the learning of the common representation by adding a position-dependent deep metric (PPDM) and middle-point distance minimisation (MPDM) constraints.

The main advantage of CFR and its improvements over the two-model approach is that they are designed to learn a shared and balanced feature representation across the treated and control subjects, to reduce bias in the CATE estimation. 
\subsection{Tailored Methods for Treatment Effect Heterogeneity Modelling or Uplift Modelling}

In this section, we survey methods that are specifically designed for treatment effect heterogeneity modelling or uplift modelling. We categorise the methods into four categories: tree-based methods, which build binary trees using designed splitting criteria; support vector machine (SVM)-based methods, which reformulate uplift modelling within the SVM framework; generative deep learning methods, which utilise a variational autoencoder or generative adversarial network to estimate the potential outcomes; and ensemble-based methods, which build upon tree-based methods.

\subsubsection{Tree-Based Methods}
\label{sec_tree}
Tree-based methods build binary tree models for estimating the CATE. Both treatment effect heterogeneity modelling and uplift modelling communities have developed tree-based methods separately.

The procedure of building a tree-based CATE estimation model is similar to that of building a normal regression/decision tree \citep{Breiman1984}, in the sense that they all build tree models using recursive partitioning, which, starting from the root node, recursively splits the node into two child nodes using a splitting criterion.
The major difference between existing decision/regression trees and tree-based CATE estimation algorithms lies in how they define their splitting criteria.
In the same fashion, the main difference among different tree-based CATE estimation algorithms also lies in their splitting criteria. 
Therefore, we focus our discussion on the difference in the splitting criteria of tree-based CATE estimation algorithms.


The main advantage of tree-based methods lies in the interpretability, which is very important for many applications of CATE estimation. Furthermore, tree-based methods naturally provide groups of subjects with heterogeneous CATEs as defined by the paths from the root to leaf nodes of a tree model. 
However, a major drawback to all tree-based methods is that the tree construction process is fundamentally greedy and does not return the ``optimal'' tree. One tree and an alternative tree (by slightly perturbing data) from the same algorithm can differ significantly. 

\paragraph{Uplift Incremental Value Modelling (UpliftIVM) \cite{Hansotia2002}}
UpliftIVM is one of the earliest tree-based methods proposed in the uplift modelling community. UpliftIVM searches for a splitting point that maximises the following criterion:
\begin{equation}
\mathcal{C}^{\textrm{UpliftIVM}} \vcentcolon = |\hat{\tau}_L - \hat{\tau}_R  \mid ,
\end{equation}
where $\hat{\tau}_L$ and $\hat{\tau}_R$ are the estimated conditional average treatment effect within the left and right child nodes, respectively. In other words, UpliftIVM aims to find the split that maximises the difference between the estimated CATEs of the two child nodes. 

Specifically, in UpliftIVM the within-node CATEs $\hat{\tau}_L$ and $\hat{\tau}_R$ are estimated as the average outcome difference between the treatment and control groups using training data within the node, i.e.,  $\hat{\tau}_L = \frac{\sum_{i=1} ^{n_L}T_iY_i}{\sum_{i=1} ^{n_L}T_i} - \frac{\sum_{i=1} ^{n_L}(1-T_i)Y_i}{\sum_{i=1} ^{n_L}(1-T_i)}$ and  $\hat{\tau}_R = \frac{\sum_{i=1} ^{n_R}T_iY_i}{\sum_{i=1} ^{n_R}T_i} - \frac{\sum_{i=1} ^{n_R}(1-T_i)Y_i}{\sum_{i=1} ^{n_R}(1-T_i)}$, where $n_L$ and $n_R$ denote the number of subjects in left and right child nodes, respectively. 

An advantage of UpliftIVM is its simplicity, and it performs well when the magnitude of CATE heterogeneity is large. However, it is prone to outliers and spurious treatment effect heterogeneities.

\paragraph{Squared t-Statistics Tree (t-stats) \citep{Su2009} } 
t-stats is an early tree-based algorithm for modelling treatment effect heterogeneity.
It builds a tree model by seeking the split with the largest value of the squared t-statistic for testing the null hypothesis that the average treatment effect is the same in the two potential child nodes. The t-stats tree maximises the following splitting criterion:
\begin{equation}
\mathcal{C}^{tstats} \vcentcolon= \frac{(\hat{\tau}_L-\hat{\tau}_R)^2}{\hat{\sigma}^2 (1/n_{1L} + 1/n_{0L} + 1/n_{1R} + 1/n_{0R})} ,
\label{TS}
\end{equation}
where $\hat{\sigma}^2 = \sum_{i \in \{0, 1\}} \sum_{j \in \{L, R\}} \frac{n_{ij}-1}{n-4} \sigma_{ij}^2$ and $n=n_{0L} + n_{0R} + n_{1L} + n_{1R}$. Here, $n_{1L}$ ($n_{1R}$) and $n_{0L}$ ($n_{0R}$) denote the number of treated and control subjects, respectively, in the left (right) child node; and $\sigma_{1L}^2$ ($\sigma_{1R}^2)$ and $\sigma_{0L}^2$ ($\sigma_{0R}^2)$ are the sample variances of the treated and control subjects, respectively, in the left (right) child node.

The within-node CATEs $\hat{\tau}_L$ and $\hat{\tau}_R$  are estimated in the training data in the same way as in UpliftIVM discussed above. 
t-stats differs from UpliftIVM in that t-stats (Equation (\ref{TS})) uses a pooled variance estimator (for estimating the common sample variances of various populations with different means) to normalise the difference of within-node CATEs.

\paragraph{Uplift Decision Tree (UpliftDT) \citep{Rzepakowski2010_DTuplift}}
UpliftDT is developed in the uplift modelling community for binary outcomes, and its motivation is different from the previously described tree-based methods. The previous tree-based methods aim to find the split that maximises the difference between the estimated CATEs of the left and right child nodes, whereas UpliftDT aims to maximise the estimated CATE within each child node. Specifically, UpliftDT maximises the following splitting criterion:
\begin{align}
\mathcal{C}^{Eu} = \frac{n_L}{n} \hat{\tau}_L^2 + \frac{n_R}{n} \hat{\tau}_R^2,
\label{upliftDT}
\end{align}
where $\hat{\tau}_L$ and $\hat{\tau}_R$ are estimated as the within-node CATEs in the training data as before.

The above splitting criterion is referred to as the Euclidean criterion by \citet{Rzepakowski2010_DTuplift}. To see this, note that for the binary outcome $Y$, the treatment effect within a node can be written as $\hat{\tau} = P(Y|T=1) - P(Y \mid T=0)$, and thus $\hat{\tau}^2$ can be viewed as the Euclidean distance between the treated and control subjects within the node. 

\citet{Rzepakowski2010_DTuplift} proposed two splitting criteria based on KL divergence and $\chi^2$ divergence. However, they argued that the Euclidean criterion is superior because it is more stable than the other criteria and has the important property of being symmetric.
It is worth noting that the $\chi^2$ splitting criterion has also been investigated by others in the uplift modelling community. In the work of \citet{Michel2017}, a similar tree-based method was proposed utilising the $\chi^2$ divergence.


A benefit of the splitting criterion in UpliftDT is that it can be extended to handle categorical outcomes and multiple branch splitting for a decision tree. 
To see this, for a $p$-way split, we can rewrite the splitting criterion in Equation (\ref{upliftDT}) as $\mathcal{C}^{Eu} = \sum_{i=1}^p \frac{n_p}{n} \hat{\tau}_p^2$. UpliftDT has also been extended to handle multiple treatments by \citet{Rzepakowski2012_DTforUpliftModel}. 

\paragraph{Balance-Based or Significance-Based Uplift Tree \citep{Radcliffe2011}}
These two types of trees are designed in the uplift modelling community for binary outcomes. 

The balance-based uplift tree aims to maximise the uplift difference in two splitting nodes while minimising the difference in size between the nodes. The following splitting criterion is used:
\begin{equation}
 \mathcal{C}^{BL} \vcentcolon = |\hat{\tau}_L - \hat{\tau}_R  \mid  (1-  \mid  \frac{n_L-n_R}{n_L+n_R} \mid ^\alpha),
\end{equation}
where $n_L$ and $n_R$ are the number of subjects in the left and right nodes, respectively, $0 \le \alpha \le 1$ is a  hyperparameter, and $\hat{\tau}_L$ and $\hat{\tau}_R$ are the within-node CATEs in the training data as before. 

The significance-based uplift tree uses the significance of the interaction between the treatment variable and a candidate splitting variable as a measure for the splitting quality. In each partition, the data in the current node are fitted with a linear model where each candidate split variable and the treatment form an interaction term. The significance of the interaction is tested by a $t$- statistic:
\begin{equation}
\mathcal{C}^{SIG} \vcentcolon= \frac{(n-4)(\tau_L-\tau_R)^2}{\textrm{SSE} \cdot  (1/n_{1L} + 1/n_{0L} + 1/n_{1R} + 1/n_{0R})},
\label{sig}
\end{equation}
where $\textrm{SSE} = \sum_{i \in \{1, 0\}} \sum_{j \in \{L, R\}} n_{ij} P_{ij}(Y=1)(1-P_{ij}(Y=1))$, and $\hat{\tau}_L$ and $\hat{\tau}_R$ are estimated as the within-node CATEs in the training data as before. 

The splitting criterion of the balance-based uplift tree is closely related to the criterion of the UpliftIVM. 
Furthermore, for the significance-based uplift tree, it can be seen that the $\textrm{SSE}$ in  Equation (\ref{sig}) is the weighted sum of the population variances. 
Contrasting this criterion with the one from t-stats (Equation (\ref{TS})), it can be seen that the two criteria are equivalent except that t-stats uses sample variances instead of population variances in the denominator.

\paragraph{Causal Inference Tree (CIT) \citep{Su2012}}
A CIT  is a tree-based method for treatment effect heterogeneity modelling, and it was proposed by the same author as the t-stats tree. The CIT assumes that the potential outcomes, $Y(0)$ and $Y(1)$, come from two Gaussian distributions with the same variance. In other words, $Y(T)\sim \mathcal{N}(T\mu_1 + (1-T)\mu_0, \sigma^2)$, where $\sigma^2$ is the variance, and $\mu_1$ and $\mu_0$ are the means of the treatment and control outcomes, respectively. Furthermore, the CIT assumes that the treatment follows a Bernoulli distribution $T\sim \textrm{Bernoulli}(\pi)$. Using these assumptions, the CIT then proposes to find the split that maximises the following log-likelihood within the node:
\begin{align}
\mathcal{C}^{CIT} \vcentcolon =  -\frac{n_L}{2} \cdot \ln (n_L \text{SSE}_{L}) - \frac{n_R}{2} \cdot \ln (n_R \text{SSE}_{R})
+ n_{1L} \ln n_{1L} + n_{0L} \ln n_{0L} + n_{1R} \ln n_{1R} + n_{0R} \ln n_{0R},
\end{align}
where $\text{SSE}_L$ and $\text{SSE}_R$ are the sum of squared errors for the left and right child nodes, respectively.  $\text{SSE}_R$ is defined as $\text{SSE}_R = \sum_{i=1}^{N_{1R}} (Y_i - \hat{Y}_1 )^2 +  \sum_{i=1}^{N_{0R}} (Y_i - \hat{Y}_0)^2 $, where $\hat{Y}_1 = \sum_{i=1}^{N_R} Y_i\cdot T_i / \sum_{i=1}^{N_R} T_i$ and $\hat{Y}_0 = \sum_{i=1}^{N_R} Y_i\cdot (1-T_i) / \sum_{i=1}^{N_R} (1-T_i)$ are the means of the treated and the control outcomes, respectively, within the right child node. $\text{SSE}_L$ is defined similarly.
The authors showed that the CIT consistently outperforms the t-stats tree. 

\paragraph{Causal Tree (CT) \citep{Athey2016}} 
A CT is a more recent tree-based algorithm developed specifically to estimate the CATE. 
A main difference between the CT and the previously mentioned tree-based methods is that the authors designed the CT to be an ``honest'' approach in the sense that, instead of using the training dataset as a whole, the CT divides the training dataset (of size $n$) into two parts, the splitting set (of size $n_s$) and the estimation set (of size $n_e$), and then uses the subjects in the splitting set to determine the split and the subjects in the estimation set to estimate the CATE within the node. 
The splitting criterion of the CT can be represented as follows:
\begin{align}
\mathcal{C}^{CT} \vcentcolon= (\frac{n_L}{n}\hat{\tau}_L^2+  \frac{n_R}{n}\hat{\tau}^2_R) - (\frac{1}{n} + \frac{1}{n_s})( \frac{S_{1L}^2}{p} + \frac{S_{0L}^2}{1-p}+  \frac{S_{1R}^2}{p} + \frac{S_{0R}^2}{1-p}),
\label{CT}
\end{align}
where $S_{1L}, S_{0L}, S_{1R}, S_{0R}$ denote sample variances of treated and control subjects in left and right nodes, respectively. $ (\frac{1}{n} + \frac{1}{n_s})$ is the weight to penalise small-sized leafs, and $p$ is the marginal treatment probability in data samples including the training dataset.   

Unlike other tree methods, in the CT, the CATEs are estimated using inverse propensity score weighting ~\citep{Rosenbaum1983} as $\hat\tau(\pmb{x})= \sum\limits_{\pmb{x}_i \in \mathcal{N}}\frac{T_i\cdot Y_i}{e(\pmb{x}_i)} / \sum\limits_{\pmb{x}_i \in \mathcal{N}}\frac{T_i}{e(\pmb{x}_i)}T
-\sum\limits_{\pmb{x}_i \in \mathcal{N}}\frac{(1-T_i)\cdot Y_i}{(1-e(\pmb{x}_i))} /\sum\limits_{\pmb{x}_i \in \mathcal{N}}\frac{1- T_i}{(1-e(\pmb{x}_i))},$
where $\mathcal{N}$ is a node, and $e(\pmb{x})=P(T=1|\pmb{x})$ denotes the propensity score for subject $\pmb{x}$.

\citet{Athey2016} also discussed an ``adaptive'' CT where the splitting criterion does not have the variance part (the second term in Equation (\ref{CT})) and the algorithm does not divide the data into two sets. Recalling the splitting criterion of UpliftDT (Equation (\ref{upliftDT})), it can be clearly seen that the splitting criteria of the adaptive version of the CT and UpliftDT are equivalent.
In their evaluation, the honest CT performed consistently better than the adaptive CT. However, a drawback to the honest CT is that it effectively only utilises a portion of the dataset, because it requires dividing the subjects into the splitting and the estimation sets.


\paragraph{Bayesian Score Tree ~\citep{Hecerman2000}}
This is an early method designed for targeted advertising and is different from the previously surveyed methods. 
The Bayesian score tree does not directly maximise the uplift difference in each split. Instead, it models the uplift as a new split by the treatment variable on a child node. Each child node is forced to be split by the treatment variable if its path does not contain the treatment variable. A child node is removed if the overall tree quality score does not increase when it is split by the treatment variable. 
The uplift is calculated by the difference between the outcome probabilities in the left and right child nodes. 
A Bayesian score, such as the one used by \citet{BayesianScoreTree1993}, is employed to calculate the score of the candidate trees.

\subsubsection{Support Vector Machine-Based Uplifting Modelling Methods}
In this section, we discuss methods that utilise a support vector machine (SVM) \citep{Cortes1995} for CATE estimation proposed in the uplift modelling community. The main benefit of these methods is that by reformulating the uplift modelling problem within the SVM framework, they enjoy the benefit of an SVM, which has been proven to be effective in many supervised learning applications. However, a problem with these methods is that none of them has implementation available online for potential users. 

\paragraph{$L_1$ and $L_p$ Uplift SVMs ($L_1$-USVM and $L_p$-USVM) \citep{Zaniewicz2013,Zaniewicz2017}}

These two SVM-based methods were proposed by the same authors, Zaniewicz and Jaroszewicz, who  recast the uplift modelling problem as a three-class classification problem. Specifically, the two methods aim to predict whether the treatment has a positive treatment effect, no treatment effect, or a negative treatment effect. They achieve this goal by using the following two parallel hyperplanes: 
\begin{equation}
H_1: \langle\pmb{w},\pmb{x}\rangle-b_1=0,\qquad H_2: \langle\pmb{w},\pmb{x}\rangle-b_2=0,
\end{equation}
where $b_1,b_2 \in \mathbb{R}$ are the intercepts, and $\pmb{x}$ denotes the coefficients of the decision boundary. The predicted treatment effect is then specified as:
\begin{equation}
\hat{\tau} (\pmb{x}) = \begin{cases}
+1 \quad \textrm{if } \langle\pmb{w},\pmb{x}\rangle > b_1 \textrm{ and }  \langle\pmb{w},\pmb{x}\rangle > b_2,\\
0 \quad \textrm{if } \langle\pmb{w},\pmb{x}\rangle \le b_1 \textrm{ and }  \langle\pmb{w},\pmb{x}\rangle > b_2,\\
-1 \quad \textrm{if} \langle\pmb{w},\pmb{x}\rangle \le b_1 \textrm{ and }  \langle\pmb{w},\pmb{x}\rangle \le b_2.
\end{cases}
\end{equation}
$L_1$-USVM \citep{Zaniewicz2013} utilises the $L_1$-norm as the regularisation for $\pmb{w}$, but it is sensitive to the parameter setting because of the discontinuity problem with the $L_1$-norm. 
Then, $L_p$-USVM  \citep{Zaniewicz2017} was proposed to utilise the $L_p$-norm in the optimisation to replace $L_1$-norm. $L_p$-USVM not only resolves the discontinuity problem, but also improves the convergence and efficiency and provides more stable results. However, it also introduces an additional hyperparameter $p$ that needs to be tuned.

The above two methods are designed for data from randomised controlled trials. The authors also proposed extended methods to observational data by adding one more regularisation term \citep{Jaroszewicz2015}. However, adding the additional regularisation term makes the objective function difficult to be optimised, because the new objective is not differentiable.

\paragraph{Lift Curve SVMs}
Another contribution from the uplift modelling community is the SVM for differential prediction ($\textrm{SVM}^{upl}$) \citep{Kuusisto2014}. In $\textrm{SVM}^{upl}$, the authors proposed to directly find the decision boundary that maximises the area under the uplift curve (AUUC). The uplift curve is an evaluation metric used by the uplift modelling community for comparing the performance of uplift models. We give the precise definition of the uplift curve (Equation (\ref{uplift})) when introducing the evaluation metrics (Section \ref{sec_metrics}). 
Intuitively, the idea of $\textrm{SVM}^{upl}$ is similar to the SVM-based methods for supervised learning that maximise the area under the ROC curve (AUC), i.e., $\textrm{SVM}^{perf}$ \citep{Joachims2005}. However, instead of maximising the AUC, $\textrm{SVM}^{upl}$ maximises the AUUC for uplift modelling.

The authors showed that maximising the AUUC is equivalent to maximising a weighted difference between the AUC for the treatment group and the AUC for the control group. Specifically,
\begin{equation}
\max(AUUC) \equiv \max(AUC_{T=1} - \lambda AUC_{T=0}),
\label{AUUC}
\end{equation}
where $\lambda= \frac{\sum\limits_{i=1}^{n} Y_i(1-T_i)\sum\limits_{i=1}^{n} (1-Y_i)(1-T_i)\sum\limits_{i=1}^{n}T_i}
{\sum\limits_{i=1}^{n} Y_iT_i\sum\limits_{i=1}^{n} (1-Y_i)T_i\sum\limits_{i=1}^{n}(1-T_i)}$. 
The difference in Equation (\ref{AUUC}) is equivalent to the sum of the AUC for the treatment group and the control group by flipping the control group outcomes:
\begin{equation}
\max(AUUC) = \max (AUC_{T=1} -\lambda(1-AUC_{T=0}^-)) = \max(AUC_{T=1} + \lambda AUC_{T=0}^-)
\end{equation}
where $AUC_{T=0}^-$ indicates the AUC of the control group with flipped outcome labels.

By showing that maximising the AUUC is equivalent to maximising the sum of two AUCs, the authors then solved the $\textrm{SVM}^{Upl}$ optimisation problem by utilising the $\textrm{SVM}^{perf}$ algorithm \citep{Joachims2005}, which is designed to optimise the AUC directly.
Since none of the methods proposed by the treatment effect heterogeneity modelling community directly maximises the AUUC, it could be beneficial to apply this approach to sociology and medical problems.

\subsubsection{Deep Learning-Based Methods}
\label{sec_deep2}
Recently, several generative learning methods for treatment effect heterogeneity modelling have been proposed. We survey them separately from the deep learning methods discussed above, since the generative approach of these methods does not fall into the single-model, two-model, X-Learner, or R-Learner approaches discussed above.

It is worth noting that the uplift modelling community seldom discusses deep learning-based methods.
Since deep learning-based methods excel at dealing with large-scale datasets, these methods may contribute to online marketing where the number of subjects is usually abundant.

\paragraph{Causal Effect Variational Autoencoders}
\label{sec_cevae}
The causal effect variational autoencoder (CEVAE) \cite{Louizos2017} is a variational autoencoder (VAE)-based treatment effect heterogeneity modelling approach. 
This method uses a VAE to learn a latent confounding set $\pmb{z}$ from the observed covariates $\pmb{x}$, and then uses $\pmb{z}$ to estimate the CATE. 
A VAE \citep{Kingma2013} is a new type of autoencoder based on variational inference that is able to approximately infer the intractable posteriors of the latent variables. 
The CEVAE assumes that the observed covariates ($\pmb{x}$) are independent of both treatment and outcome conditioning on the latent confounders $\pmb{z}$. It adopts a VAE to infer the posterior of the latent confounders $p(\pmb{z}|\pmb{x})$ and estimate the CATE as:
\begin{equation}
\hat{\tau}(\pmb{x}) = \int_{\pmb{z}}p(Y|\pmb{z},T=1)p(\pmb{z} \mid \pmb{x})d\pmb{z} - \int_{\pmb{z}}p(Y \mid \pmb{z},T=0)p(\pmb{z} \mid \pmb{x})d\pmb{z}
\end{equation}

CEVAE can infer unobserved confounders that are difficult to measure. For example, the income of a patient is rarely available from electronic medical records but can be inferred from the patient's postcode and occupation. 
However, a drawback to the CEVAE is that there is no guarantee that the inferred latent posterior $p(\pmb{z}|\pmb{x})$ will converge to the true posterior, because the CEVAE relies on variational approximations.
Another drawback to the CEVAE is that it assumes that all the factors in $\pmb{z}$ are confounders, which is often not satisfied by data.

A recently proposed improvement to the CEVAE is the disentangled variational autoencoder (TEDVAE) \citep{Zhang2020}. The TEDVAE learns three disentangled sets of latent factors: the instrumental factors $\pmb{z}_t$, which affect only the treatment; the confounding factors $\pmb{z}_c$, which affect both the treatment and the outcome; and the risk factors $\pmb{z}_y$, which affect only the outcome.
Disentangling the latent factors facilitates accurate CATE estimations, and alleviates the user's burden of choosing the appropriate set of covariates, since users can safely include all observed variables without the implication that including variables unrelated to the outcome may increase the bias and the variance of the CATE estimations.

\paragraph{Generative Adversarial Network for Individualised Treatment Effects (GANITE) \citep{Yoon2018}}
GANITE utilises  a generative adversarial network (GAN) to model treatment effect heterogeneity.
GANITE consists of two components, a counterfactual block that generates the counterfactual outcome $Y^{cf}$ with input $(\pmb{x},T,Y)$, and an ITE block that generates the CATE $\hat{\tau}(\pmb{x})$ for the subjects.

Specifically, the counterfactual block contains a generator $\boldsymbol{G}$ paired with a discriminator $\boldsymbol{D_G}$. $\boldsymbol{G}$ takes input $(\pmb{x},T,Y)$ and generates a counterfactual $Y^{cf}$ for the treatment $1-T$, while $\boldsymbol{D_G}$ takes $(\pmb{x},Y,Y^{cf})$ as input and outputs whether the outcome is generated by $\boldsymbol{G}$. During training, $\boldsymbol{G}$ is trained to maximise the probability of $\boldsymbol{D_G}$ incorrectly identifying whether $Y^{cf}$ is factual or counterfactual, while $\boldsymbol{D_G}$ is trained to maximise the probability of correctly distinguishing $Y^{cf}$ from $Y$.  
After training the counterfactual block, the counterfactual outcome $Y^{cf}$ generated by $\boldsymbol{G}$ along with $(\pmb{x},T,Y)$ are fed into the ITE block, which consists of a generator $\boldsymbol{I}$ paired with an discriminator $\boldsymbol{D_I}$. The generator $\boldsymbol{I}$ takes the covariates $\pmb{x}$ as input and generates the potential outcomes $Y$ and $Y^{cf}$. The discriminator $\boldsymbol{D_I}$ aims to discriminate whether the outcomes generated by $\boldsymbol{I}$ are the inputs from the counterfactual block (generated by $\boldsymbol{G}$).
After training, only the generator $\boldsymbol{I}$ in the ITE block is used for predicting the CATE for new subjects.

\subsubsection{Ensemble-Based Methods}
\label{sec_ensemblemethods}
Ensemble-based CATE estimation methods have been proposed to address the high variance problem of tree-based methods. 
Most ensemble methods use tree-based methods as base learners.
Generally speaking, ensemble-based methods perform better than a single tree-based model; however, ensemble-based methods lose the interpretability possessed by the tree-based methods and have higher time complexity than tree-based methods.

\paragraph{Uplift Bagging}
Bagging \citep{Breiman1996} is a simple and popular ensemble method in supervised learning. When using bagging for uplift modelling, a set of bootstrap training datasets is randomly sampled from the original training dataset with replacement. A bootstrap training dataset has the same size as the original training dataset. An uplift model is built on each bootstrap training dataset. The final prediction for a test subject is the average of predicted uplifts of all models in the example.


For CATE estimation, \citet{Radcliffe2011} stated that they used bagging in real-world applications, but they did not provide any experimental results. \citet{Soltys2015}  implemented and compared two bagging methods. The base learners used were UpliftDT (using the Euclidean distance \citep{Rzepakowski2010_DTuplift} and the two-model decision tree (using C4.5~\citep{Quinlan1993_C45}, as discussed in Section \ref{sec_two}). 
Based on their evaluations \citep{Soltys2015}, the uplift bagging methods performed significantly better than the uplift decision trees, and were competitive with the uplift random forest method discussed below. 

\paragraph{Uplift Random Forest (UpliftRF) \citep{Guelman2012, Guelman2015}}
UpliftRF is an uplift ensemble-based on the idea of random forests \citep{Breiman2001}. It consists of four steps. Firstly, a set of bootstrap training datasets is randomly sampled from the original training dataset with replacement. Secondly, each bootstrap training dataset is projected to a fixed number of $k$ randomly selected covariate spaces. Thirdly, UpliftDT \citep{Rzepakowski2010_DTuplift} is built on every training dataset from the above two steps. Fourthly, the set of uplift trees is used to predict the uplift for a new subject by using the average predictions of all trees.     


\citet{Soltys2015} implemented an ensemble-based algorithm called double uplift random forests (DURF). DURF is a bagged ensemble of the two-model approach using randomised trees from Weka~\citep{Hall2009}. In their evaluation, both UpliftRF (using Euclidean distance) and DURF performed better than the two-model decision trees, but were not significantly different from the bagged UpliftDT (using the Euclidean distance) \citep{Rzepakowski2010_DTuplift} or the bagged two-model decision tree (using C4.5~\citep{Quinlan1993_C45}).    

\paragraph{Causal Conditional Inference Forest (UpliftCCIF) \citep{Guelman2014} } 
UpliftCCIF is based on tree models and uses a similar strategy as UpliftRF to construct the ensemble of trees, i.e., by randomly sampling the training subjects and covariates with replacement. The major difference between UpliftCCIF and UpliftRF lies in the tree splitting procedure. Whether to split the node in a base learner tree model of UpliftCCIF is determined by testing the null hypothesis that there are no interactions between the treatment $T$ and any of the covariates in $X$. Specifically, the null hypothesis is formulated as $H_0 = \cap_{j=1}^k H_0^j$ with $H_0^j: \mathbb{E}[Y^*|x_j] = \mathbb{E}[Y^{*}]$, where $Y^{*}$ is the transformed outcome, as discussed in Section \ref{sec_OutcomeTransform}, and $x_j$ is a covariate value in $\pmb{x}$. The authors used Bonferroni-adjusted $p$-values for handling the multiplicity in the statistical tests. If the null hypothesis is rejected, the splitting covariate is selected as the one with the smallest $p$-value. After the trees are built, the uplift of a subject is estimated as the average of the uplifts predicted by the individual trees.

%
%

\paragraph{Causal Forests (CF) \citep{Wager2018} }
CF is a random forest-like algorithm for treatment effect heterogeneity modelling. 
CF uses the Causal Tree (CT) algorithm (discussed in Section \ref{sec_tree}) as its base learner, and constructs the forest from an ensemble of $k$ causal trees where each tree provides a CATE estimation $\hat{\tau}_b(\pmb{x})$ for a subject. The forest then uses the average of the predicted CATEs from $k$ trees as its prediction, i.e., $\hat{\tau}(\pmb{x}) = \frac{1}{k}\sum_{b=1}^{k}\hat{\tau}_b(\pmb{x})$.

An important advantage of CF over the other surveyed ensemble methods is that the estimations of CF are asymptotically Gaussian and unbiased for the true CATE $\tau(\pmb{x})$. In other words, $(\hat{\tau}(\pmb{x}) - \tau(\pmb{x})) /\sqrt{Var(\tau(\pmb{x}))}\rightarrow \mathcal{N}(0,1)$. Furthermore, the authors provides a way to estimate the asymptotic variances.
CF is a general framework in the sense that its theoretical properties are valid as long as the trees used as base learners are ``honest'', i.e., that the outcome of any sample is not used for both selecting the split and estimating the within-node CATE $\hat{\tau}$. Based on this property, the authors proposed a CF instantiated using propensity tree, which completely ignores the outcome $Y$ when choosing the splits and builds the tree using the Gini criterion \citep{Breiman1984} of the treatment $T$.

\section{Applications}
In this section, we discuss the main applications of targeted advertising from the uplift modelling community, and applications in personalised medical treatment and social sciences from the treatment effect heterogeneity modelling community. 
While applications in the uplift modelling community mostly utilise data from randomised experiments, those in the treatment effect heterogeneity community often make use of observational data, since ethical and cost concerns often prohibit controlled trials in many medical and sociology studies. 

\subsection{Applications in Marketing}
\label{Marketing}
Driven by the need by companies for increased sales and minimal advertisement costs, targeted advertising has long been the focus of the uplift modelling community. Most of the work in this area is based on real-world datasets obtained by randomised experiments, and some of the results have been monetised by companies according to the literature. The results in these applications confirm that using uplift modelling to target customers is more effective than random targeting or using standard supervised learning methods.

\citet{Hecerman2000} applied the Bayesian score tree (discussed in Section~\ref{sec_tree}) to an MSN advertising experimental dataset. Registrants of Windows 95 were randomly divided into treatment and control groups, where the treated subjects were mailed advertisements and the control subjects received nothing. The outcome was an MSN sign-up within a time period. 110,000 subjects were involved in the experiment, where the treatment subjects accounted for 90\% of the total. 70\% of the data were used for training and the rest were used for evaluation. 
Their results showed that the uplift model achieved more revenue when compared to the mail-to-all strategy. 

\citet{Hansotia2002} applied the UpliftIVM algorithm (discussed in Section~\ref{sec_tree}) to analyse customer responses to a promotion for \$10 off of a \$100 purchase. The data were obtained from a holiday promotion of a major national retailer in the United States. Promotional mail (treatment) was randomly sent to 50\% of the total 282,277 customers. 
They compared a two-model approach using logistic regression with UpliftIVM. 
The evaluation criterion was the uplift in the 50\% reserved evaluation dataset (and the other 50\% was used for building models). 
Their results showed that UpliftIVM performed better than the two-model approach when targeting the top 10\% of customers, while their overall performance was similar. 

\citet{Radcliffe2007} discussed three real-world applications of uplift modelling in marketing: ``deep-selling'', where the goal is to use a promotion to increase the frequency or size of customer transactions;
customer retention, which aims to mitigate customer attrition; 
and cross-selling, which involves selling new products to existing customers. 
The deep-selling experiment included 100,000 subjects, and the treatment and control groups were split by 50:50. 
The sample sizes in two other examples were not reported. 
They used a balance-based uplift tree~\citep{Radcliffe2011} (discussed in Section~\ref{sec_tree}).
With deep-selling, the authors showed that the uplift modelling approach was better at increasing revenue than standard supervised methods.
For customer retention, they considered a problem where a mobile service provider was experiencing an annual churn rate of 9\%. 
Originally, the company targeted its entire customer base with a retention offer and the churn rate increased to 10\%. Using uplift modelling, they targeted 30\% of customers, and were able to reduce the customer churn from 9\% to 7.8\%.
Using an estimated average revenue per user of \$400/year, the increase in revenue was around \$8.8 million. 
For cross-selling, they tackled a banking problem where banks wanted to sell new banking products to their existing customers and found that by using uplift modelling, banks could achieve between 80\% and 110\% increases in sales while reducing the volume of mailing offers by 30\% to 80\% when compared to a standard supervised learning approach. 

\citet{Guelman2012} applied uplift modelling methods to a customer retention dataset from a Canadian insurance company. A randomised experiment was conducted with a treatment group of 8249 subjects and a control group of 3719 subjects whose policies were due for renewal. 
The customers in the treatment group received a letter explaining that there would be an increase in their insurance premiums and a phone call from an insurance advisor. 
The customers in the control group received no retention effort. 
Four methods were applied to the dataset: UpliftRF (discussed in Section \ref{sec_ensemblemethods}), the two-model approach with logistic regression, the single-model approach using logistic regression with interaction terms between the treatment and all covariates~\citep{Lo2002_TrueLiftModel}, and the uplift decision tree~\citep{Rzepakowski2010_DTuplift}  (discussed in Section~\ref{sec_tree}).
Based on the uplift curves obtained from 10-fold cross validation, all methods performed better than the baseline retention rate by random targeting. 
Furthermore, UpliftRF performed better than the others, especially in top-ranked customer subgroups, but did not dominate other methods in the other customer subgroups.   

Hillstrom's Email Advertisement dataset is an open-access uplift modelling dataset \citep{Hillstrom2008} that contains 64,000 samples collected from a randomised experiment of an email advertisement campaign. 
The customers were evenly distributed in two treatment groups and a control group, where the first treatment group was sent a ``Men's Advertisement Email'' and the second treatment was sent a ``Women's Advertisement Email''. 
The control group received no email. 
The outcome variables were the visit and conversion status of the customers. 
Radcliffe \citep{Radcliffe2008} conducted analysis using the Hillstrom dataset, in which it was shown that the ``Men's Email'' was more effective than the ``Women's Email'' and the best customer subgroup to target was the subgroup of those who had visited the store using both a phone and web browser, and had spent more than \$160. 
The analysis identified that some customers were negatively affected by the ``Women's Email''.

Another recent dataset from the uplift modelling community is the Criteo Uplift modelling dataset \citep{DiemertEustacheBetleiArtem}. It contains 25,309,483 subjects where each row represents the behaviour of a customer. There are 11 anonymised pre-treatment covariates for each customer, a treatment indicator representing whether the customer received a promotional email, and two outcomes indicating the visiting and conversion status of the customer. 
Recently, several uplift modelling methods \citep{Rahier2020,Betlei2020} have been proposed to maximise the AUUC metric (as discussed in Equation \ref{AUUC}) for the dataset.

\subsection{Applications in Social Science}
Social science is one of the main applications areas of the treatment effect heterogeneity modelling community. In the following, we present some examples of the applications.

\citet{Imai2013} analysed the get-out-the-vote (GOTV) \citep{Gerber2000} randomised experiment where 69 different voting mobilisation methods, including canvassing (by in-person direct contact to encourage voting), phone calls and mailing were randomly given to registered voters in New Haven and Connecticut during the 1998 U.S. presidential election. The goal of their analysis was to select the best voting mobilisation strategy for individuals. To avoid the interference among voters within the same household (which violates the SUTV assumption), they focused on a subset of 14,774 voters in single-voter households, of which 5269 voters belonged to the control group. They used a single-model approach (discussed in Section \ref{sec_single}) with a modified SVM algorithm \citep{Cortes1995} as the base learner. Specifically, they used two separate L1-regularisation terms for covariates that only affected the outcome and covariates that interacted with both the outcome and the treatment (the type of covariates were manually identified by the authors). 
Their analysis showed that canvassing was the most effective treatment. When canvassing was used, any additional treatment such as phone calling and mailing in combination with canvassing reduced the effectiveness of canvassing. In addition, when canvassing was absent, the most effective treatment was contact by mail with a civic duty appeal. Any other treatment was found to be less effective or have negative effects. 

\citet{Imai2013} analysed a dataset for a national supported work (NSW) program \citep{LaLonde1986,Dehejia1999} to determine whether a job training program increased the income of workers. The dataset contained 297 and 425 workers randomly assigned to the treatment and control groups, respectively, and the 1978 panel study of income dynamics workers (PSID) from low-income subjects. 
In other words, the dataset consisted of two components, where the first component was obtained from a randomised controlled trial (the NSW sample), and the second component was obtained from an observational study (the PSID sample). The method used was the same as the one described in the GOTV application.
They built the model on the randomised NSW samples and applied it to the PSID samples. Their analysis showed that, overall, the training program was beneficial. However, it benefited educated Hispanics and low-income non-Hispanics the most, and it did not help employed white workers with high-school degrees.

\citet{Kuenzel2019} analysed a field experiment for the effect of canvassing (an in-person direct contact by conversation) on reducing transphobia (i.e., prejudice against transsexual and transgender people). The dataset was originally used by \citet{Broockman2016} where the analysis showed that brief but high-quality canvassing significantly reduced prejudice against transgender individuals for at least three months. The dataset consisted of three groups: a treatment group of 913 individuals who were canvassed on the topic of reducing transphobia, a placebo group of 912 individuals who were canvassed by an irrelevant conversation (i.e., about recycling), and a control group of 68,278 individuals who had not been canvassed at all. The outcomes of interest were the results of an online survey that measured the subjects' attitudes towards transphobia at 3 days after treatment. 
Using covariates such as religion, ideology, and demographics, the authors applied the single-model approach, the two-model approach, and the X-Learner (as discussed in Section \ref{sec_meta}) using random forests as the base learner. 
The results showed that the treatment effects estimated by the single-model approach were mostly 0 or almost 0, despite the fact that the average effect of the treatment was 0.22 \citep{Broockman2016}. The two-model approach and the X-Learner produced similar results, but the estimates of the two-model approach had increased variance.

\subsection{Applications in Personalised Medical Treatments}
Personalised medical treatment is another major application area of the treatment effect heterogeneity community, although some uplift modelling literature has also considered it \citep{Nassif2012_BNforUpliftModel,Jaskowski2012}. However, research in the treatment effect heterogeneity modelling community is more focused on the medical implications of the results, whereas research from the uplift modelling community uses medical datasets to compare different models. 

\citet{TreatmentSElection2011} studied the personalised treatment problem with the two-model approach. They used a clinical trial dataset from the AIDS Clinical Trials Group (ACTG), which studied the effect of a protease inhibitor for treating human immunodeficiency virus \citep{Hammer1997}. 
A total of 1156 patients were included in the trial and were randomly divided into treatment and control groups. The control group was given a two-drug combination, while the treatment group was given a three-drug combination. 
Previous studies showed that the treatment was significantly more effective than the control. However, it was also noted that some patients did not respond to the treatment and instead suffered from toxic side effects. 
The authors considered three covariates and used the change of CD4 count at week 24 from the baseline level as the continuous outcome. They used linear regression as the base learner for the two-model approach.
The results showed that the treatment effects of patient groups defined by different covariates were significantly heterogeneous. 

\citet{Weisberg2015} applied CATE estimation algorithms to clinical trial data from the Randomized Aldactone Evaluation Study. Aldactone is medicine for treating fluid build-up due to heart failure, liver scarring, or kidney disease. The trial was designed to test whether the medicine could reduce the mortality of patients who had suffered from severe heart failure. With Aldactone as the treatment, the study used 63 variables describing the demographics, history, and concomitant medications as covariates, and the maximum potassium level (continuous) within the first 12 weeks of treatment as the outcome. The authors used a transformed outcome approach (discussed in Section \ref{sec_OutcomeTransform}). The model was built on 80\% of the total 1632 subjects and tested on the remaining 20\% of the subjects. It showed that the treatment, although highly significant in the original study \citep{Pitt1999}, was no longer significant when the characteristics of the subjects were considered.

Some works extend CATE estimation to censored survival outcomes. \citet{Zhang2017} extended the causal tree \citep{Athey2016} (discussed in Section \ref{sec_tree}) to censored outcomes and used it to study the heterogeneous treatment effects of radiotherapy on the survival outcomes of breast cancer patients and glioma cancer patients using patients' gene profiles as covariates. The breast cancer dataset contained 964 subjects and the expression profile of 11,535 genes, and the glioma cancer dataset contained 632 subjects and 11,543 genes. For each dataset, about 50\% of the subjects were treated with radiotherapy while the rest were not.
The datasets can be accessed at \href{https://gdc.cancer.gov}{TCGA} (The Cancer Genome Atlas). 
The proposed method was compared with popular cancer subtype clustering methods including semi-supervised clustering and the L1-regularised COX proportional hazard model. Results on individual test sets showed that the CATE estimation method was better at finding subgroups with treatment effect heterogeneity. 

Recently, \citet{Tabib2020} proposed a random forest-based method to censored outcomes and showed the effectiveness of their algorithm on a breast cancer dataset and a colon cancer dataset. It is also possible to extend other surveyed methods to censored survival outcomes.

\newcolumntype{L}{>{\centering\arraybackslash}m{3cm}}
\begin{table}[!t]
	\footnotesize
	\centering
	\setlength{\tabcolsep}{3pt}
	\caption {Software packages for CATE estimation and uplift modelling. B, C, and N denote binary, categorical, and numerical variables, respectively. 
	}
	\resizebox{0.9\linewidth}{!}
	{
		\begin{tabular}{|c|c  c c |c c |c |c| p{5 cm} |}
		\hline
		Package	& \multicolumn{3}{c|}{\begin{tabular}{c c c}\multicolumn{3}{c}{Covariates}  \\
				B & C & N  \end{tabular}} & \multicolumn{2}{c|}{\begin{tabular}{c c}\multicolumn{2}{c}{Outcome} \\B  & N  \end{tabular}} & Methods & Language & URL\\
		\hline
		\textsf{causalTree} &\multicolumn{3}{c|}{\begin{tabular}{c c c} \checkmark & \checkmark & \checkmark \end{tabular}} &  \multicolumn{2}{l|}{\begin{tabular}{c c} \checkmark &\enskip\checkmark \end{tabular}} &
		\small \begin{tabular}{c}Causal Tree \\ t-stats Tree\\ TO$^{\#}$ Tree \end{tabular} &  \scriptsize R &\scriptsize\url{https://github.com/susanathey/causalTree} \\
		\hline
		\textsf{causalToolbox} & \multicolumn{3}{c|}{\begin{tabular}{c c c} \checkmark & \checkmark & \checkmark\end{tabular}} &  \multicolumn{2}{l|}{\begin{tabular}{c c}\checkmark &\enskip\checkmark \end{tabular}}  & \small \begin{tabular}{c} Single-model\\ Two-Model \\ X-Learner \\ TO$^{\#}$  \end{tabular} & \scriptsize R&\scriptsize\url{https://github.com/soerenkuenzel/causalToolbox} \\ 
		\hline
		\textsf{grf} & \multicolumn{3}{c|}{\begin{tabular}{c c c} \checkmark & $\times$ & \checkmark\end{tabular}}  & \multicolumn{2}{l|}{\begin{tabular}{c c}\checkmark &\enskip\checkmark \end{tabular}} & \small Causal Forest &  \scriptsize R&\scriptsize\url{https://CRAN.R-project.org/package=grf}
		\\\hline
		\textsf{Uplift} & \multicolumn{3}{c|}{\begin{tabular}{c c c} \checkmark & \checkmark & \checkmark\end{tabular}} &  \multicolumn{2}{l|}{\begin{tabular}{c c}\checkmark & \enskip$\times$ \end{tabular}} & \small \begin{tabular}{c}UpliftDT \\ UpliftRF\\ UpliftCCIF \end{tabular} & \scriptsize R&\scriptsize\url{https://cran.r-project.org/package=uplift}
		\\\hline
		\textsf{CausalML} & \multicolumn{3}{c|}{\begin{tabular}{c c c} \checkmark & \checkmark & \checkmark\end{tabular}} &  \multicolumn{2}{l|}{\begin{tabular}{c c}\checkmark &\enskip\checkmark$^*$ \end{tabular}} & \small \begin{tabular}{c}UpliftDT \\ UpliftRF\\ UpliftCCIF\\ Single-model\\ Two-Model\\ X-Learner \end{tabular}  & \scriptsize Python&\url{https://github.com/uber/causalml}\\
		\hline
		\textsf{pylift}  & \multicolumn{3}{c|}{\begin{tabular}{c c c} \checkmark & \checkmark & \checkmark\end{tabular}} &  \multicolumn{2}{l|}{\begin{tabular}{c c}\checkmark & \enskip$\times$ \end{tabular}} & \small \begin{tabular}{c}TO$^{\#}$  \end{tabular}& Python &\scriptsize\url{https://github.com/wayfair/pylift}\\
		\hline
	\end{tabular}
}
	\caption*{*: the uplift modelling methods in CausalML are implemented for binary outcomes only. \#: TO refers to the transformed outcome approach.}
	\label{tab_package}
\end{table}

\begin{table}[!t]
	\footnotesize
	\centering		
	\setlength{\tabcolsep}{3pt}
	\caption{Source codes for deep learning-based CATE estimation algorithms. B, C, and N denote binary, categorical, and numerical variables, respectively. }
	\begin{tabular}{|c|c  c c |c c |c | c | p{5 cm} |}
		\hline
		& \multicolumn{3}{c|}{\begin{tabular}{c c c}\multicolumn{3}{c}{Covariates} 
				\\ B & C & N  \end{tabular}} &  \multicolumn{2}{c|}{\begin{tabular}{c c}\multicolumn{2}{c}{Outcome} \\B  & N  \end{tabular}} & Method & Language & URL\\
		\hline
		\textsf{CFR} & \multicolumn{3}{c|}{\begin{tabular}{c c c} \checkmark & $\times$ & \checkmark\end{tabular}}  & \multicolumn{2}{l|}{\begin{tabular}{c c} $\times$ &\enskip\checkmark \end{tabular}} & CFR & Python & \scriptsize\url{https://github.com/clinicalml/cfrnet}\\
		\hline
		\textsf{SITE} &  \multicolumn{3}{c|}{\begin{tabular}{c c c} \checkmark & $\times$ & \checkmark\end{tabular}}  & \multicolumn{2}{l|}{\begin{tabular}{c c} $\times$ &\enskip\checkmark \end{tabular}}  & SITE & Python & \scriptsize\url{https://github.com/Osier-Yi/SITE}\\
		\hline
		\textsf{CEVAE} & \multicolumn{3}{c|}{\begin{tabular}{c c c} \checkmark & $\times$ & \checkmark\end{tabular}}  & \multicolumn{2}{l|}{\begin{tabular}{c c} $\times$ & \enskip\checkmark \end{tabular}}  & CEVAE & Python & \scriptsize\url{https://github.com/AMLab-Amsterdam/CEVAE}\\ 
		\hline
		\textsf{TEDVAE} & \multicolumn{3}{c|}{\begin{tabular}{c c c} \checkmark & $\times$ & \checkmark\end{tabular}} & \multicolumn{2}{l|}{\begin{tabular}{c c} $\times$ &\enskip\checkmark \end{tabular}} & TEDVAE & Python & \scriptsize\url{https://github.com/WeijiaZhang24/TEDVAE}\\
		\hline
	\end{tabular}
	\label{tab_newMethods}
\end{table}

\section{Software, Metrics, Demonstration, and Discussion}

In this section, we summarise the available software packages and source codes for CATE estimation. We also illustrate the methods on synthetic, semi-synthetic, and real-world datasets, and discuss their modelling behaviour, usability, interpretability, and scalability. We did not aim to find the best performing method, since this is beyond the scope of this survey. Rather, we provide discussion based on empirical evaluations of the methods in the available packages and recently proposed methods from both the causal effect heterogeneity modelling and uplifting modelling communities. For an empirical evaluation focused on comparing the accuracy of some of the algorithms for treatment effect heterogeneity modelling, we refer the reader to Dorie et. al. \citep{Dorie2019}. 
For empirical evaluations focused on comparing the performances of some uplift modelling algorithms, see Devriendt et al. \citep{Devriendt2018} and Gubela et al. \citep{Gubela2019}.

\subsection{Software Packages and Source Codes}
Table \ref{tab_package} presents a summary of the available software packages for uplifting modelling and CATE estimation. We include those which are well documented and are easy for end-users to use.



The \textsf{causalTree} package implements the causal tree \citep{Athey2016} (discussed in Section \ref{sec_tree}), the t-stats tree \citep{Su2009} (discussed in Section \ref{sec_tree}), and the transformed outcome approach (discussed in Section \ref{sec_OutcomeTransform}) using a regression tree (RT) as the base learner. 
The \textsf{causalToolbox} package implements the single-model, two-model, X-Learner, and transformed outcome approaches (discussed in Section \ref{sec_single}, \ref{sec_two}, \ref{sec_X}, and \ref{sec_OutcomeTransform}) using random forests (RF) and BART as base learners. 
The \textsf{grf} package implements the causal forest algorithm, as discussed in Section \ref{sec_ensemblemethods}.
The \textsf{Uplift} package was developed by the uplift modelling community and provides implementations of the uplift decision trees (UpliftDT) (discussed in \ref{sec_tree}), uplift random forests (UpliftRF), and uplift causal conditional inference forest (UpliftCCIF) algorithms (discussed in Section \ref{sec_ensemblemethods}). 
The \textsf{CausalML} package implements the same uplift modelling methods as the \textsf{Uplift} package and the same CATE approaches as the \textsf{causalToolbox} package. However, the base learners used in \textsf{CausalML} are linear regression (LR), XGBoost, and multi-layer perceptron, which are different from those provided in \textsf{causalToolbox}. 
Finally, the \textsf{pylift} is a uplift modelling package that implements the transformed outcome approach using XGBoost and random forests as base learners.

Table \ref{tab_newMethods} summarises the source codes for some recently proposed deep learning-based algorithms, including CFR \citep{Shalit2016}, SITE \citep{Yao2018_Twin}, CEVAE \citep{Louizos2017}, and TEDVAE \citep{Zhang2020}, as discussed in Sections \ref{sec_cfr} and \ref{sec_cevae}.
These methods have shown potential for CATE estimation tasks. However, the main difference between these source codes and the software packages is that the former do not provide ready-to-use interfaces for users to apply the methods to their own datasets or to tune parameters.

Recently, a few packages have been proposed specifically for generating synthetic datasets in order to evaluate treatment effect estimation algorithms: the \textsf{OPOSSUM} package \cite{Winkel2017} provides a Python interface for generating synthetic datasets with ground-truth CATEs, and \textsf{aciccomp} \cite{Dorie2017} is an R package which contains the datasets used in the Atlantic Causal Inference Competitions.
\subsection{Illustration}
\begin{figure}[!t]
	\begin{adjustbox}{minipage=\linewidth,scale=0.75}
		\begin{subfigure}{.32\textwidth}
			\includegraphics[width=\linewidth]{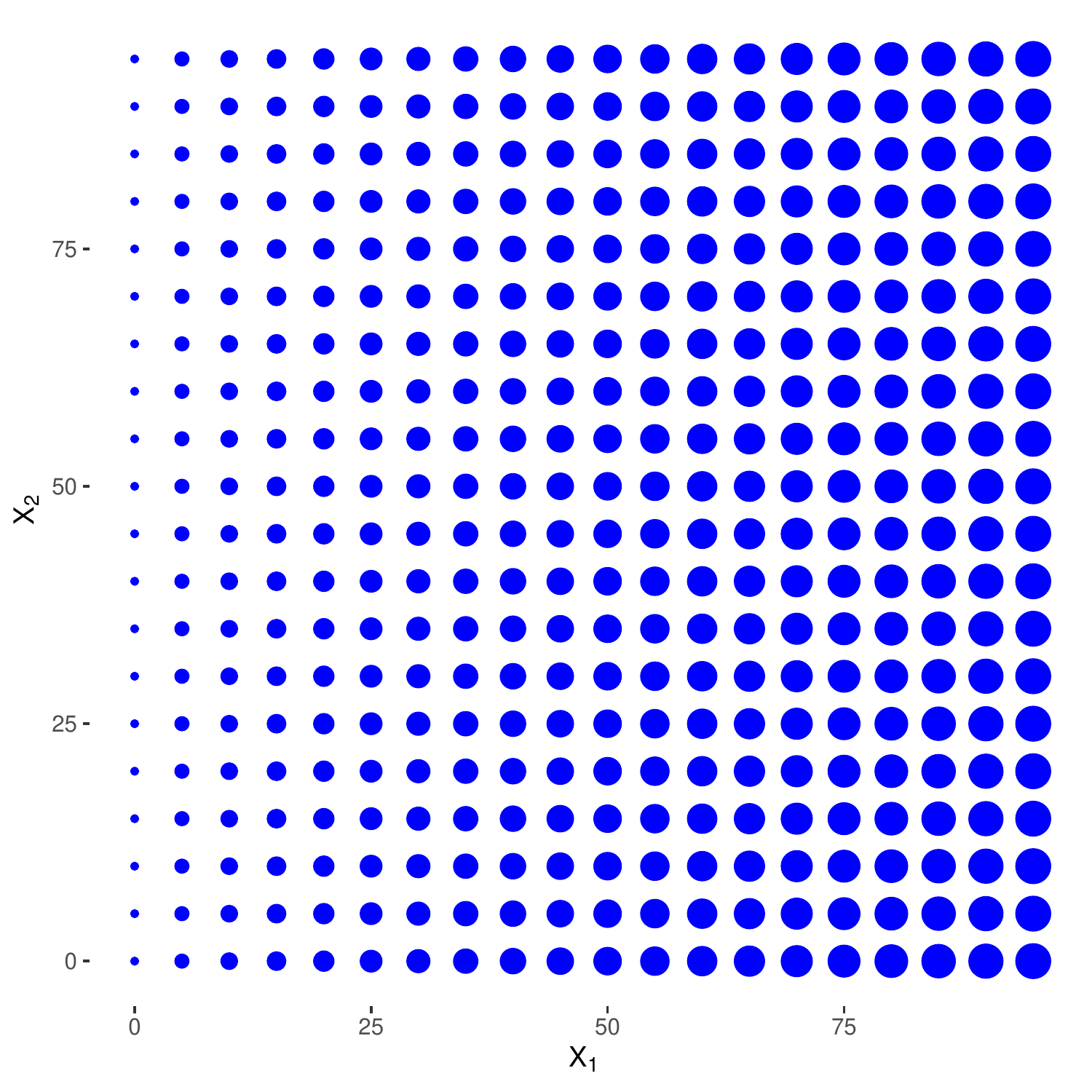}
			\caption{Treatment}
		\end{subfigure}
		\begin{subfigure}{.32\textwidth}
			\includegraphics[width=\linewidth]{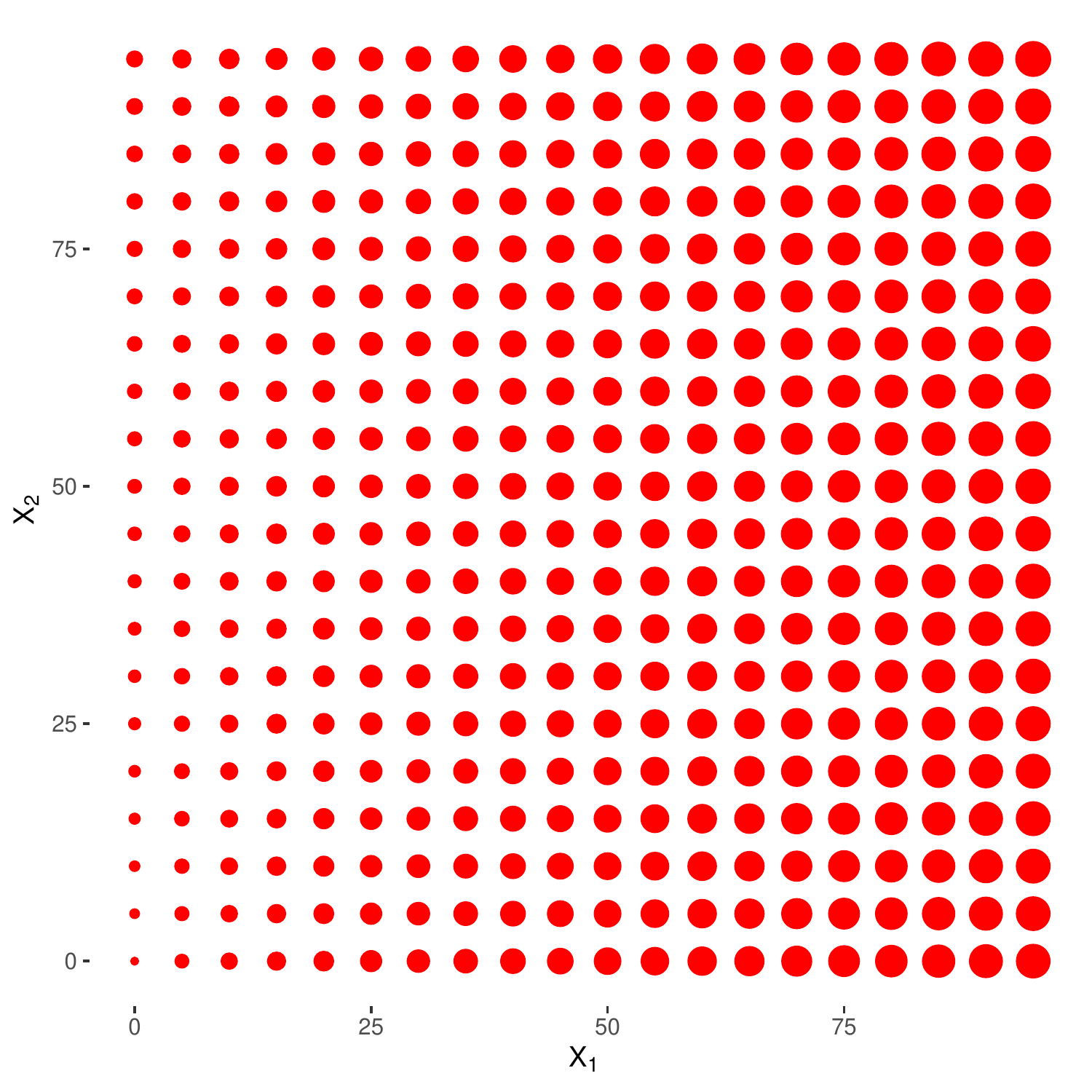}
			\caption{Control}
		\end{subfigure}
		\begin{subfigure}{.32\textwidth}
			\includegraphics[width=\linewidth]{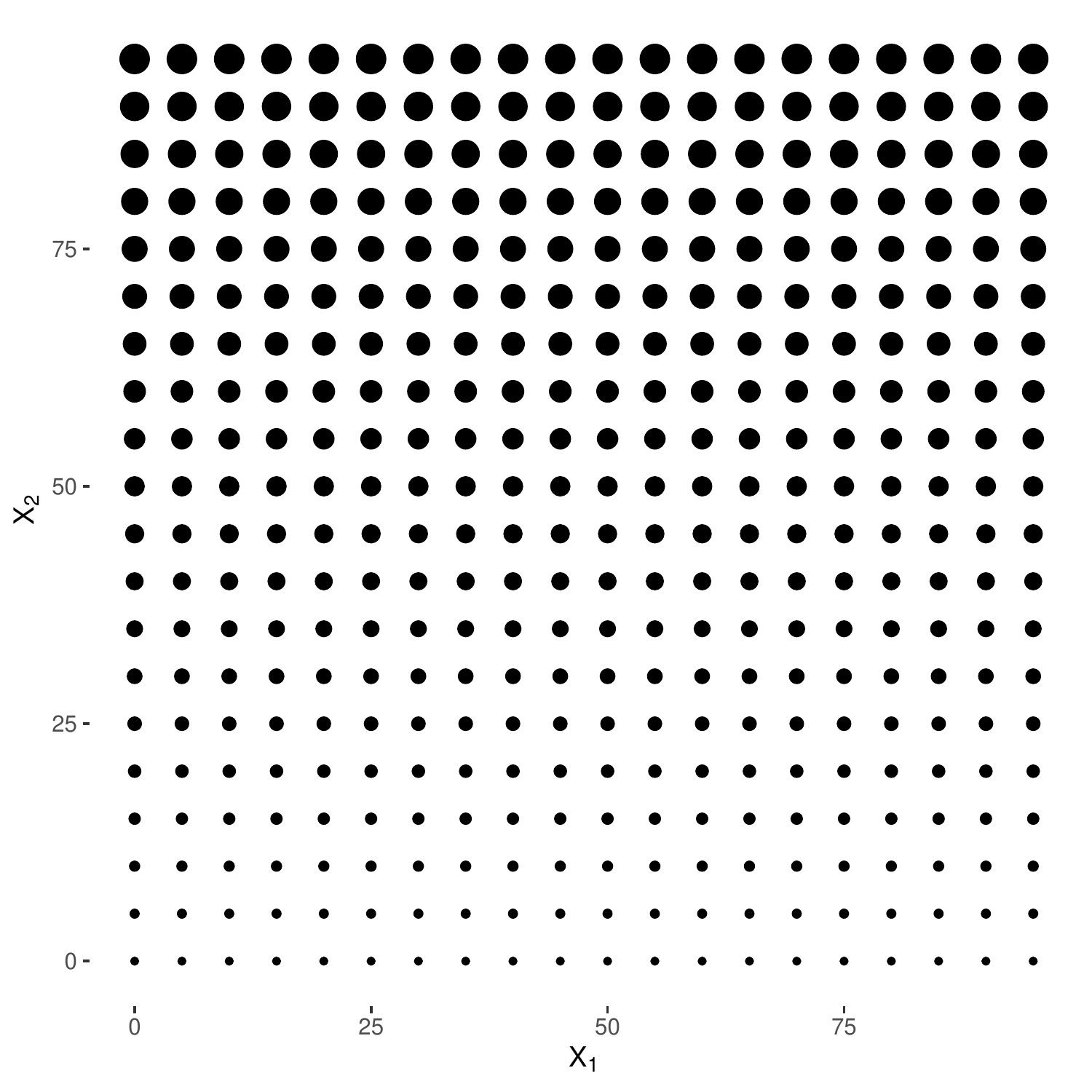}
			\caption{Theoretical CATE}
		\end{subfigure}
	\end{adjustbox}
	\caption{Disc plots of the treatment, control, and ground-truth CATEs of the synthetic dataset. Values in covariates $X_1$ and $X_2$ are grouped in bins of size $5$. The disc areas are proportional to the average outcome values / CATEs in the squares. 
		Note that the areas of discs in Subfigure (c) are in a different scale from Subfigures (a) and (b). The scale is around seven times smaller, since the magnitude of CATEs is small. Disc plots in Figures~\ref{fig_Results-Representative} and \ref{fig_differentTreatmentRate} use the same scale as in Subfigure (c).} 
	\label{fig_IllustrativeDataSet}
\end{figure}

Firstly, we use a synthetic dataset to provide a visual illustration for the behaviour in the CATE estimation of some representative methods. 
For practical examples that apply the Single-Model approach (discussed in Section 3.1.1), the Two-Model approach (discussed in Section 3.1.2), X-Learner (discussed in Section 3.1.3), R-Learner (discussed in Section 3.1.4), CFR (discussed in Section 3.1.6), and Causal Forest (discussed in Section 3.2.4) to a semi-synthetic dataset, please refer to \citep{Carvalho2019}.
The dataset we used was first introduced in \citep{Radcliffe2011} to provide an illustration of uplift modelling methods. It consists of 100,000 subjects of two covariates, a binary treatment and a continuous outcome. 
The data generating procedure is described as follows. First, two covariates, $X_1$ and $X_2$, are uniformly sampled from integers ranging from $0$ to $99$. Then, the two potential outcomes are generated based on the covariates and the treatment.
The outcomes in the control group are uniformly sampled from the interval defined by $[0, x_1)$, and the outcomes in treatment groups are uniformly sampled from the interval defined by $[0,x_1) + [0,x_2)/10 + 3$. In other words, the ground truth of the CATE has a linear relationship with the covariates as $\tau(\pmb{x}) = x_2/20 + 3$. 
We generate three different sets of data with different treatment assignment ratios. Specifically, the treatment assignments are sampled from a Bernoulli distribution with $p=0.5$, $p=0.1$, and $p=0.9$ for assigning a subject to the treatment. Visualisations of the outcomes in both the treatment and control groups, along with the ground truth CATEs, are illustrated in Figure~\ref{fig_IllustrativeDataSet}. 

\begin{figure}[!t]
	\centering
	\begin{adjustbox}{minipage=\linewidth,scale=0.7}
		\begin{subfigure}{.32\textwidth}
			\includegraphics[width=\linewidth]{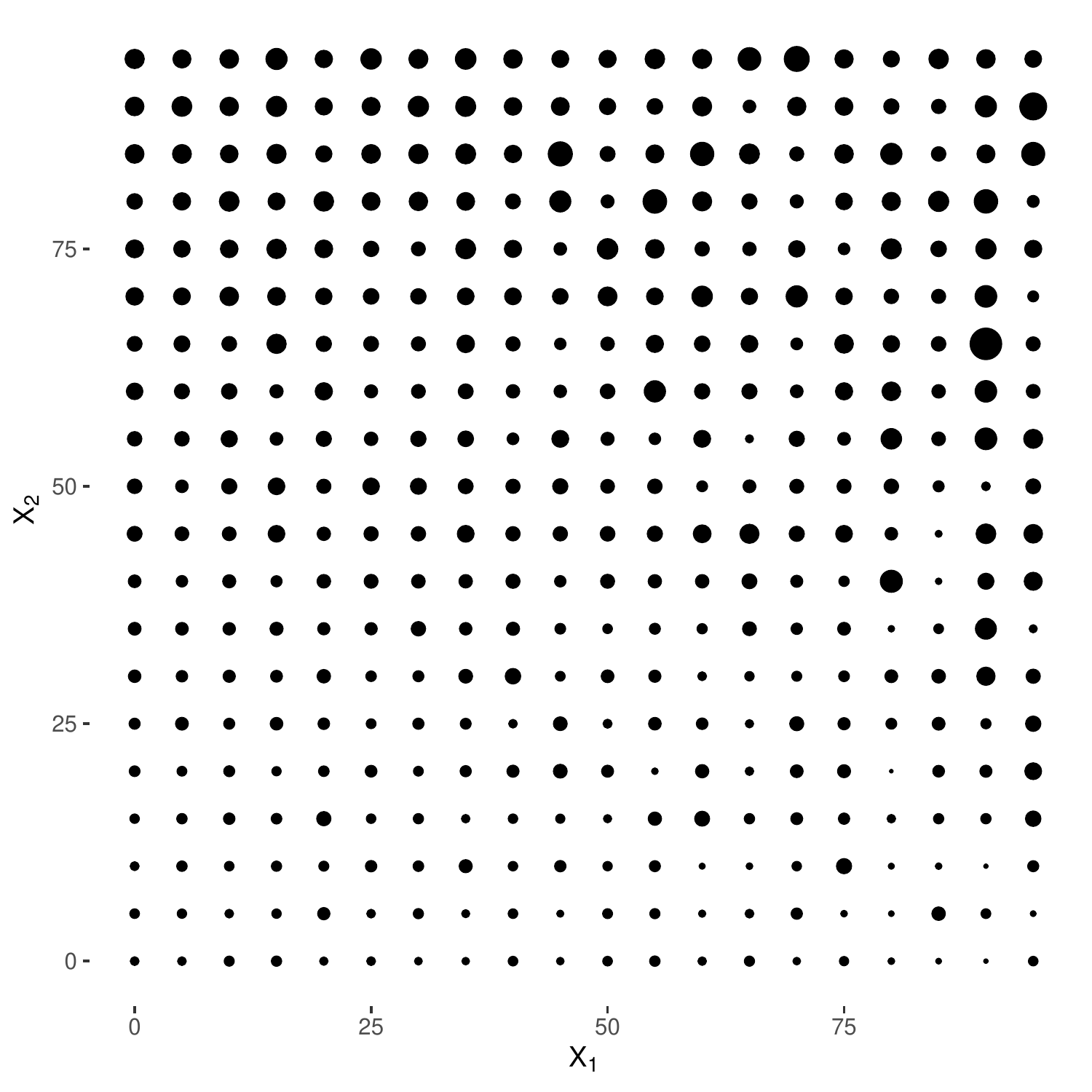}
			\caption{Two-Model LR.}
		\end{subfigure}
		\begin{subfigure}{.32\textwidth}
			\includegraphics[width=\linewidth]{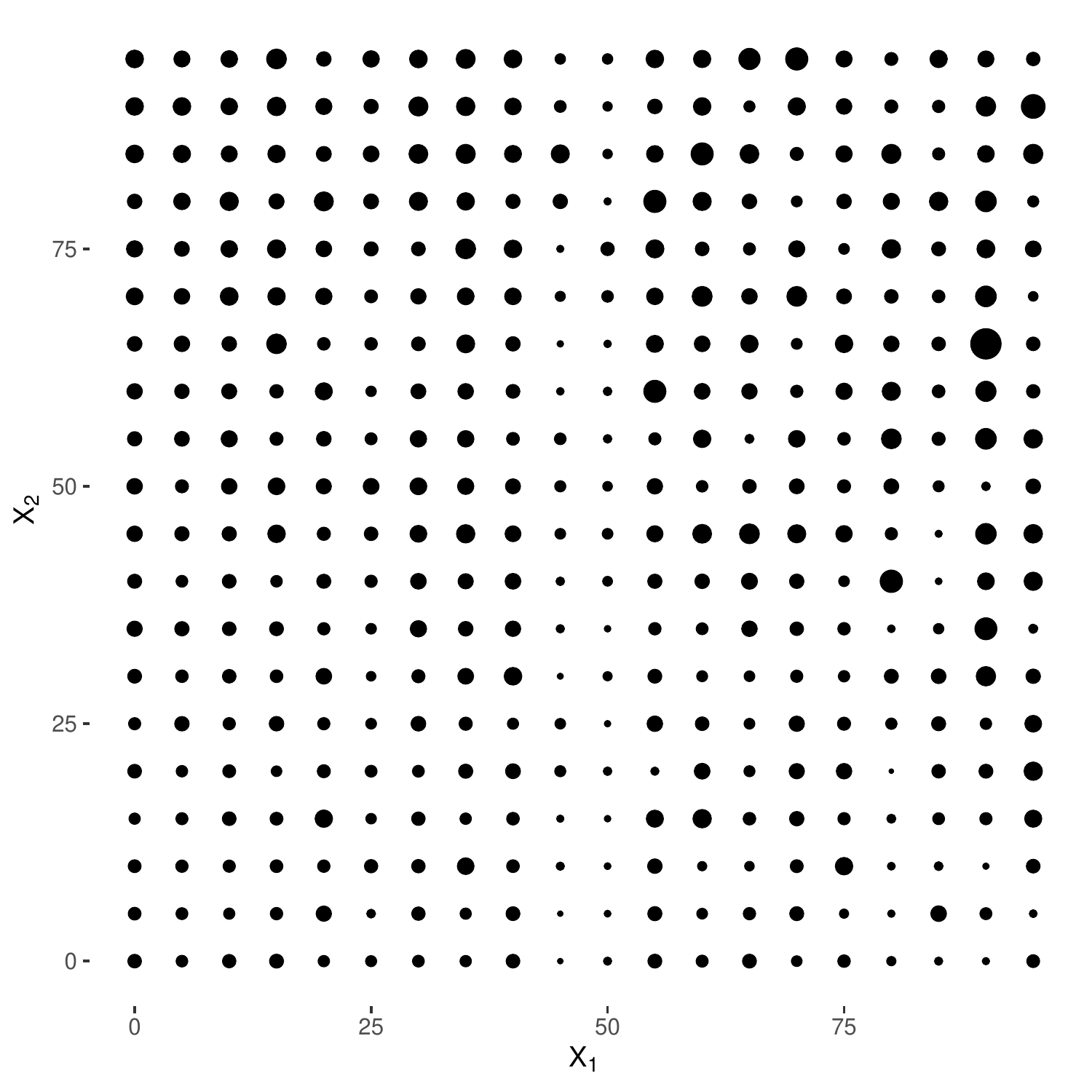}
			\caption{Two-Model RT.}
		\end{subfigure}
		\begin{subfigure}{.32\textwidth}
			\includegraphics[width=\linewidth]{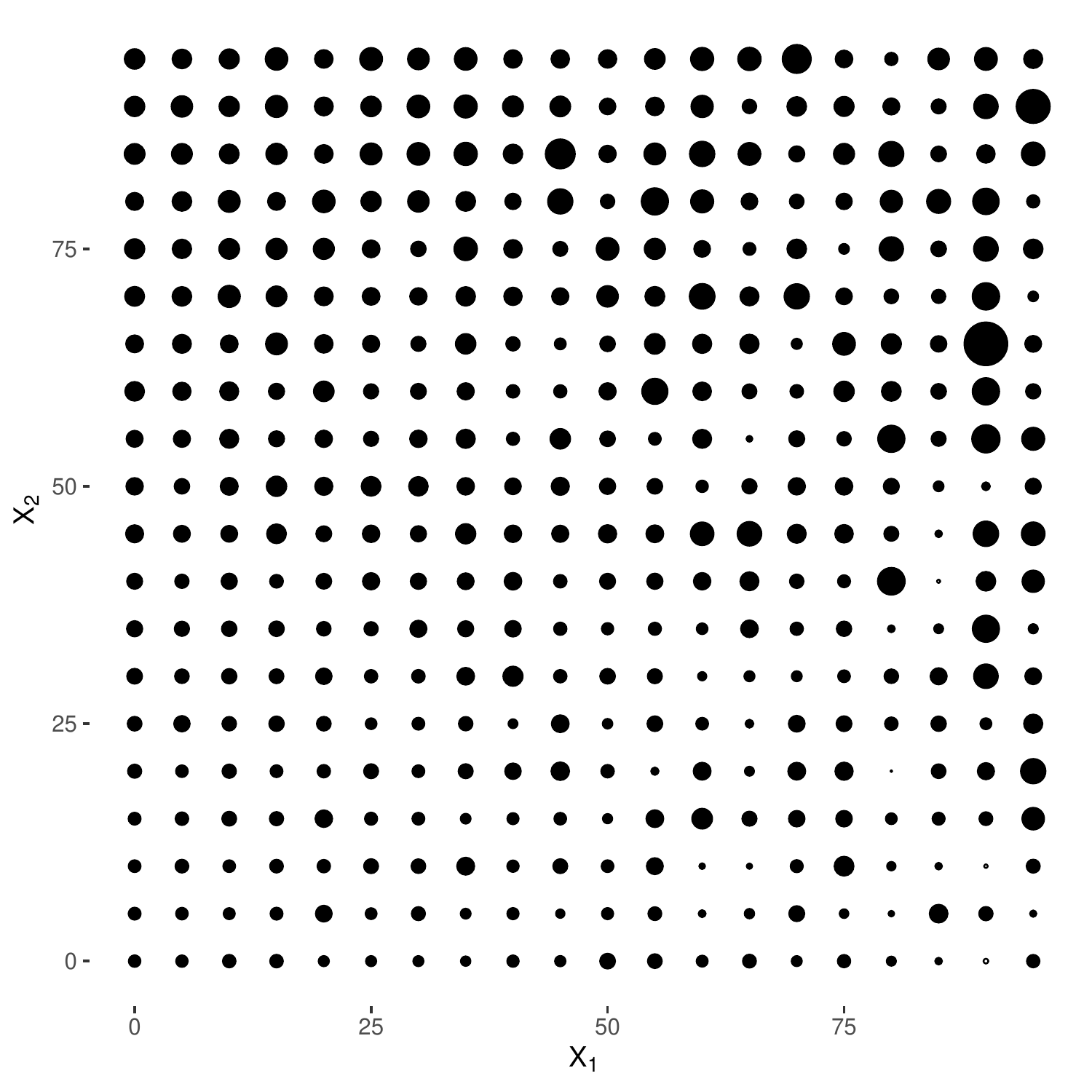}
			\caption{Two-Model RF.}
		\end{subfigure}
		\\
		\begin{subfigure}{.32\textwidth}
			\includegraphics[width=\linewidth]{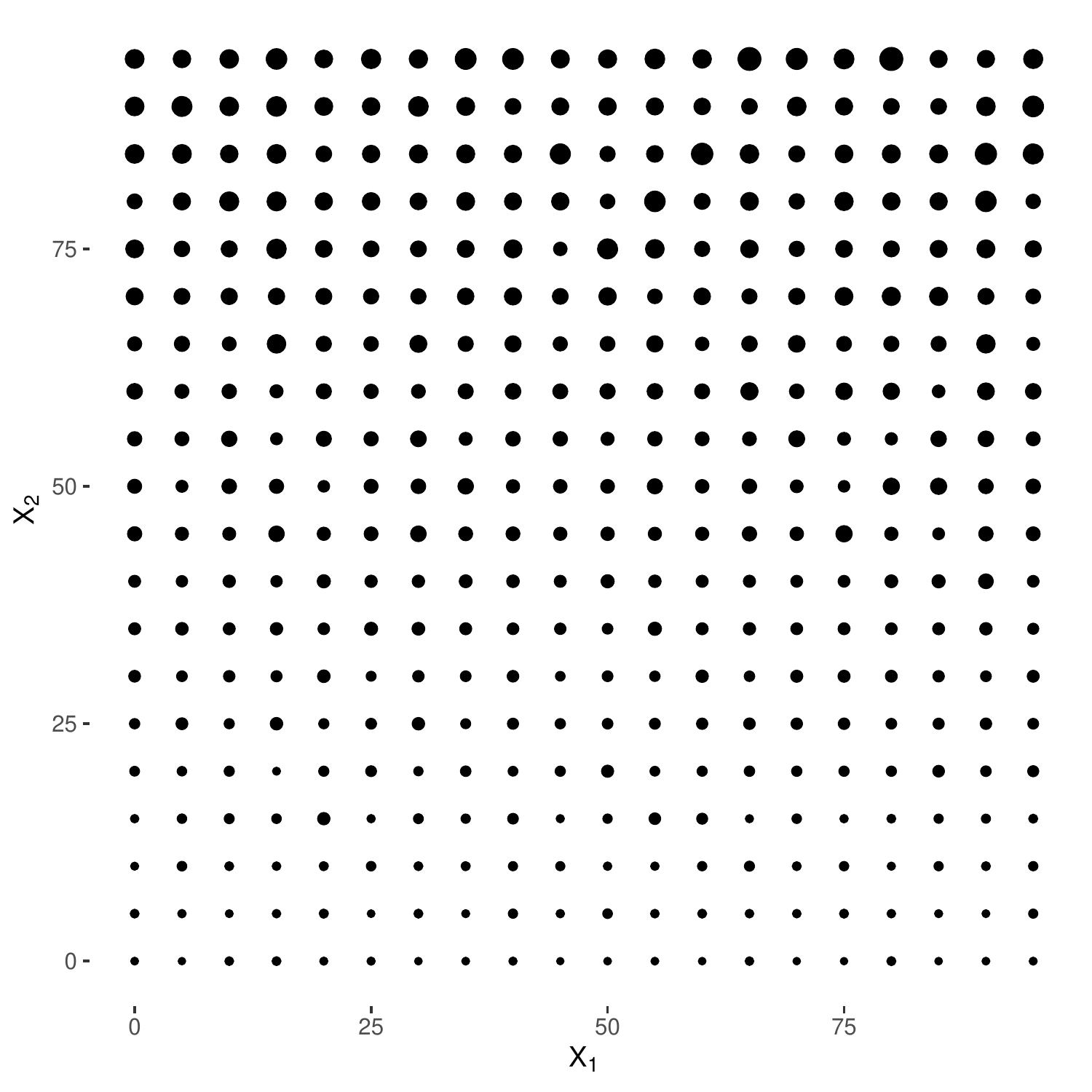}
			\caption{X-Learner LR.}
		\end{subfigure}
		\begin{subfigure}{.32\textwidth}
			\includegraphics[width=\linewidth]{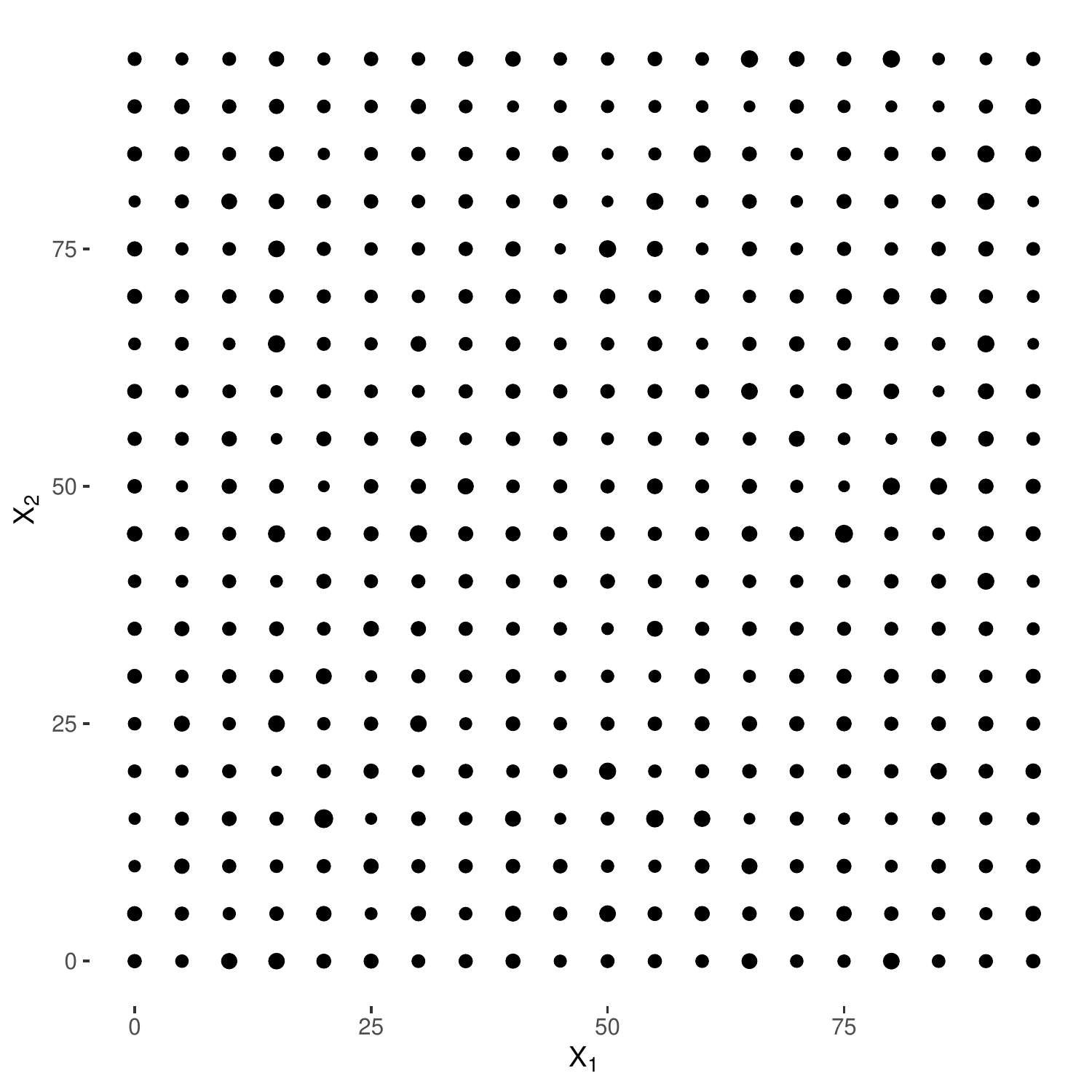}
			\caption{X-Learner RT.}
		\end{subfigure}
		\begin{subfigure}{.32\textwidth}
			\includegraphics[width=\linewidth]{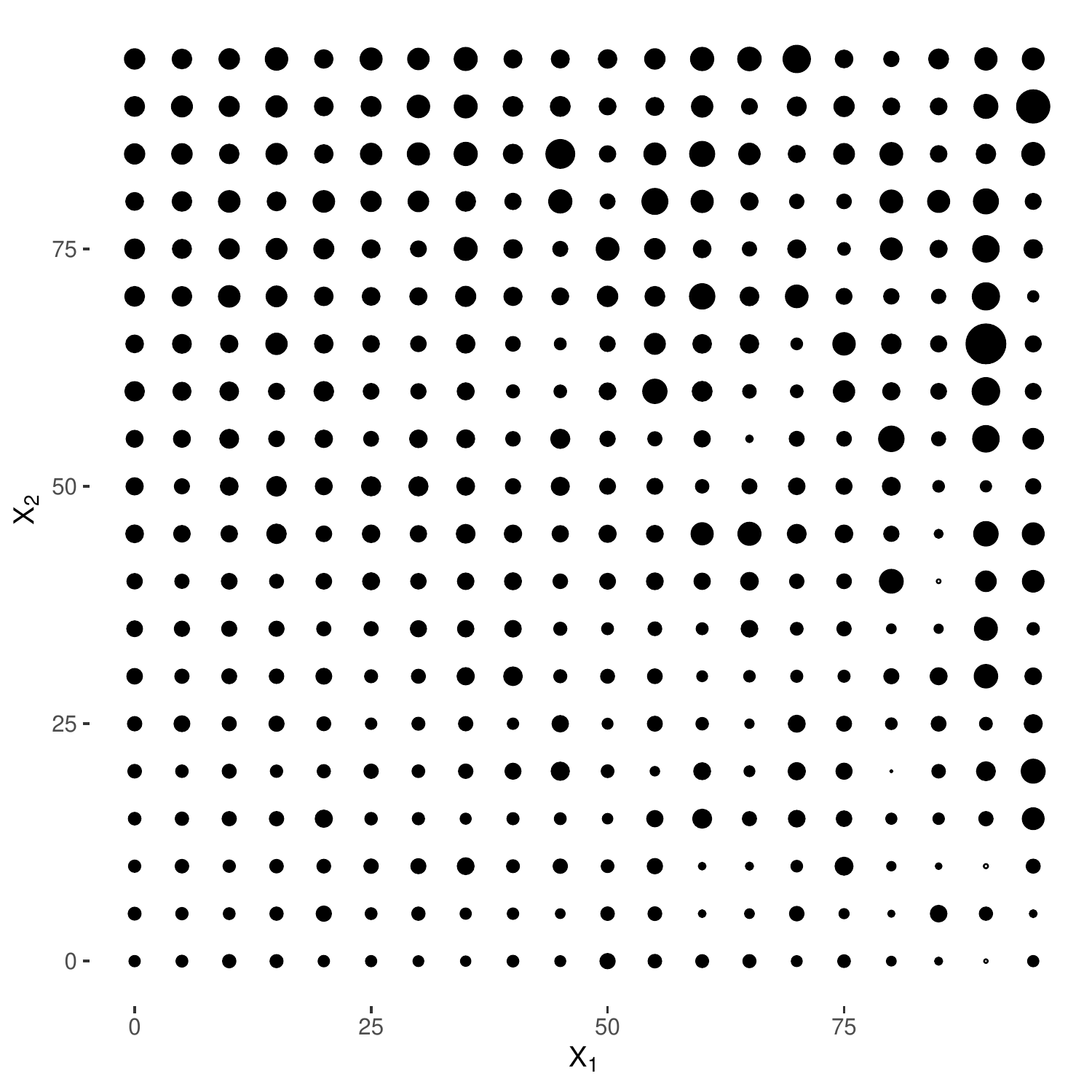}
			\caption{X-Learner RF.}
		\end{subfigure}
		\\
		\begin{subfigure}{.32\textwidth}
			\includegraphics[width=\linewidth]{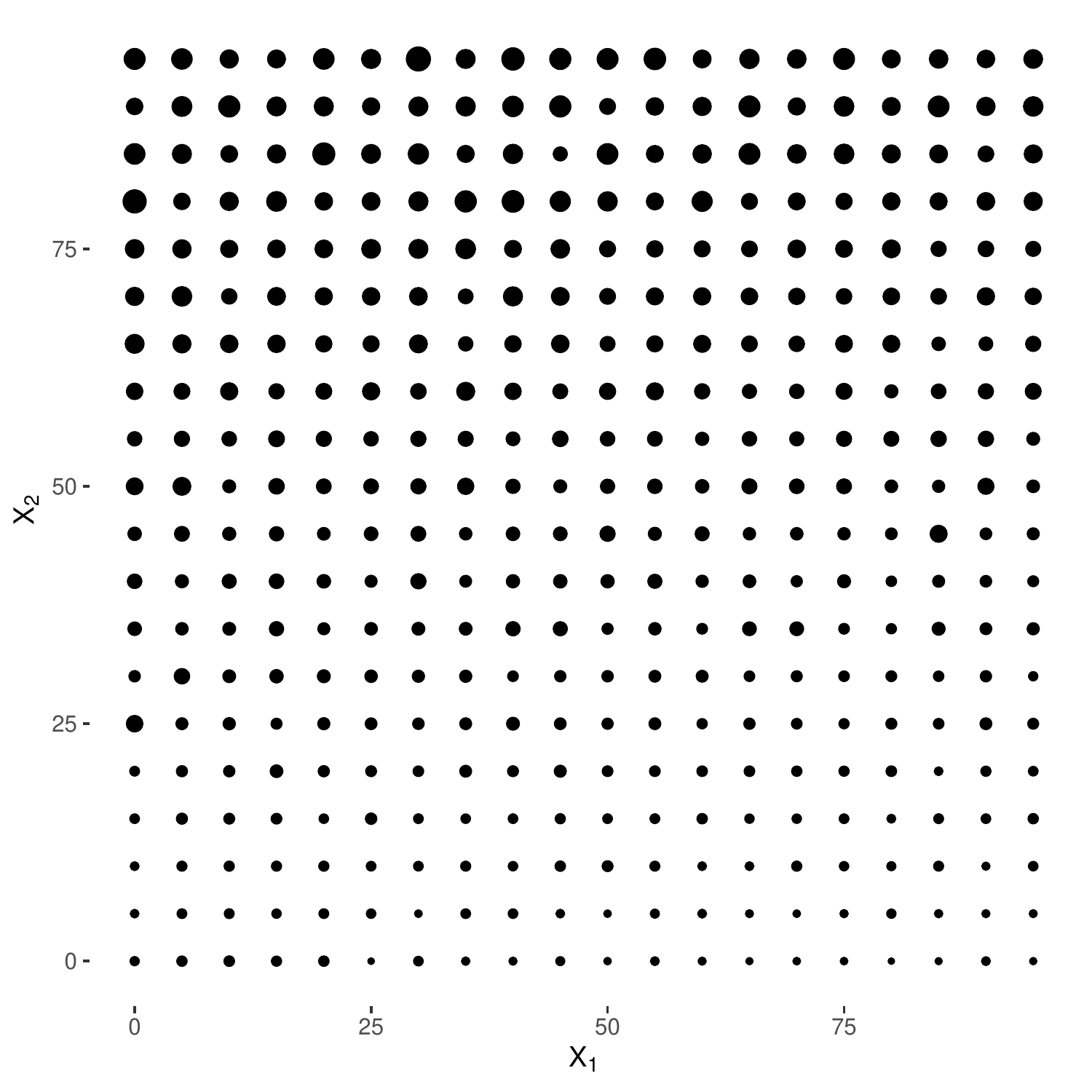}
			\caption{Transformed Outcome LR.}
		\end{subfigure}	
		\begin{subfigure}{.32\textwidth}
			\includegraphics[width=\linewidth]{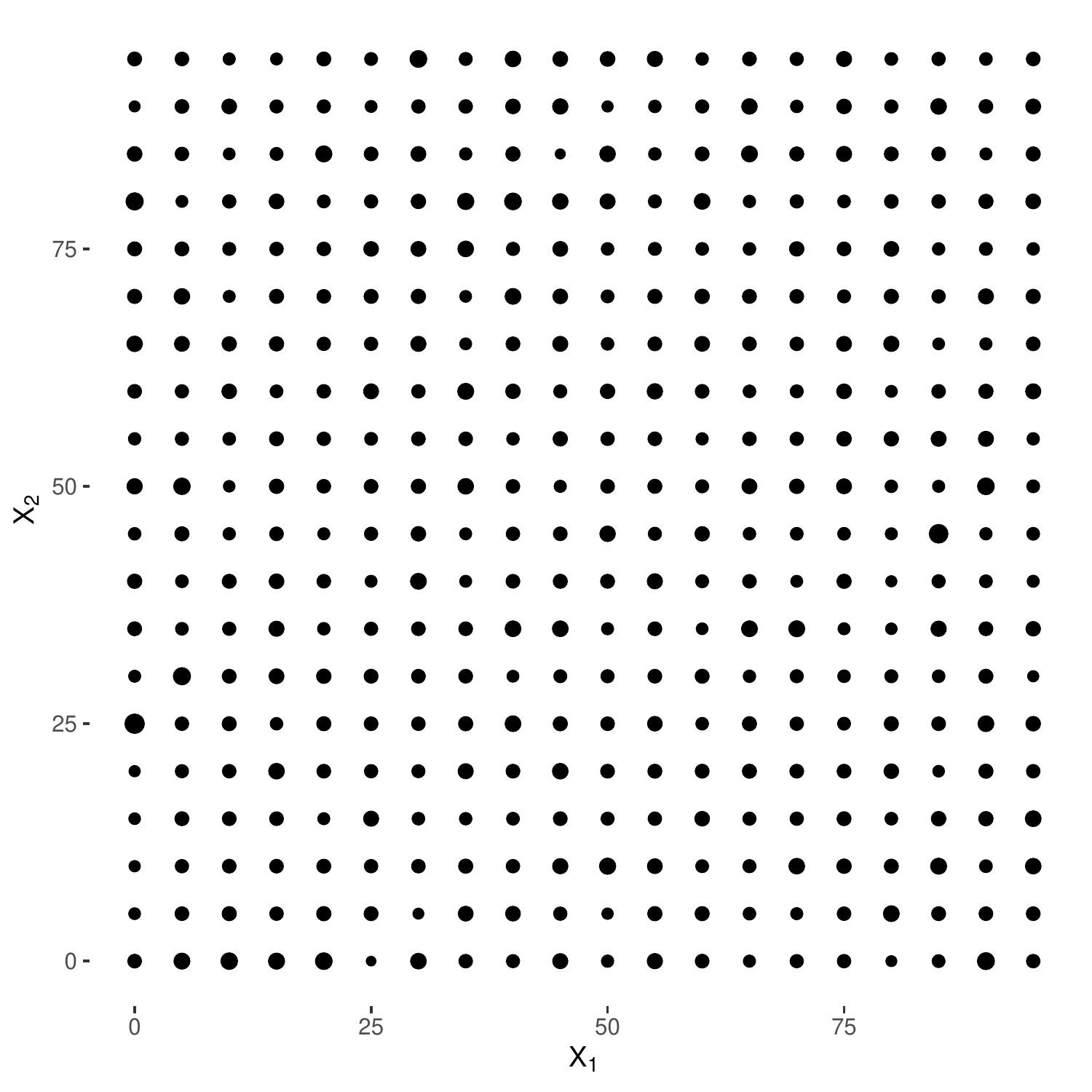}
			\caption{Transformed Outcome RT.}
		\end{subfigure}	
		\begin{subfigure}{.32\textwidth}
			\includegraphics[width=\linewidth]{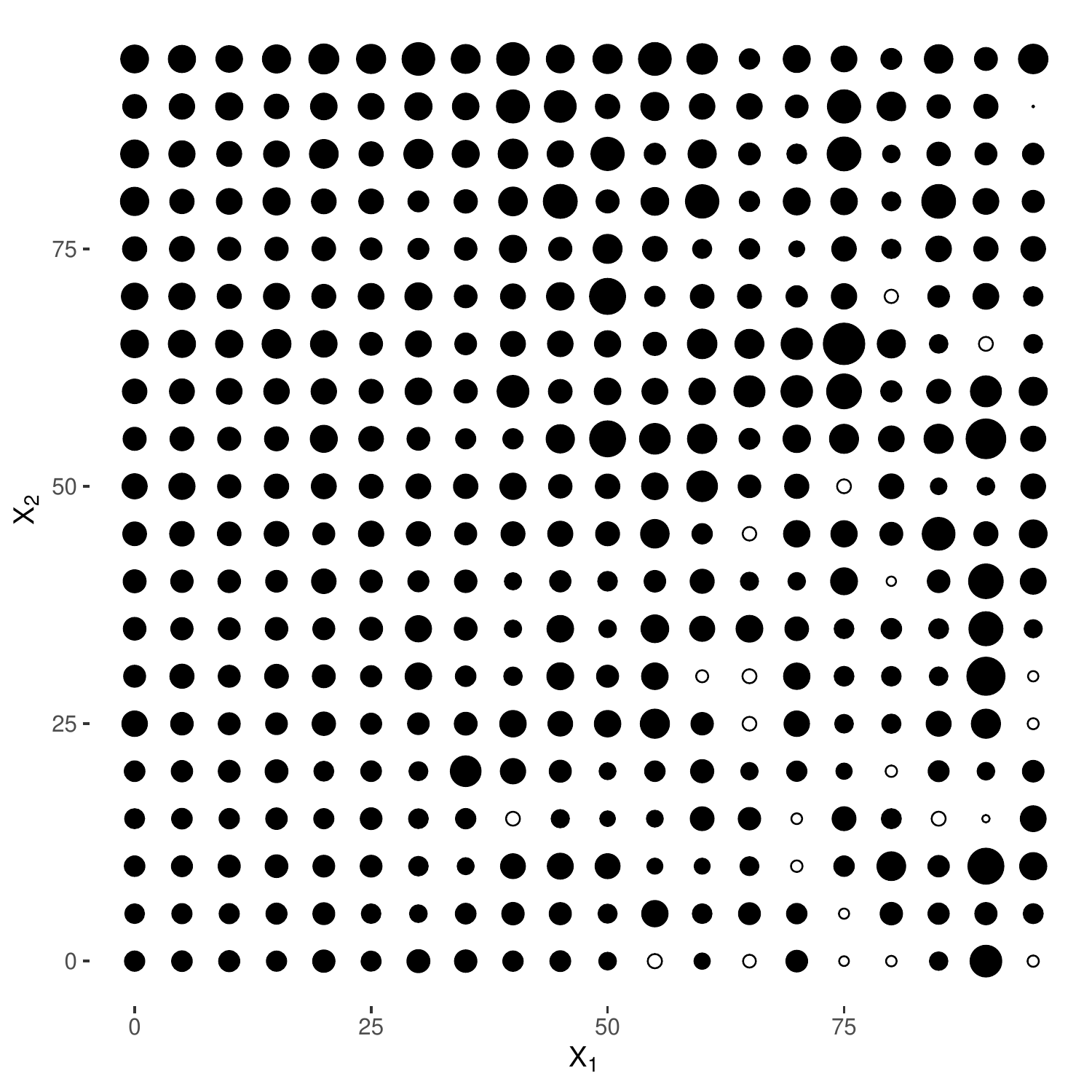}
			\caption{Transformed Outcome RF.}
		\end{subfigure}
		\\
		\begin{subfigure}{.32\textwidth}
			\includegraphics[width=\linewidth]{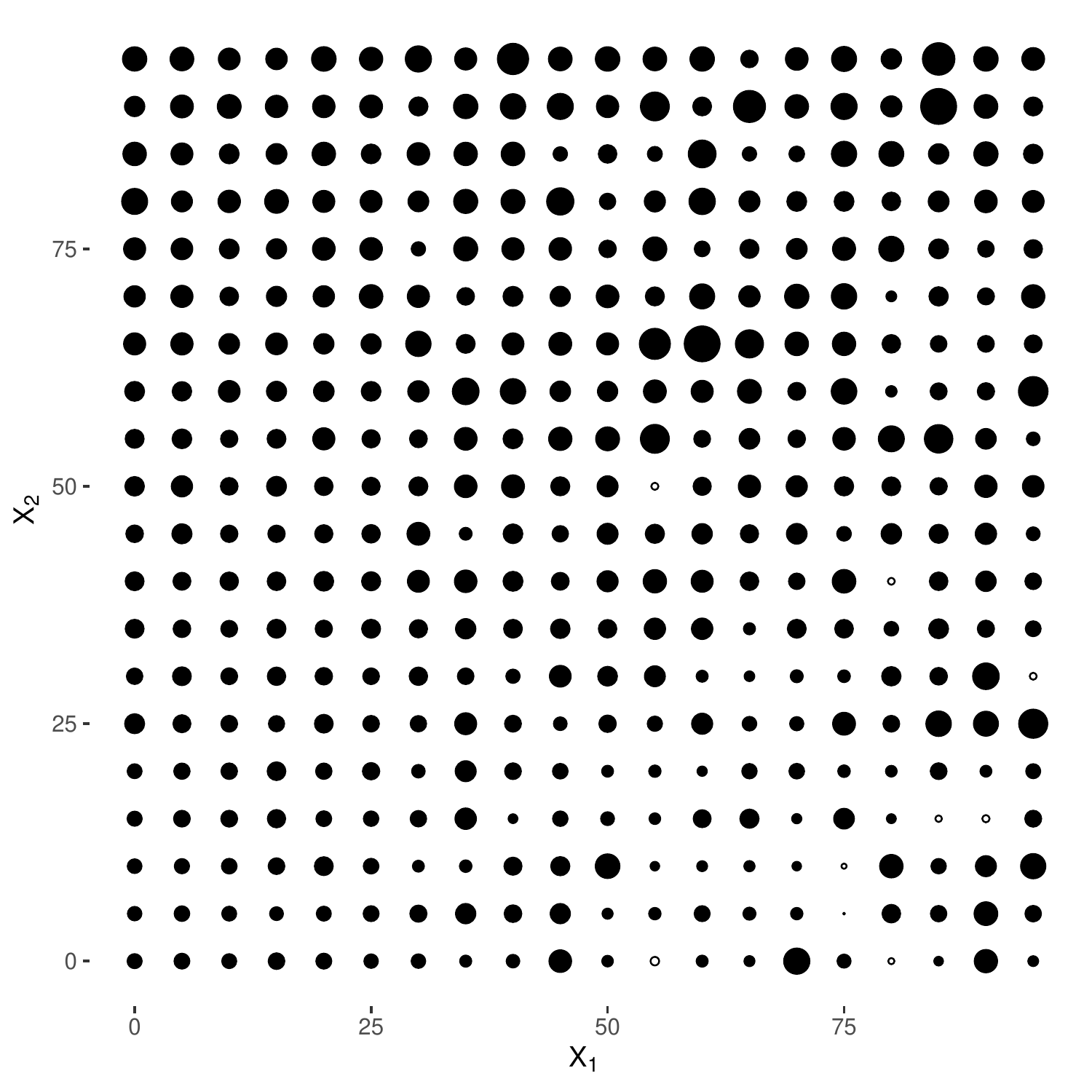}
			\caption{Squared t-Statistics Tree.}
		\end{subfigure}
		\begin{subfigure}{.32\textwidth}
			\includegraphics[width=\linewidth]{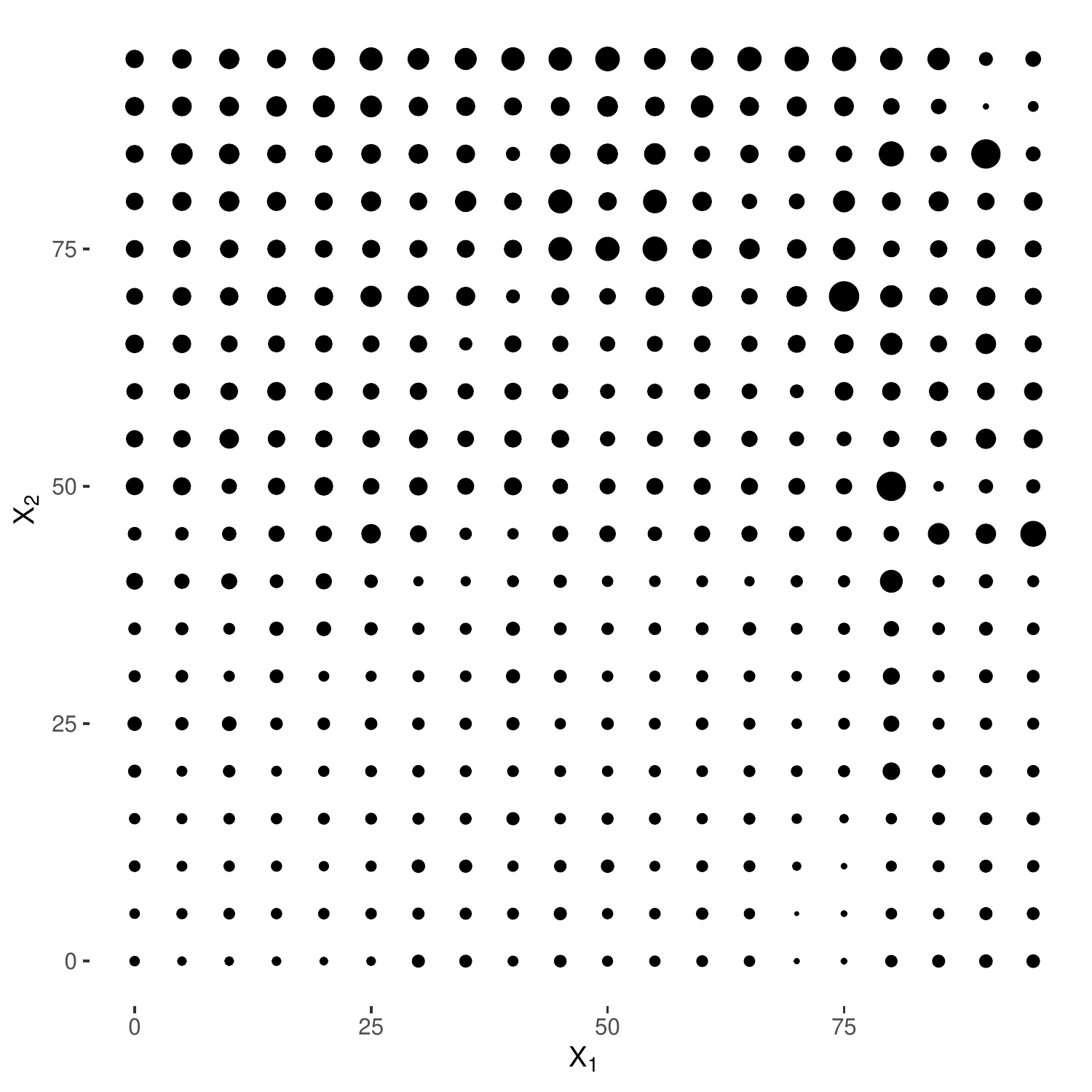}
			\caption{Causal Tree.}
		\end{subfigure}
		\begin{subfigure}{.32\textwidth}
			\includegraphics[width=\linewidth]{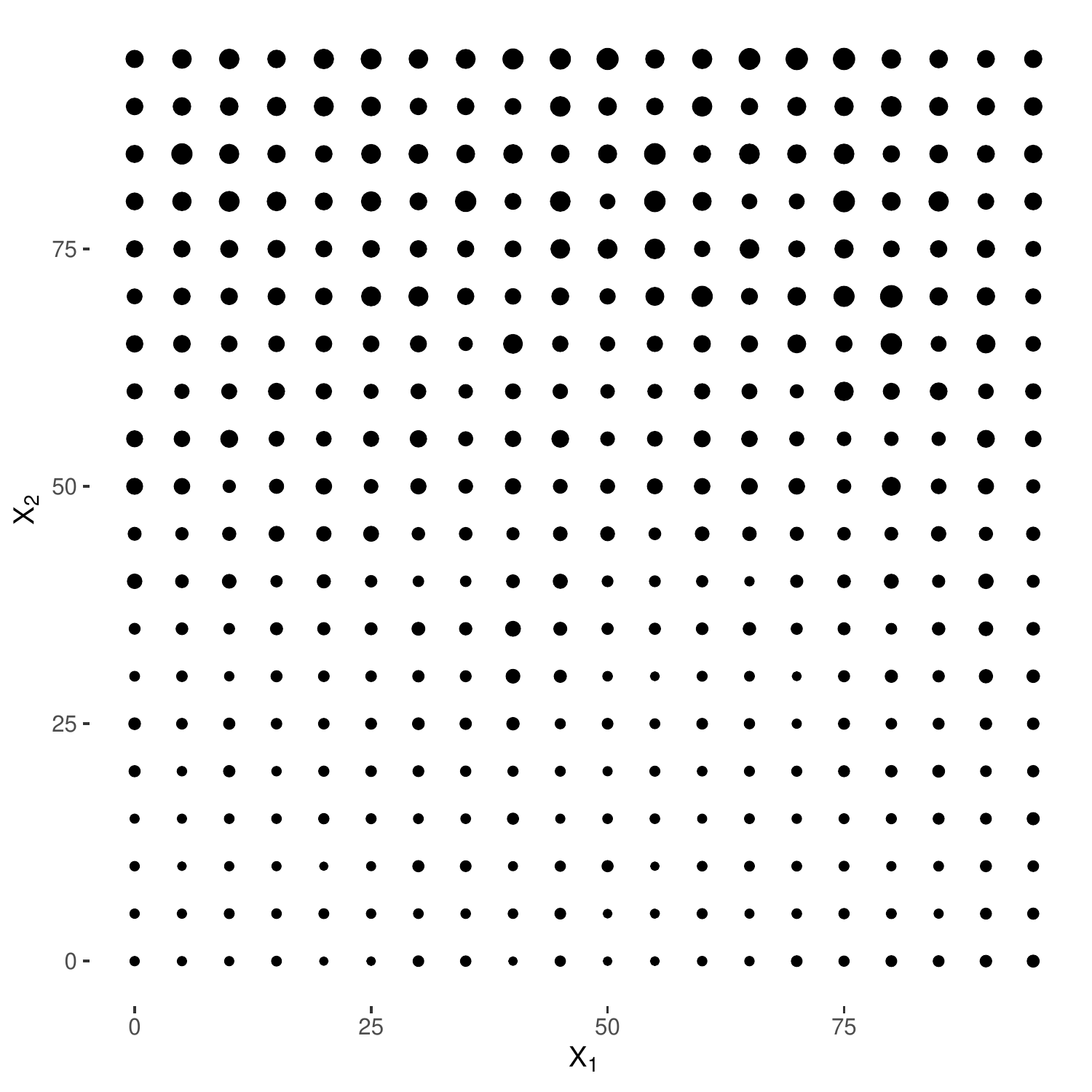}
			\caption{Causal Forest.}
		\end{subfigure}
	\end{adjustbox}
	\caption{Estimated CATEs of representative methods on the balanced synthetic dataset. The disc plots use the same scale as  in Subfigure \ref{fig_IllustrativeDataSet}(c).}
	\label{fig_Results-Representative}	
	
\end{figure}

\begin{figure}[!t]
	\centering
	\begin{adjustbox}{minipage=\linewidth,scale=0.7}
	\begin{subfigure}{.32\textwidth}
		\includegraphics[width=\linewidth]{balanced_two_random_forest.pdf}
		\caption{Two-Model RF (balanced).}
	\end{subfigure}	
	\begin{subfigure}{.32\textwidth}
		\includegraphics[width=\linewidth]{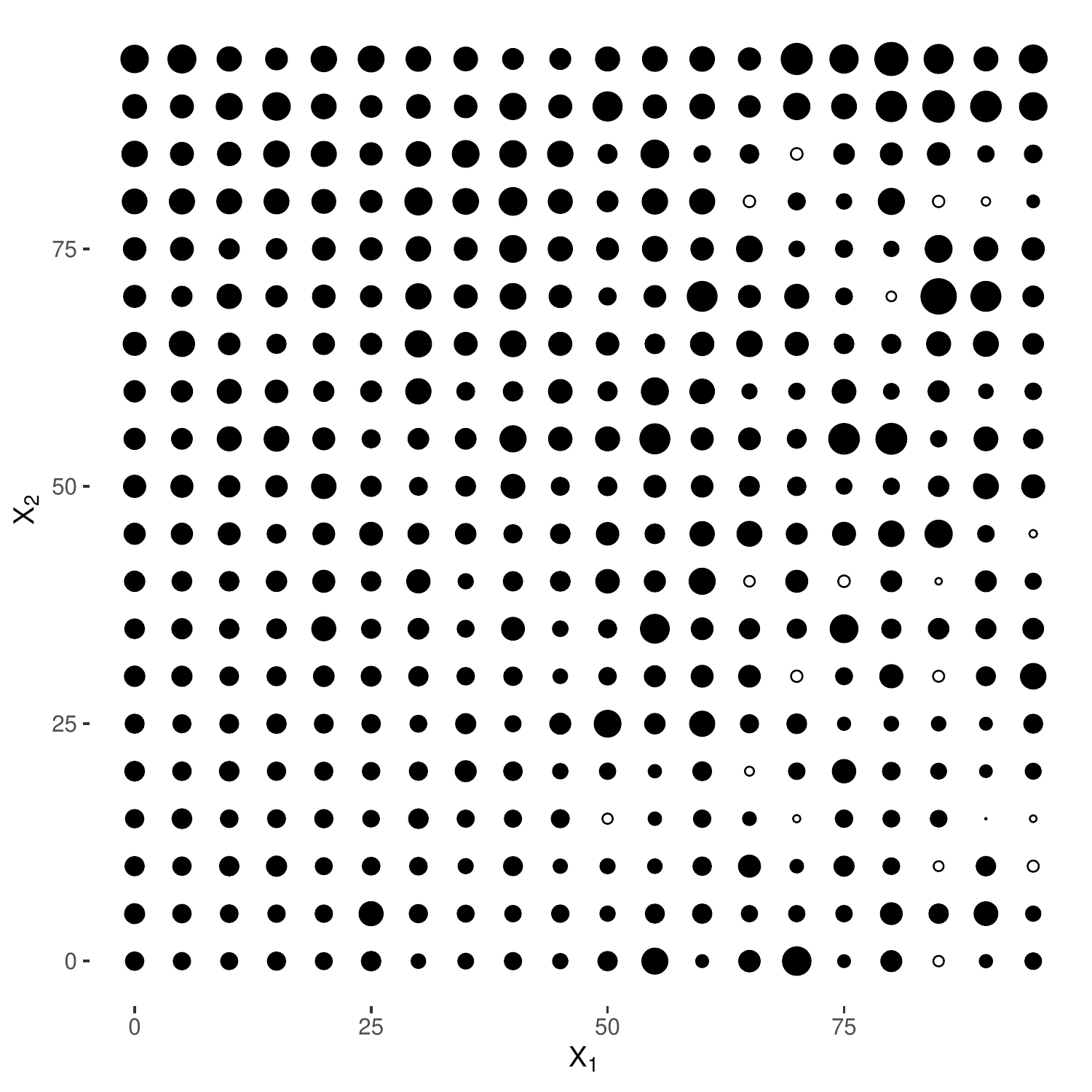}
		\caption{Two-Model RF (p=0.1).}
	\end{subfigure}	
	\begin{subfigure}{.32\textwidth}
		\includegraphics[width=\linewidth]{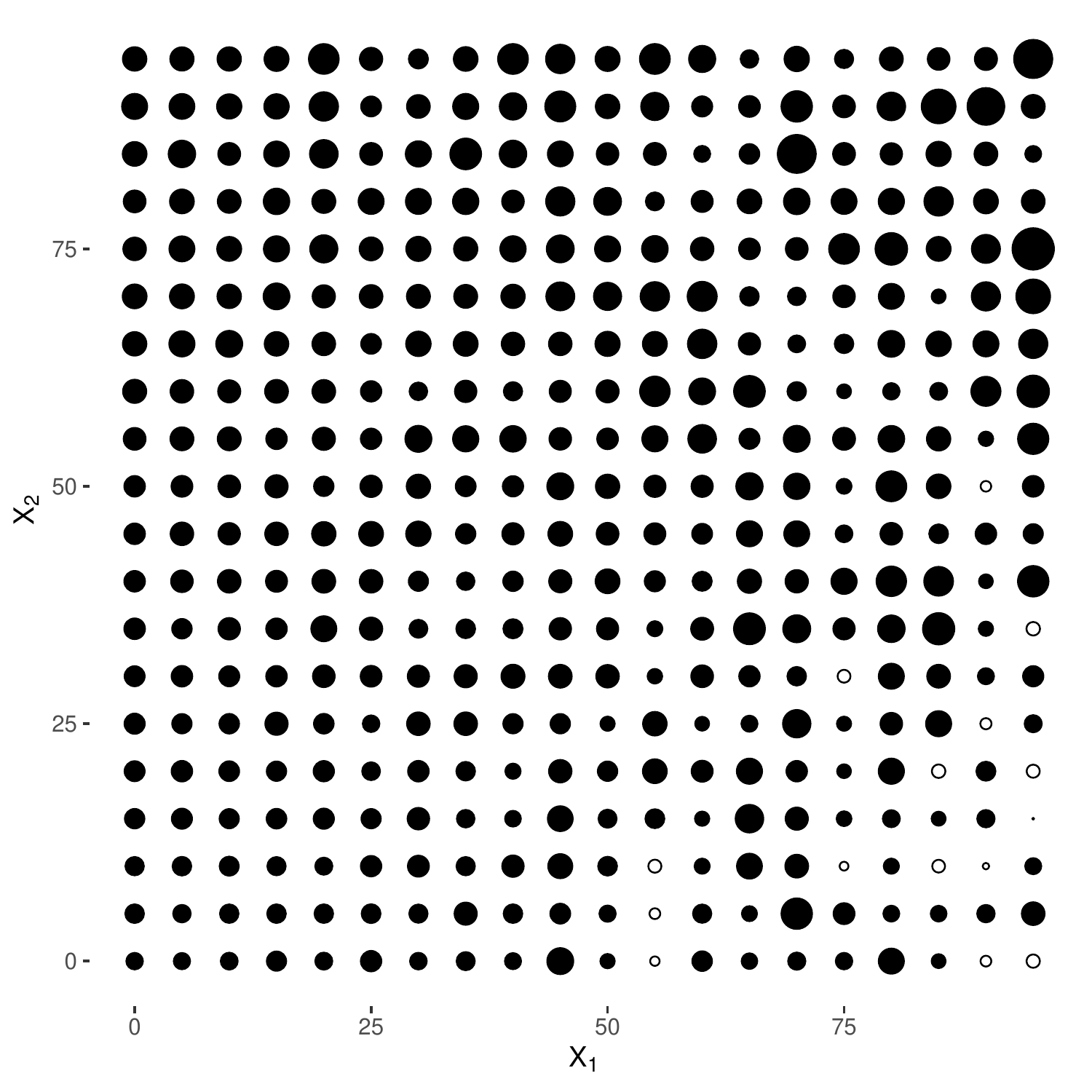}
		\caption{Two-Model RF (p=0.9).}
	\end{subfigure}
	\\
	\begin{subfigure}{.32\textwidth}
		\includegraphics[width=\linewidth]{balanced_ct.pdf}
		\caption{Causal Tree (balanced).}
	\end{subfigure}	
	\begin{subfigure}{.32\textwidth}
		\includegraphics[width=\linewidth]{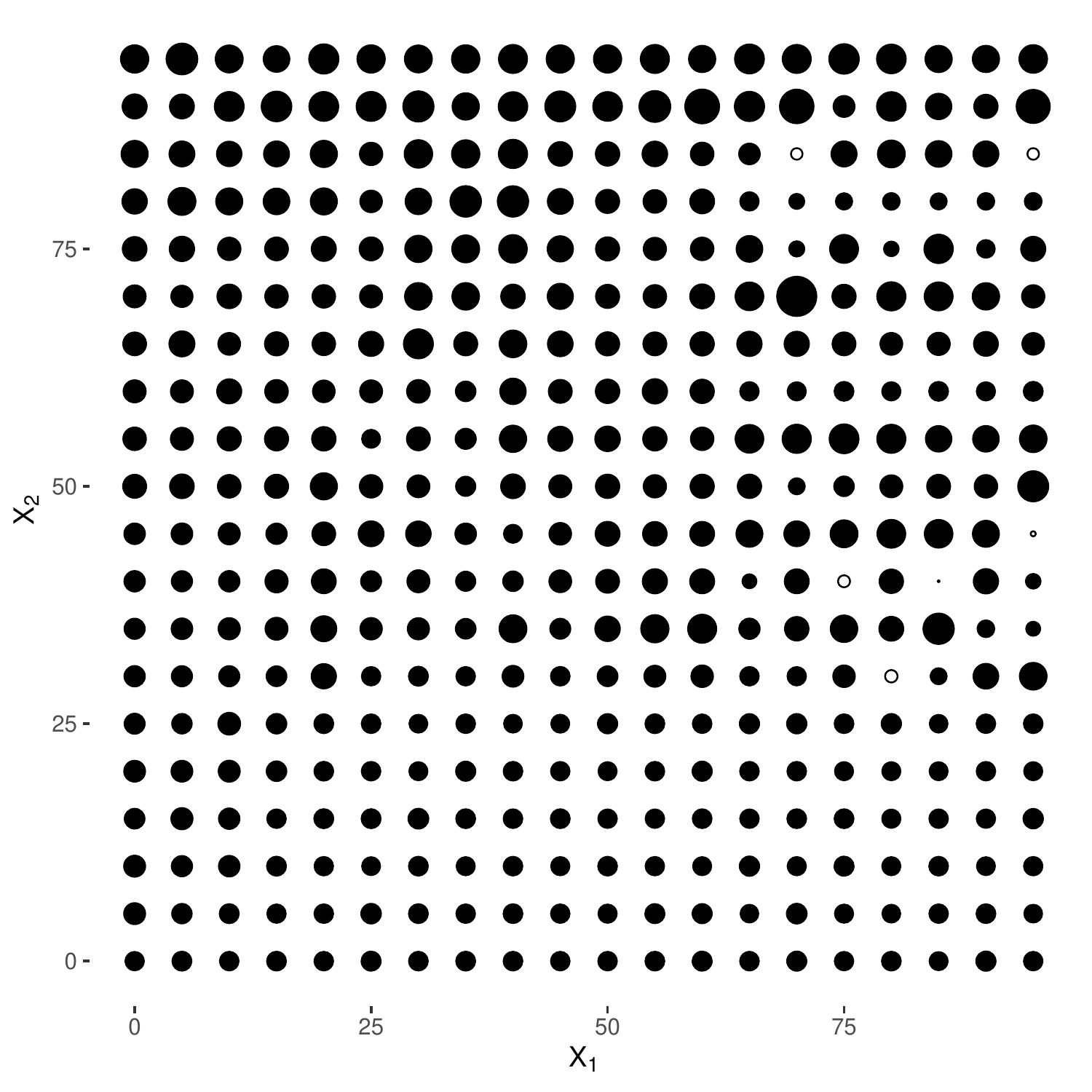}
		\caption{Causal Tree (p=0.1).}
	\end{subfigure}	
	\begin{subfigure}{.32\textwidth}
		\includegraphics[width=\linewidth]{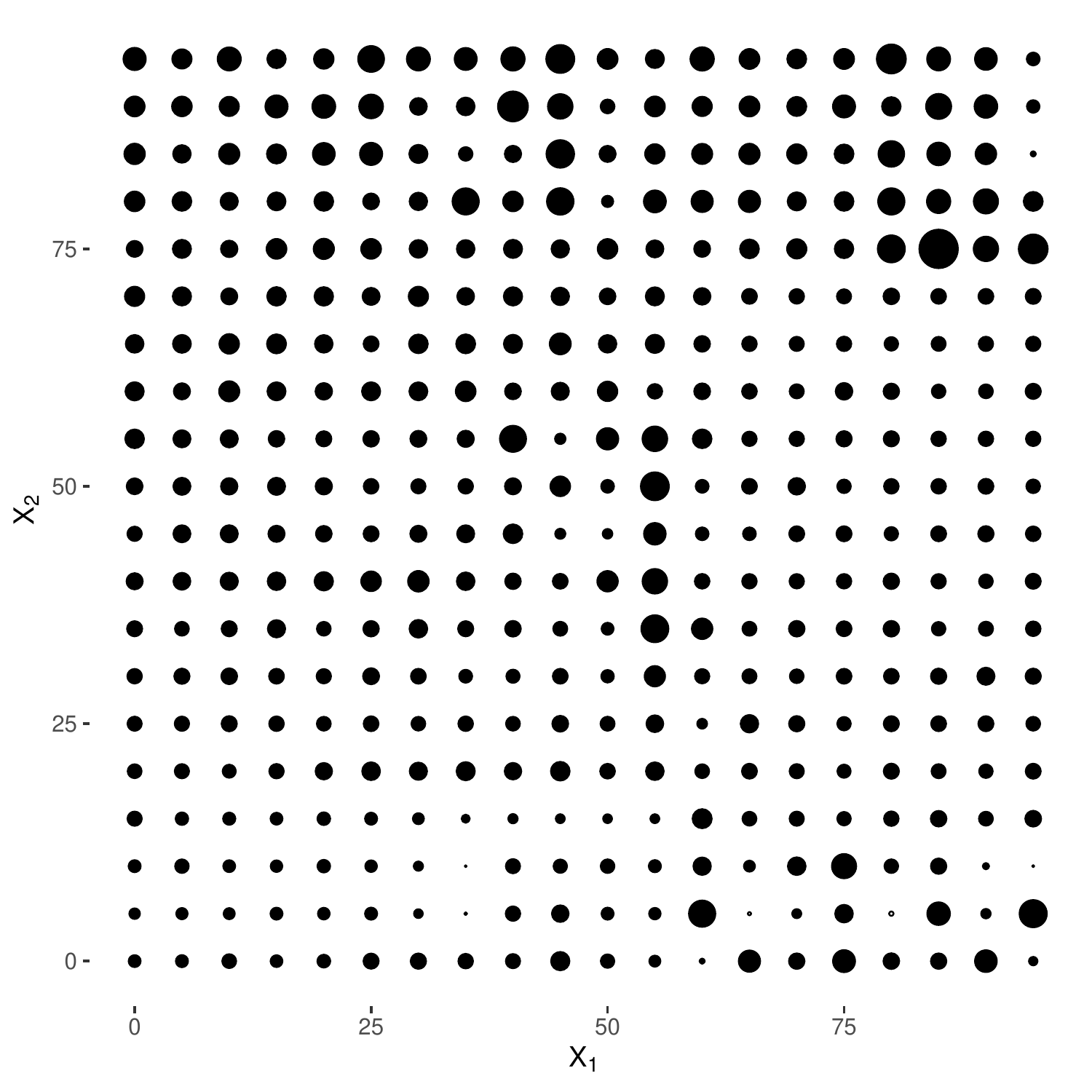}
		\caption{Causal Tree (p=0.9).}
	\end{subfigure}
	\\
	\begin{subfigure}{.32\textwidth}
		\includegraphics[width=\linewidth]{balanced_x_linear_regression.pdf}
		\caption{X-Learner RF (balanced).}
	\end{subfigure}
	\begin{subfigure}{.32\textwidth}
		\includegraphics[width=\linewidth]{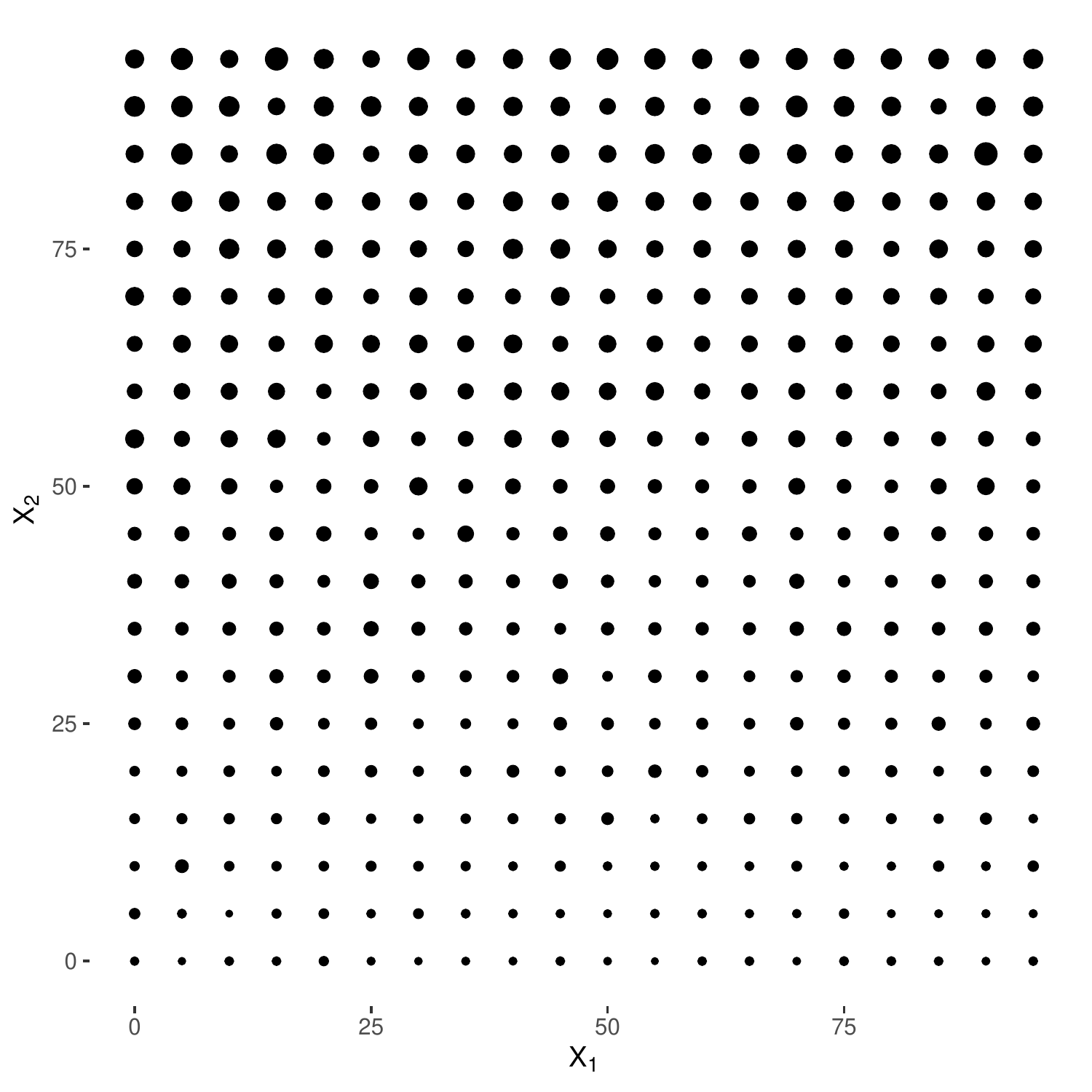}
		\caption{X-Learner RF (p=0.1).}
	\end{subfigure}
	\begin{subfigure}{.32\textwidth}
		\includegraphics[width=\linewidth]{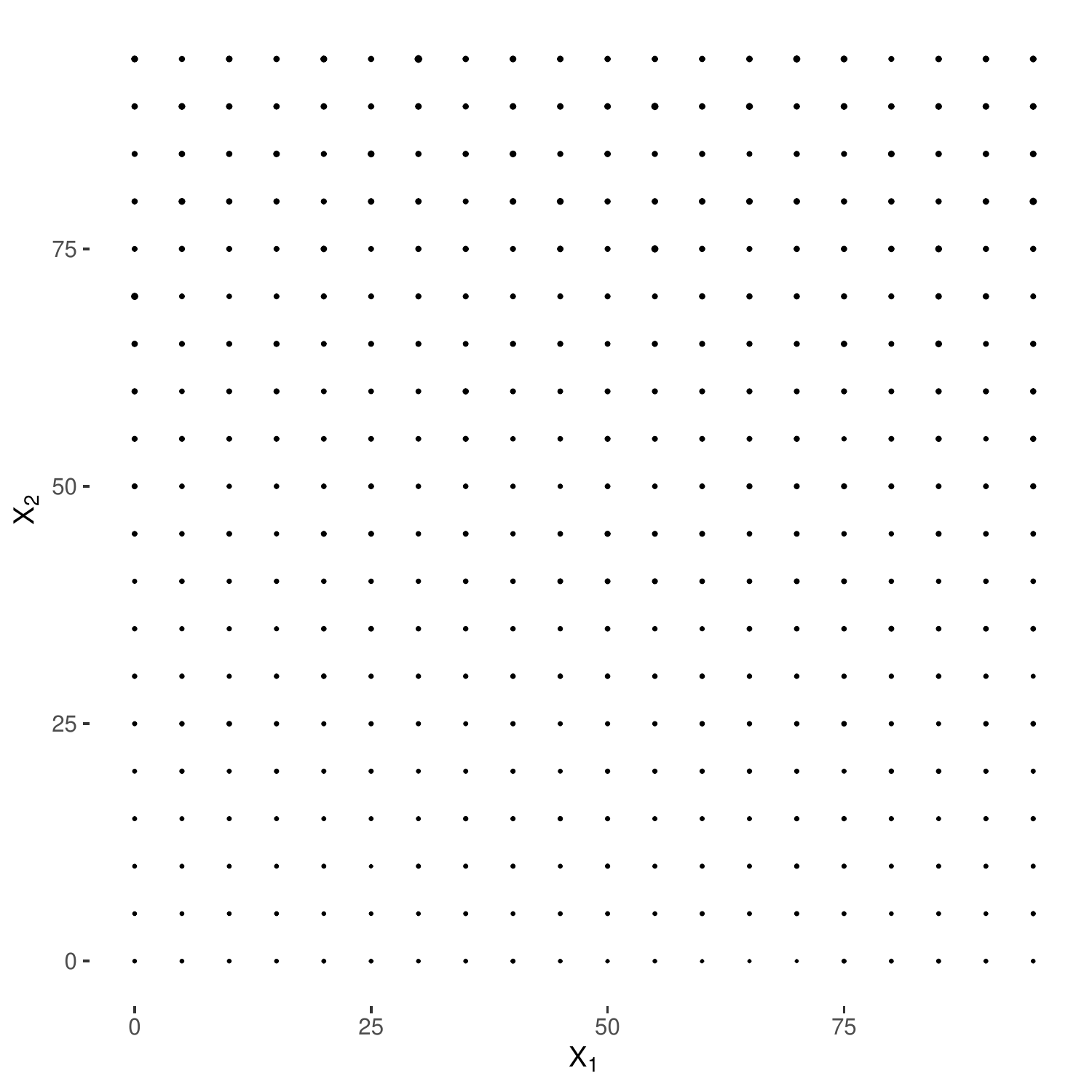}
		\caption{X-Learner RF (p=0.9).}
	\end{subfigure}
	\end{adjustbox}
	\caption{Illustration of  X-Learner linear regression, transformed outcome linear regression, and causal forest in datasets of different treatment and control sample ratios. Hollow discs indicate that the estimated CATE is negative. The disc plots use the same scale as  in Subfigure \ref{fig_IllustrativeDataSet}(c).}
	\label{fig_differentTreatmentRate}

\end{figure}

We use disc plots for visualisation. In the disc plots, the horizontal and vertical axes represent the covariates $X_1$ and $X_2$, respectively. The size of a disc is proportional to the average value of the outcomes (Y in Figure \ref{fig_IllustrativeDataSet}(a) and Figure 3(b)) ) or CATEs (Figure \ref{fig_IllustrativeDataSet}(c)) for the subjects binned within the two-dimensional space with a width of 5.  
For example, the size of the lower-left-most disc in Figure \ref{fig_IllustrativeDataSet}(a) represents the average treated outcomes for subjects in the space of $x_1 \in [0,5)$ and $x_2 \in [0,5)$. The lower-left disc in Figure \ref{fig_IllustrativeDataSet}(c) indicates the average CATE for subjects in that square. 
We keep the scale of the discs the same for Figure \ref{fig_IllustrativeDataSet}(a) and Figure \ref{fig_IllustrativeDataSet}(b), but use a different scale for Figure \ref{fig_IllustrativeDataSet}(c), since the average CATEs are much smaller than the outcome values. 
The estimated CATEs of some representative methods are shown in Figure~\ref{fig_Results-Representative} on the balanced dataset with $p=0.5$. Furthermore, we also illustrate the behaviour of the selected methods (the best from the previous illustration) on datasets with an imbalanced treatment ratio $p=0.1$ and $p=0.9$ in Figure~\ref{fig_differentTreatmentRate}. 

When the models are specified in the same way as the data generating procedure, i.e., using linear regression (LR) as the base learner, the CATE modelling behaviour of the algorithms is similar to the ground-truth CATEs. This can be seen from the disc plots of the two-model LR (Figure \ref{fig_Results-Representative}(a)), X-Learner LR (Figure \ref{fig_Results-Representative}(d)), and transformed outcome LR (Figure \ref{fig_Results-Representative}(g)), where the trends of the disc plots are similar to the trends in Figure~\ref{fig_IllustrativeDataSet}(c). However, it is worth noting that in most applications the ground-truth relationships between the CATE and the covariates are unknown to the users, and thus specifying the parametric form of the base learner is difficult. 

When the models are specified differently from the data generating procedure, i.e., using regression trees (RT) or random forests (RF) as base learners, we observe that the modelling behaviour of different methods is significantly different.
In the second and the third columns of Figure~\ref{fig_Results-Representative}, we can hardly see any two methods which produce the same CATE estimations. Furthermore, the disc plots of the two tailored tree-based methods---the squared t-statistics tree (Figure \ref{fig_Results-Representative}(j)) and causal tree (Figure \ref{fig_Results-Representative}(k))---are also different from each other. Even with the same method, the performance changes significantly when datasets change with different treatment and control sample ratios, as shown in Figure~\ref{fig_differentTreatmentRate}. 
We can see that X-Learner RF performs well when the data are balanced and when the number of treated subjects is smaller than the number of control subjects ($p=0.1$ in Figure~\ref{fig_differentTreatmentRate}(h)). However, its performance significantly worsens when the treated subjects are dominant in the data ($p=0.9$ in Figure~\ref{fig_differentTreatmentRate}(i)). 
An explanation for this can be seen from Equation (\ref{formula_xlearner}). When $p=0.9$, the dominating component of the equation is $\hat{\tau}_0(\pmb{x})$, which is estimated from the control subjects and may underestimate the treatment effects. We observe that when the weight for $\hat{\tau}_0(\pmb{x})$ is reduced, X-Learner RF performs better. However, in practice, the weight can be difficult to determine when the ground truth is unavailable.

CATEs are difficult to model. The synthetic dataset has only two covariates with a sufficiently large number of subjects, and the ground-truth CATEs follow a linear relationship with one covariate ($X_2$). It is reasonable to expect that most methods should present nearly perfect estimations of the datasets. However, the results are far from perfect. 
When a method is not correctly specified, i.e., when a non-liner modelling method is used for the data generated by a linear relationship, the CATE estimations change significantly when the proportion of treated samples changes, as shown in Figure ~\ref{fig_differentTreatmentRate}.
Even when a method is correctly specified (two-model LR, X-Learner LR, and transformed outcome LR), there is still a significant degree of underestimation of the CATEs. 

A major reason for the difficulties with CATE estimation is the counterfactual problem where we can only observe one treatment/control outcome for any subject.
Another reason is that the average values of the CATEs are weak in the data, since they are only at a magnitude of around 1/20 of an outcome value. Unfortunately, such weak signals are common in many real-world applications. In marketing promotion and personalised medicine, the scale of treatment effects is commonly quite small. Therefore, treatment effect heterogeneity modelling and uplift modelling are challenging and will require a significant amount of research in the future.

\subsection{Evaluation Metrics}
\label{sec_metrics}
Before advancing to further discussions, we formally introduce the metrics commonly used when evaluating causal effect heterogeneity modelling or uplift modelling methods. 
The evaluation of CATE estimation methods on real-world datasets is a difficult task. 
Several metrics have been proposed from both communities depending on whether ground-truth CATEs are available.

\subsubsection{Metrics With Known Ground-Truth Treatment Effects}
When the ground-truth CATEs are known, i.e., in synthetic datasets or semi-synthetics where the potential outcomes are simulated based on real-world covariates, the precision in the estimation of heterogeneous effects (PEHE) \citep{Hill2011} is a straightforward metric used for datasets with ground-truth individual treatment effects. Given the ground-truth CATE, the PEHE is defined as:
\begin{align}
\text{PEHE} = \frac{1}{n} \sum\limits_i^{n}(\hat{\tau}(\pmb{x}_i)-\tau(\pmb{x}_i))^2.
\end{align}
In other words, the PEHE measures the mean squared error between the estimated treatment effects and the ground-truth treatment effects.

Another performance metric is the mean absolute percentage error (MAPE) of the CATE estimation, which is defined as:
\begin{align}
\text{MAPE} = \frac{1}{n} \sum\limits_{i}^{n} \vert \frac{\hat{\tau}(\pmb{x}_i) - \tau(\pmb{x}_i)}{\tau(\pmb{x}_i)} \vert \times 100\%
\end{align}

\subsubsection{Metrics Without Ground-Truth Treatment Effects}
\label{Sec_curves}
The uplift curve and the Qini curve can be used for the evaluation when the outcome variable is binary without ground-truth CATEs.

The uplift and Qini curves are two closely related metrics proposed in the uplift modelling literature \citep{Radcliffe2007,Gutierrez2017,DiemertEustacheBetleiArtem}. The intuition behind these metrics is that when the subjects are ranked in descending order by their estimated CATEs, with an accurate CATE estimation, subjects with positive outcomes in the treatment group should be ranked higher than those with negative outcomes in the treatment group. Likewise, subjects with negative outcomes in the control group should be ranked higher than those with positive outcomes in the control group.  

To formerly define these metrics, we introduce some notation. For a given CATE estimator $\hat{\tau}$ and subjects $\pmb{x_i}$, let $\pi$ be a descending ordering of subjects according to their estimated treatment effects, i.e., $\hat{\tau}^\pi(\pmb{x}_i) \geq \hat{\tau}^\pi(\pmb{x}_j)$,$\forall i<j$. 
We use $\pi(k)$ to denote the first $k$ subjects. 
Let $R_	{\pi(k)}$ be the count of positive outcomes in $\pi (k)$, i.e., $R_{\pi(k)} = \sum_{i\in \pi(k)}\mathbbm{1}[Y_i=1]$, where $\mathbbm{1}$ denotes the indicator function. 
Furthermore, let $R_{\pi(k)}^{T=1}$ and $R_{\pi(k)}^{T=0}$ be the number of positive outcomes in the treatment and control groups, respectively, from $\pi (k)$. Finally, let $N_{\pi(k)}^{T=1}$ and $N_{\pi(k)}^{T=0}$ be the number of subjects in the treatment and control groups from $\pi (k)$. Now we can define the values of the two curves as:
\begin{align}
&\textrm{uplift}(k) = (\frac{R_{\pi(k)}^{T=1}}{N_{\pi(k)}^{T=1}} -\frac{R_{\pi(k)}^{T=0}}{N_{\pi(k)}^{T=0}}) \cdot (N_{\pi(k)}^{T=1} + N_{\pi(k)}^{T=0})\\
&\textrm{Qini}(k) = R_{\pi(k)}^{T=1} - R_{\pi(k)}^{T=0}\frac{N_{\pi(k)}^{T=1}}{N_{\pi(k)}^{T=0}}.
\label{uplift}
\end{align}
The uplift and Qini curves can be drawn by varying $k$ in the above equations. The uplift and Qini curves are similar in terms of their shape \citep{Gutierrez2017}.

Despite the prevalence of the uplift and Qini curves in the uplift modelling community, the treatment effect heterogeneity modelling community has not adopted these metrics. Since the ground truth CATEs are often not known in many sociology and medical applications, the uplift and Qini curves could be utilised to alleviate some of the evaluation difficulties encountered in the treatment effect heterogeneity modelling community.

Furthermore, due to ethical and cost concerns, evaluations of CATE estimation methods have been difficult. The current literature often relies on synthetically generated data to obtain ground-truth CATEs. Due to different data generating procedures, such evaluations may produce biased conclusions. Therefore, a standardised and comprehensive set of benchmark datasets would be instrumental in providing fair comparisons of methods and advancements in the field. 
Recently, both communities have made efforts in this direction. For example, the treatment effect heterogeneity modelling community has proposed several benchmarks \citep{Winkel2017,Dorie2019}, and the uplift modelling community has contributed a large-scale online advertising benchmark \cite{DiemertEustacheBetleiArtem}.


\subsection{Demonstration of Semi-Synthetic and Real-World Datasets}
In this section, we show how the methods work using three datasets: two semi-synthetic datasets frequently used in CATE estimation literature, 
and a real-world advertisement campaign dataset frequently used in the uplift modelling literature. Our purpose here is to show how the methods are used in real-world problems, how they are evaluated, and their limitations, instead of assessing which method is better. Since not all methods are applicable to all types of datasets---e.g., most uplift modelling methods cannot be applied to a dataset with  continuous outcomes---we only apply the applicable methods to each of the datasets.

\paragraph{Infant Health Development Program (IHDP)}
The IHDP dataset comes from a randomised study designed to evaluate the effects of home visits from doctors on the cognitive scores of premature infants~\citep{BrooksGunn1992}. 
The program began in 1985, and the subjects of the program were low-birth-weight, premature infants. Subjects in the treatment group were provided with intensive, high-quality childcare and home visits from a trained health-care provider. The program was effective at improving the cognitive function of the treated subjects when compared with the control subjects.

A version of this dataset was first used as a semi-synthetic dataset for evaluating CATE estimation in \citep{Hill2011}, where the outcomes were synthetically generated according to the original covariates, and selection bias was introduced by removing all subjects with non-Caucasian mothers. 
The resulting dataset contained 747 subjects (608 control and 139 treated) with 25 covariates (6 continuous and 19 binary covariates) that described both the characteristics of the infants and the characteristics of their mothers. The methods for generating the synthetic outcomes are described below.

We followed the same procedure as that described in \citep{Hill2011,Johansson:2016:LRC:3045390.3045708,Louizos2017} to replicate two settings of this semi-synthetic dataset, in which the counterfactual outcomes were simulated using the non-parametric causal inference (NPCI) package \citep{Dorie2016}.
Since the covariates are the same as in the original IHDP dataset, the difference between the two settings lies in how the outcomes are simulated. That is, ``Setting A" simulates a linear relationship between the outcome and the covariates, whereas ``Setting B" simulates an exponential relationship. The reported performance was calculated by averaging over 100 replications with a training/validation/test split proportion of 60\%/30\%/10\%.


\newcolumntype{C}{>{\centering\arraybackslash}m{1cm}}
\begin{table}[!t]
	\centering
	\caption{Means and standard errors of PEHE  and MAPE (smaller is better) across 100 replications for training and test sub-datasets of IHDP. The first group of methods are tree-based. ``T-" stands for instantiations of the two-model approach, ``X-" for instantiations of X-Learner, and ``TO-" for instantiations of the transformed outcome approach. The last group of methods are deep learning-based. Tailored uplift modelling methods are not reported, since their implementations are restricted to datasets with binary outcomes. }
	\resizebox{1\linewidth}{!}{
		\begin{tabular}{l | c c c c| c c c c}
			\hline
			&	\multicolumn{4}{c|}{Setting A} & \multicolumn{4}{c}{Setting B} \\
			\hline
			& $\text{PEHE}^{tr}$ & $\text{PEHE}^{te}$ & $\text{MAPE}^{tr}$ (\%) & $\text{MAPE}^{te}$ (\%) & $\text{PEHE}^{tr}$ & $\text{PEHE}^{te}$ & $\text{MAPE}^{tr}$ (\%) & $\text{MAPE}^{te}$ (\%) \\
			\hline
			t-stats &  1.48 $\pm$ 0.12 & 1.56 $\pm$ 0.13 & 48.3 $\pm$ 2.5 & 113.8 $\pm$ 25.3 & 6.92 $\pm$ 0.10 & 5.68 $\pm$ 0.09 & 771.3 $\pm$ 193.4 & 867.7 $\pm$ 127.6 \\
			CT &  1.48 $\pm$ 0.12 & 1.56 $\pm$ 0.13 & 56.5 $\pm$ 4.8 & 148.5 $\pm$ 54.8 & 6.92 $\pm$ 0.10 & 5.70 $\pm$ 0.10 & 631.6 $\pm$ 81.4 & 841.4 $\pm$ 113.5 \\ 
			CF	& 1.01 $\pm$ 0.08 & 1.09 $\pm$ 0.16 & 34.5 $\pm$ 4.1 & 65.6 $\pm$ 16.9 & 2.77 $\pm$ 0.03 & 3.02 $\pm$ 0.03 & 331.9 $\pm$ 87.8 & 436.6$\pm$ 103.0\\
			
			\hline
			T-RF & 0.86 $\pm$ 0.69 & 0.99 $\pm$ 0.09 & 30.5 $\pm$ 3.4  & 56.5 $\pm$ 18.1 & 2.89 $\pm$ 0.03 & 3.15 $\pm$ 0.04 & 426.7 $\pm$ 94.5 & 516.3 $\pm$ 214.9 \\
			T-BART & 0.60 $\pm$ 0.02 &  0.68 $\pm$ 0.04 & 19.1 $\pm$ 1.7 & 34.1 $\pm$ 10.7 & 2.30 $\pm$ 0.03 & 2.51 $\pm$ 0.04 & 302.9 $\pm$ 50.7 & 320.5 $\pm$ 110.8  \\
			\hline
			X-RF & 0.98 $\pm$ 0.08 & 1.09 $\pm$ 0.15 & 42.6 $\pm$ 5.1  & 109.5 $\pm$ 40.3 &  3.50 $\pm$ 0.04 & 3.59 $\pm$ 0.06 & 421.3 $\pm$ 92.3 & 516.3 $\pm$ 216.0 \\
			X-BART& 0.58 $\pm$ 0.02 & 0.66 $\pm$ 0.04 & 18.7 $\pm$ 1.9  & 32.5 $\pm$ 9.6 & 2.29 $\pm$ 0.03 & 2.49 $\pm$ 0.04 & 301.2 $\pm$ 52.3 & 313.5 $\pm$ 114.8\\
			\hline
			TO-RF & 1.01 $\pm$ 0.09& 1.07 $\pm$ 0.10 & 32.8 $\pm$ 3.8 & 80.2 $\pm$ 29.0 & 2.93$\pm$ 0.03 & 3.15 $\pm$ 0.05 & 378.5 $\pm$ 95.3 & 450.3 $\pm$ 98.6 \\
			TO-BART & 0.86 $\pm$ 0.02 &  0.91$\pm$ 0.04 & 25.0 $\pm$ 2.7 & 45.8 $\pm$ 16.0 & 2.40 $\pm$ 0.03 & 2.60 $\pm$ 0.04  & 338.2 $\pm$ 68.9 & 336.6 $\pm$ 138.4  \\
			\hline
			
			CFR	 & 0.67 $\pm$ 0.02 & 0.73 $\pm$ 0.04 & 23.4 $\pm$ 2.5 & 40.6 $\pm$ 13.8  & 2.60 $\pm$ 0.04 & 2.76 $\pm$ 0.04 & 338.2 $\pm$ 68.9 & 385.6 $\pm$ 161.9 \\
			SITE & 0.65 $\pm$ 0.07 & 0.67 $\pm$ 0.06 & 22.3 $\pm$ 2.4 & 38.5 $\pm$ 11.2    & 2.65 $\pm$ 0.04 & 2.87 $\pm$ 0.05 & 338.2 $\pm$ 68.9 & 390.7 $\pm$ 168.0\\
			
			CEVAE & 1.13 $\pm$ 0.07 & 1.37 $\pm$ 0.19 & 46.6 $\pm$ 4.2 & 86.7 $\pm$ 20.8  & 3.06 $\pm$ 0.03 & 3.42 $\pm$ 0.05 & 400.5 $\pm$ 110.8 & 490.8 $\pm$ 135.8 \\
			\hline
	\end{tabular}}
	\label{IHDP}
\end{table}

For the parameters of neural network-based methods, we used a grid search to search for an optimal parameter set that achieved the minimum loss on the validation dataset, which consisted of 30\% of the whole dataset. We then trained the network using the selected parameters on the entire set. For CFR, the parameters we grid-searched included combinations of representation layers ($\{3,4,5\}$), regression layers ($\{3,4,5\}$), pre-representation layer dimensions ($\{100,200,300,400,500\}$), post-representation layer dimensions ($\{100,200,300,400,500\}$), and imbalance regularisation parameters ($\{0,0.001,0.01,0.1, 0.316, 1,3.16, 10\}$). The grid consisted of $1800$ parameter combinations. For CEVAE, we grid-searched combinations of hidden layers ($\{3,4,5\}$), hidden layer dimensions ($\{100, 200, 300, 400, 500\}$), latent factor dimensions ($\{5,10,15,20,25,30\}$), and learning rates (\{0.0001, 0.005, 0.001, 0.05, 0.01\}). This grid consisted of $450$ parameter combinations. 
For all methods based on an ensemble of trees (i.e., random forests and BART-based algorithms) the best number of trees was selected from $100$ to $1000$ in intervals of $100$. 
For tree-based methods, pruning was conducted by cross-validation implemented within the software packages.



The results of the methods on the two settings of the IHDP datasets are shown in Table \ref{IHDP}.
For Setting A, we note that at least one method in each group performs well, and we cannot categorically say which group of methods is better than others. 
From the application viewpoint, about half of the methods produce some useful models, since their MAPEs are around 50\% or less, and a quarter of methods do not produce useful models, since their MAPEs are more than 100\%. This further shows that CATE estimation is very challenging, considering that the underlying relationships between CATEs and the covariates are linear.
For Setting B, we can see that the performance of all the methods is not satisfactory, since the smallest MAPE is larger than 300\%. 
This can be explained by the fact that the sample size of the IHDP semi-synthetic datasets is rather small (747 samples), and thus it would be difficult for the non-parametric recovery of the non-linear exponential relationships between the CATEs and the covariates.

\begin{figure}[!t]
	\begin{adjustbox}{minipage=\linewidth,scale=1}
		\begin{subfigure}{.48\textwidth}
			\includegraphics[width=\linewidth]{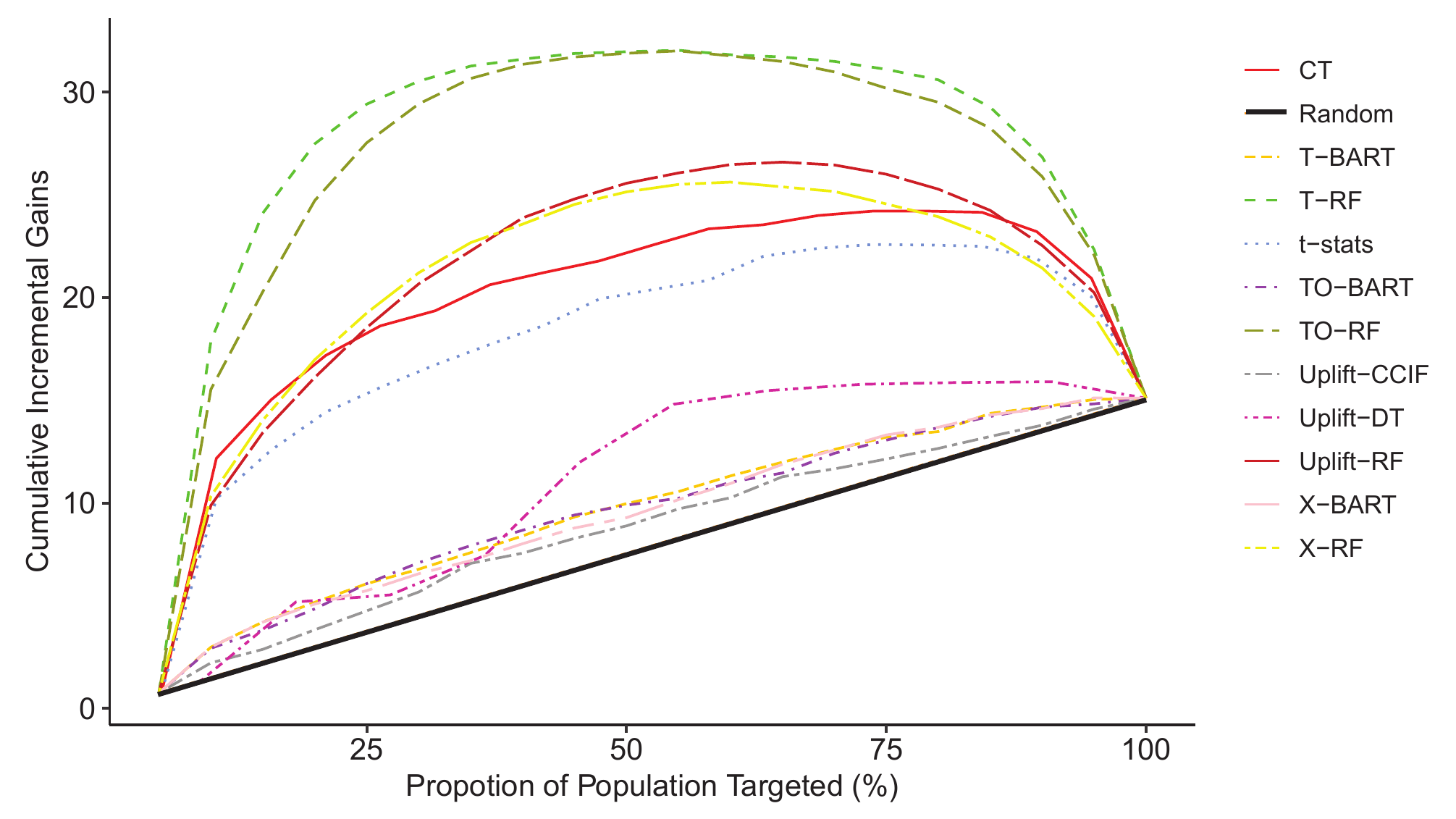}
			\caption{Uplift curves on the training data.}
		\end{subfigure}	
		\begin{subfigure}{.48\textwidth}
			\includegraphics[width=\linewidth]{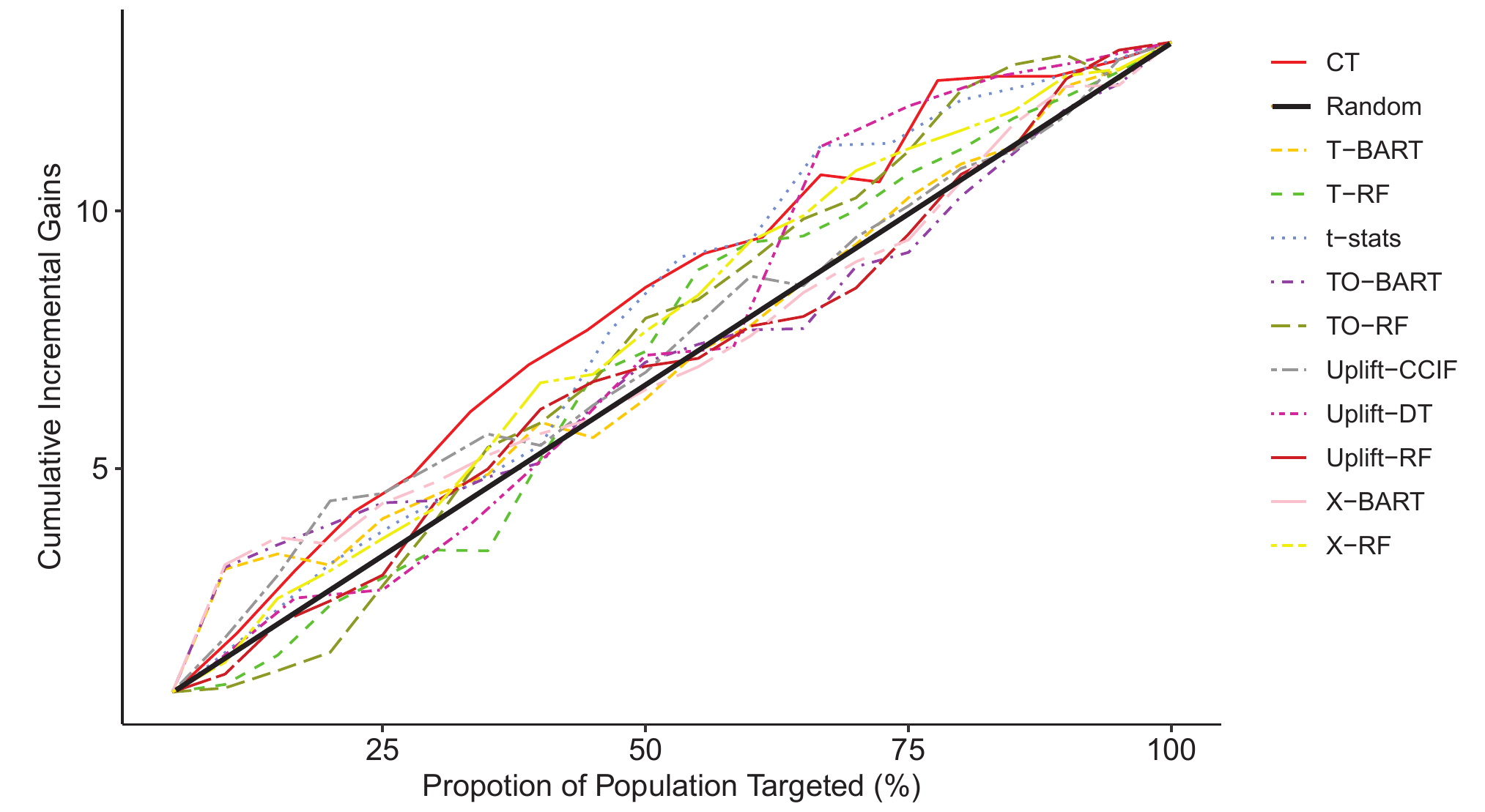}
			\caption{Uplift curves on the test data.}
		\end{subfigure}
		\\		
		\begin{subfigure}{.48\textwidth}
			\includegraphics[width=\linewidth]{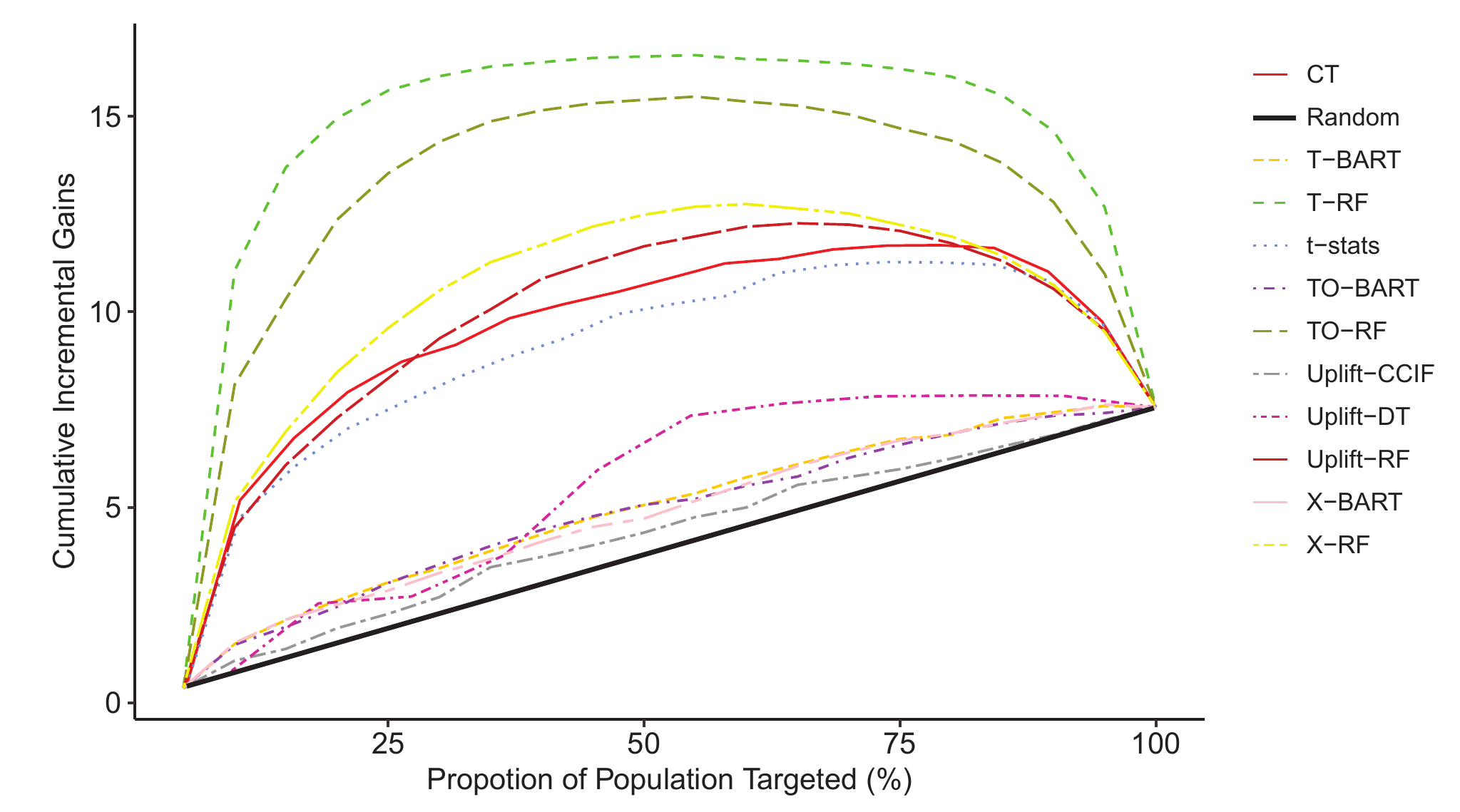}
			\caption{Qini curves on the training data.}
		\end{subfigure}	
		\begin{subfigure}{.48\textwidth}
			\includegraphics[width=\linewidth]{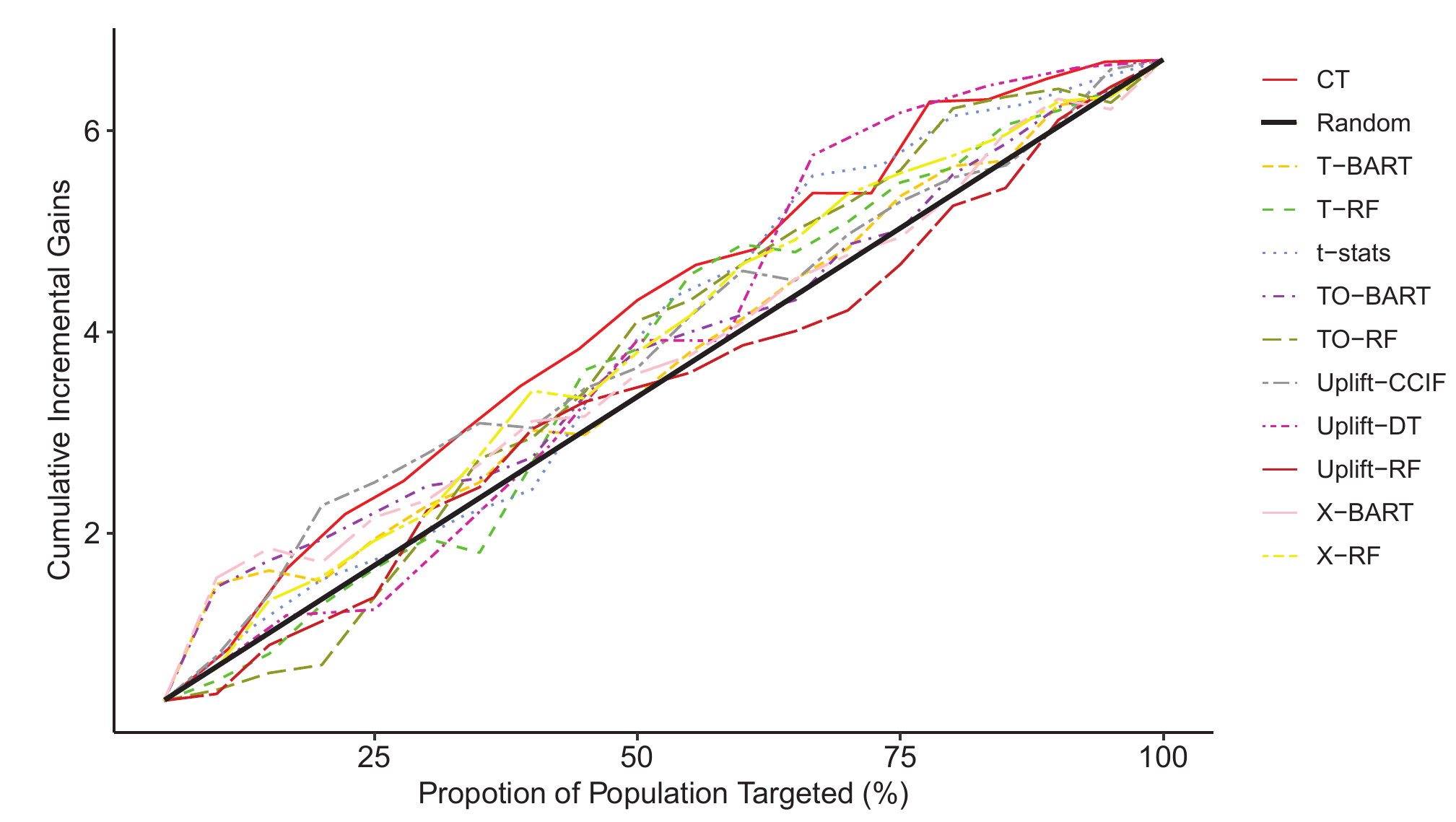}
			\caption{Qini curves on the test data.}
		\end{subfigure}
	\end{adjustbox}
	\caption{Uplift and Qini curves for the compared methods on the training and test sets of the Hillstrom's Email Advertisement dataset. The black lines indicate the curves by random predictions. ``T-" stands for instantiations of the two-model approach, ``X-" for instantiations of X-Learner, and ``TO-" for instantiations of the transformed outcome approach. Figures are best viewed in colour. }
	\label{fig_email}
\end{figure}

\paragraph{Hillstrom's Email Advertisement Dataset}

Here we use the Hillstrom's Email Advertisement dataset \cite{Hillstrom2008} (as discussed in Section \ref{Marketing}) to illustrate an uplift modelling example. We used the ``men's advertisement email'' as the treatment and the visit status as the outcome. 
The algorithms showed similar performance when using the ``women's advertisement'' as the treatment.
 
On average, the ``men's advertisement'' treatment increased the ``visit'' outcome by 7.6\%. 
We ran the algorithms using a 70\%/30\% split of training and test sets without repeats, and we used the same parameter selection procedure as that described in the last section.
Note that the source codes for CF, CFR, and CEVAE are not designed to handle categorical covariates. We tried transforming the categorical covariates into binary codes with one-hot encoding. However, the results were poor, and thus we have not included them here.


The uplift curves and the Qini curves of the evaluated methods on the Hillstrom Email dataset are illustrated in Figure ~\ref{fig_email}. 
Firstly, the majority of  models do predict uplifts in the test dataset, but the uplifts in the test dataset are much smaller than the uplifts in the training dataset. 
The rankings of the methods based on the uplifts between the training dataset and the test dataset are also inconsistent. The above two observations indicate difficulty in evaluating the performance of uplift modelling when the ground truth is unknown (which is common in practice). 
Cross-validation is often used in evaluating an uplift model, but it is an open question whether cross-validation is a valid means, since the uplifts are unobserved in the test datasets. In contrast, for evaluating a supervised method, the outcomes are observed in the test dataset. 
Secondly, the trends of uplift curves and Qini curves differ slightly, although they are largely consistent. For example, X-RF is better than Uplift-RF in the uplift curves, but worse than Uplift-RF in the Qini curves. 

The illustrations of the IHDP and Hillstrom's Email Advertisement datasets serve as a good example of cross-fertilisation between the two communities.
Almost all methods developed in the uplift modelling community are designed for binary outcomes, whereas most methods from the treatment effect heterogeneity modelling community are designed for continuous outcomes. 
On the one hand, suppose the uplift needs to be modelled for the amount of sales increase instead of purchasing or not purchasing, methods in the treatment effect heterogeneity modelling community can be directly applied to solve this extended uplift modelling problem.
On the other hand, uplift modelling methods can also be applied to sociology and medical problems when the outcomes of interest are binary, e.g., whether a medicine is effective for patients with specific gene mutations.
\begin{figure}[!t]
	\centering
	\begin{adjustbox}{minipage=\linewidth,scale= 1}
		\begin{subfigure}{.48\textwidth}
			\includegraphics[width=\linewidth]{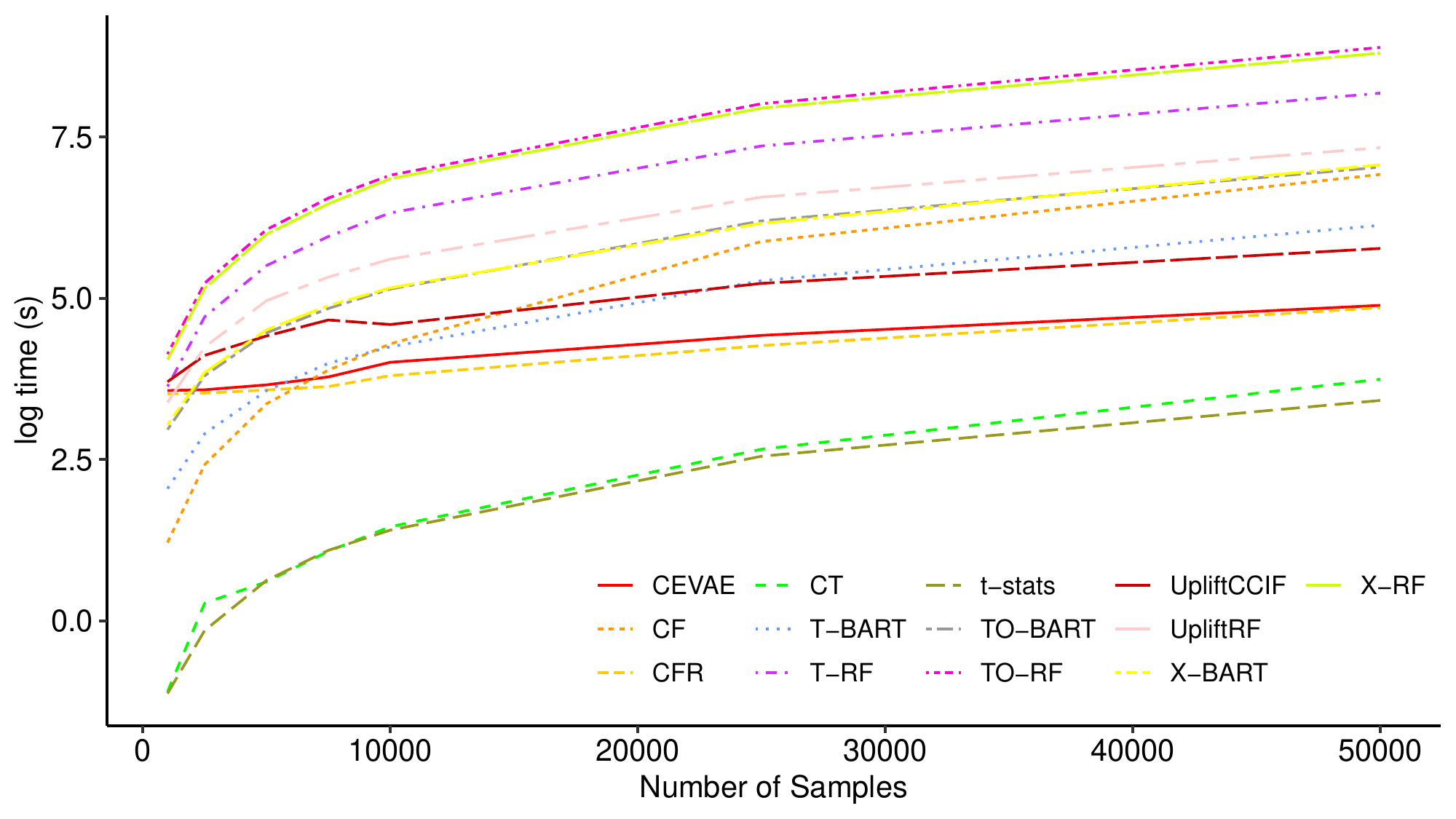}
			\caption{Number of subjects.}
		\end{subfigure}	
		\begin{subfigure}{.48\textwidth}
			\includegraphics[width=\linewidth]{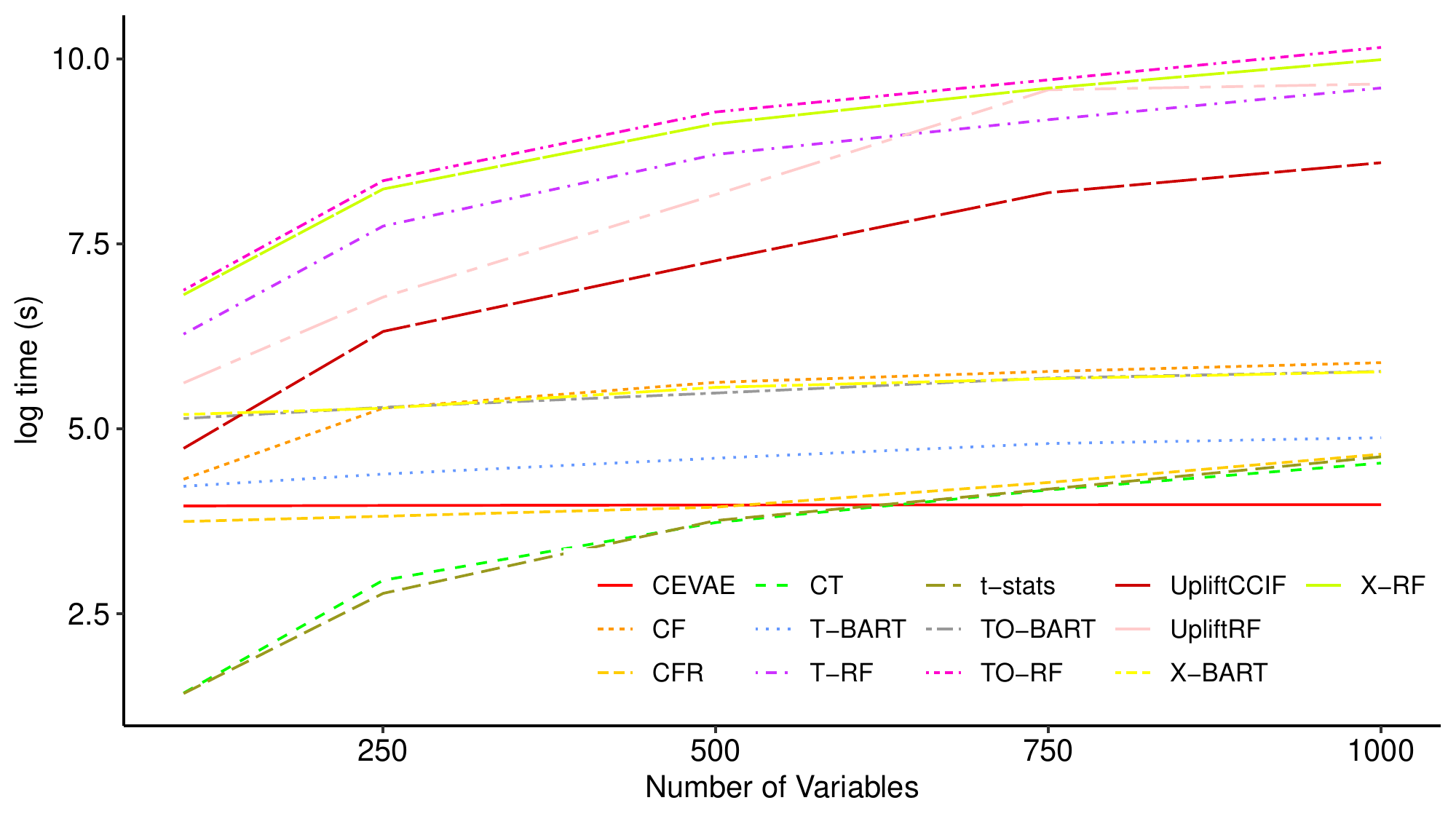}
			\caption{Number of covariates.}
		\end{subfigure}
	\end{adjustbox}
	\caption{Running time of the compared methods. The curves for CFR and SITE are similar. We only show CFR, since one is hidden behind the other. The $Y$ axis is in logarithmic form. ``T-" stands for instantiations of the two-model approach, ``X-" for instantiations of X-Learner, and ``TO-" for instantiations of the transformed outcome approach. Figures are best viewed in colour. }
	\label{fig_scalability}
\end{figure}
\subsection{Scalability}
As datasets get larger, it becomes imperative to understand how the CATE estimation and uplift modelling methods scale.
We compared the running time of different methods by varying the numbers of subjects and covariates. The comparison was conducted using an AMD Ryzen 3700x CPU and 32 GB of RAM. For deep learning-based methods, the running time was obtained using a single Nvidia GeForce GTX 1080 Ti GPU with 11 GB of RAM. For GPU training, the batch sizes were set to one-fifth the number of subjects, and a total of $200$ epochs of training were performed. 
Each experiment was repeated 10 times and the average running time was reported.

We utilised a synthetic dataset proposed in \citep{Haeggstroem2017}. In this dataset, the causal structure among the $10$ covariates, the treatment 
$T$, and the outcome $Y$ is depicted in \cite{Haeggstroem2017}. 
To evaluate the scalability with regard to the number of variables, additional variables (which are not related to the treatment or the outcome) were randomly sampled from Gaussian distributions and added to the covariates. As a result, the number of variables varied from 10 to 1000 with the number of subjects fixed at 10,000. 
For evaluating scalability with regard to the number of subjects, we fixed the number of variables at 100, and varied the number of subjects from 1000 to 50,000.
Multiple CPU parallelisation was not used for the ensemble-based algorithms. In other words, the time efficiency for ensemble-based methods can be improved by utilising multiple CPUs.


The running time comparison for different numbers of subjects is shown in Figure \ref{fig_scalability}(a). The two tree-based methods, causal tree (CT) and t-statistics tree (t-stats), are the fastest among all the compared algorithms. The representative deep learning-based algorithms CEVAE and CFR use almost constant time regardless of the number of subjects. This means that they are efficient when the number of subjects is large. 
The slowest running methods are those based on BART, including X-BART, T-BART, and TO-BART. Methods based on random forests, i.e., causal forest (CF), X-RF, T-RF, and TO-RF, are faster than the deep learning-based algorithms when sample sizes are small, but become slower when the numbers of subjects increase. 

The comparison of different numbers of covariates is shown in Figure \ref{fig_scalability}(b). The two tree-based methods, CT and t-stats, are the fastest. Deep learning-based algorithms CEVAE and CFR also use almost constant time regardless of the number of variables. They can handle datasets with a large number of variables very well. 
The slowest running methods are those based on random forests, including X-RF, T-RF, TO-RF, and UpliftRF. The BART-based algorithms take similar time, and they are faster than RF-based algorithms but slower than deep learning-based algorithms.

\begin{table}[!t]
	\centering
	\caption{Summary of the usability, interpretability, and scalability of the uplift modelling and CATE estimation packages and algorithms. }
	\resizebox{0.98\linewidth}{!}{\begin{tabular}{|p{2.5 cm} |p{3 cm} | p{3.5 cm} | p{4.5 cm}| }
			\hline
			\multicolumn{1}{|c|}{\textbf{Methods}} & \multicolumn{1}{c|}{\textbf{Usability}} & \multicolumn{1}{c}{\textbf{Interpretability}} & \multicolumn{1}{|c|}{\textbf{Scalability}} \\
			\hline
			\textbf{Single-model} & \multirow{4}{3 cm}{Depends on the base method, and can be very easy to use. } & \multirow{4}{3.3 cm}{Treatment effect is not given and needs to be derived. Results are interpretable.} & \multirow{4}{4 cm}{Scalability to the \#covariates and \#subjects can be very good. } \\
			
			\textbf{Two-Model}  &  &  &  \\
			
			\textbf{X-Learner} &  &  & 	\\
			{\textbf{Transformed outcome}} &  &  & 	\\
			\hline
			\textbf{Causal Tree} &  \multirow{3}{3 cm}{Easy to use with few parameters to set. } & \multirow{3}{3 cm}{Treatment effect is directly given and interpretable.} & \multirow{3}{4 cm}{Scalability to the \#covariates*  and \#subjects is very good.}	\\ 
			
			\textbf{t-stats Tree} &  &  & 	\\
			
			\textbf{Uplift Tree} &  &  &  \\
			\hline
			\multirow{2}{3 cm}{\textbf{Causal RF}} & \multirow{4}{3 cm}{Easy to use with few parameters to set. } & \multirow{4}{3 cm}{Treatment effect is directly given. Results are not interpretable. } & \multirow{4}{4 cm}{The time for building a model can be long, and the scalability to \#covariates and \#subjects is not good. } \\
			& & & \\	
			\multirow{1}{3 cm}{\textbf{Uplift RF}} &  & 	&  \\
			\multirow{1}{3 cm}{\textbf{Uplift CCIF}} &  & 	&  \\
			\hline
			\multirow{2}{3 cm}{\textbf{CFR}} & \multirow{5}{3 cm}{The network structure and parameters are difficult to set. } & \multirow{5}{3 cm}{Treatment effect is directly given. Results are not interpretable. } & \multirow{5}{4 cm}{It takes a long time to train a model, but the training time is constant and unaffected by \#covariates or \#subjects. Good for big  data.} \\
			& & & \\	
			\textbf{CEVAE} &  &  &   \\
			\multirow{2}{3 cm}{\textbf{SITE}} & & & \\	
			&  &  &   \\
			\hline
			\multicolumn{4}{l}{$^*$When using propensity score matching, the scalability with \#covariates is not good. }
	\end{tabular}}
	\label{tab_summary}
\end{table}

\subsection{Discussions}

Table~\ref{tab_summary} summarises the usability, interpretability, and scalability of the methods implemented in the packages and codes listed in Tables 1 and 2. Usability is about the ease of use of the implementation of a method in the packages or codes, i.e., parameter setting and tuning.  Interpretability refers to the level of explanation provided for a prediction, for example, why a predicted outcome may be linked to some specific covariate value. 
Scalability is about the speed of a method in relation to the dataset size and number of covariates.   
Based on our experience with various uplift modelling and treatment effect heterogeneity modelling methods, we have the following observations:
\begin{itemize}
	\item The accuracy of the methods is data-specific. For example, X-Learner BART performed well with the IHDP dataset, but only performed marginally better than the baseline for Hillstrom's Email Advertisement dataset. Currently, there is no good understanding of which method suits the best type of data.    
	\item The accuracy of ensemble methods is stable across different datasets, although they may not be the best all the time. This is expected, as discussed in \citep{Soltys2015} where the authors observed that ensemble methods frequently outperformed methods that build a single model. However, ensemble methods suffer from a lack of interpretability and their training time is long. 
	\item The accuracy of the single-model approach is low, while the accuracies for the two-model approach and the X-Learner are competitive. Based on their design, the two-model approach is good for datasets with balanced treatment and control subjects, and the X-Learner is good for unbalanced data. A strength of both the X-Learner and the two-model approaches is that they provide good flexibility for using a rich set of existing supervised methods.
	\item The performance of deep learning-based methods is determined by dedicated parameter tuning for each specific dataset. Neural networks are known to be prone to data variability and parameter selection~\citep{Novak2018}. This is more problematic in CATE estimation than in supervised learning, since there are no ground-truth treatment effects available in most cases. 
\end{itemize}

In this survey, we focus our discussion on a single binary treatment and exclude the discussion of multiple treatments and non-binary treatments (i.e., ordinal or continuous treatments).
We focus on a single treatment, since almost all of the surveyed methods can be extended to multiple treatments, in a manner similar to extending binary classification to multiple classes. There are some methods specifically designed for handling multiple treatments. We refer the reader to Olaya, Coussement, and Verbeke \citep{Olaya2020} for a recent survey on uplift modelling with multiple treatments. 
We exclude a discussion of non-binary treatments, since most existing methods are designed for binary treatments. Recently, several methods for non-binary treatments have been proposed from the treatment effect heterogeneity modelling community, and we refer the reader to Zhao, Dyk, and Imai \citep{Zhao2020} for an overview of these methods. 


\section{Conclusions}
Estimating the heterogeneous effects of an action on the outcomes of individual subjects is an important problem with a wide range of applications. 
Motivated by different applications, researchers from the treatment effect heterogeneity modelling and the uplift modelling communities have both contributed to the problem.
In this article, we provided a unified survey of the methods proposed by the two communities. Using the potential outcome framework, we showed that, with the overlap, SUTV, and unconfoundedness assumptions, the objectives of treatment effect heterogeneity modelling and uplift modelling are the same: they both aim to estimate the conditional average treatment effect (CATE) from data. 
With a unified objective and notation, we systematically reviewed the methods developed by the two communities, focusing on the inherent connections between them. 
Although most methods are strongly linked to supervised machine learning methods, we should stress here that CATE estimation is not supervised learning, and that it is crucial to understand the assumptions to ensure the correct use of the methods.
We discussed applications of the methods in targeted marketing, personalised medicine, and social studies. Finally, we summarised existing open-source software packages and source codes, and demonstrated their implementation with synthetic, semi-synthetic, and real-world datasets. We showed that CATE estimation and evaluation are challenging tasks, and we offered some general guidelines for method and software-tool selection based on our experience.  

An important direction for future work is to find balance between the interpretability, ease of use, and accuracy of CATE estimation algorithms. 
On the one hand, black-box methods such as deep learning-based methods are accurate at CATE estimation with carefully tuned parameters. However, their parameter tuning procedures are difficult, and the models are generally not interpretable. 
On the other hand, tree-based methods are easy to interpret and require minimal parameter tuning, although their performance is often limited. 
An emerging direction for bridging this gap is to construct tree-based models on top of the results from other CATE estimators \citep{Lee2020}.


\bibliographystyle{ACM-Reference-Format}
\bibliography{mybib,RefCausalClassification}

\end{document}